# Heterogeneous superconductivity emerging from a charge-stripe state in kagome lattices

Hongyu Li[1,*], Shuikang Yu[2,*], Huachen Rao[1], Zhuying Wang[2], Xingchen Zhou[2], Houpu Li[2], Yunmei Zhang[2], Zhao Gong[1], Xin Yu[3], Rui Wu[3], Tao Wu[1,4], Zhenyu Wang[2,4], Ziji Xiang[1,4], Xianhui Chen[1,2,4]

[1] *Hefei National Research Center for Physical Sciences at the Microscale, University of Science and Technology of China, Hefei, 230026, China*

[2] *Department of Physics, University of Science and Technology of China, Hefei, 230026, China*

[3] *Beijing National Laboratory for Condensed Matter Physics and Institute of Physics, Chinese Academy of Sciences, Beijing 100190, China*

[4] *Hefei National Laboratory, University of Science and Technology of China, Hefei, 230088, China*

* These authors contributed equally: Hongyu Li, Shuikang Yu

e-mail: zywang2@ustc.edu.cn, zijixiang@ustc.edu.cn, chenxh@ustc.edu.cn

**Charge density wave (CDW) and superconductivity have long been recognized as competing orders. However, recent achievements establish that the interplay between these two collective electronic phenomena can be more perplexing, with emergent phenomena arising from their cooperation. Here we report a distinctive case of the kagome superconductors $CsV_3Sb_{5-x}Sn_x$, in which the characteristics of superconductivity evolve in accordance with the form of CDW order. As the triple-*Q* CDW in $CsV_3Sb_5$ transforms to a short-range stripelike CDW at $x_c = 0.07$, a sudden change in superconducting properties occurs concomitantly. A combination of thermodynamic, electrical transport and scanning tunneling microscopy experiments reveals a heterogeneous superconducting state beyond $x_c$: the superconducting gap exhibits pronounced spatial variation, in conformity with a distribution of critical temperature and the resulted superconducting precursor effects. Moreover, abundant normal fluid persists deep into the superconducting state, as evidenced by remnants of electronic specific heat and zero-energy quasiparticle interference. These observations are unrelated to the chemical inhomogeneity and residual strain effect introduced by Sn substitution. Instead, they emulate the phenomenology that a sign-changing superconducting order parameter is subject to strong pair-breaking scattering with the short-range charge stripes and hence reflect the unconventional nature of superconductivity developing inside the charge-stripe state.**

## Introduction

The phase diagrams of quantum materials are often delineated by the presence of multiple electronic orders. Charge density wave (CDW) and superconductivity, two typical symmetry-breaking electronic instabilities, compete with each other according to conventional wisdom. Recent studies, however, have revealed more complicated relationships between these orders in many categories of superconductors [1-6]. In high-temperature cuprate superconductors, short-range CDW ubiquitously coexists with superconductivity [2-4], and dynamic CDW correlations may be involved in *d*-wave pairing [7]; in certain members such as underdoped $YBa_2CuO_{6+\delta}$ (YBCO), long-range CDW order can be induced by intense magnetic fields ($H$) [8,9] and an anomalous vortex liquid phase with "resilient" superconducting behavior arises from its coupling with superconductivity [10,11]. Moreover, the intertwinement between CDW and superconductivity promotes exotic pairing states such as pair density wave (PDW) in both high-temperature [1,12,13] and conventional [5,14] superconductors. The recently discovered kagome-lattice superconductor $CsV_3Sb_5$ [15] serves as an ideal candidate for exploring the complex interplay between electronic orders. It exhibits a triple-$Q$ CDW state with an onset temperature $T_{CDW}$ = 94 K [15]; the CDW adopts an in-plane 2×2 superlattice and mixing 2×2×2 and 2×2×4 stacking along $c$ axis [16-18]. Superconductivity develops inside the CDW state with a transition temperature $T_c \approx$ 3.0 K, characterized by an anisotropic full superconducting gap ($\Delta$) [19-21].

The coexisting superconductivity and CDW in $CsV_3Sb_5$ can be effectively manipulated via applying pressure ($p$) or chemical doping. The $T$-$p$ phase diagram is marked by two critical pressures of $P_{c1} \approx 0.7$ GPa and $P_{c2} \approx 2.0$ GPa [22,23]; at $P_{c1}$, the triple-$Q$ CDW is replaced by a unidirectional (single-$Q$) form thus called a stripelike order, which is fully suppressed at $P_{c2}$ [23,24]. $T_c$ is enhanced at each transition, rendering a double-peak superconducting dome in the $T$-$p$ phase diagram [22,23]. Substituting the V atoms in the kagome plane with Ti, Nb or Ta can monotonically suppress the triple-$Q$ CDW and boost $T_c$, eventually leading to an isotropic superconducting gap in the undistorted kagome phase [25-27]. By contrast, hole doping via substitution of Sn onto the in-plane Sb sites produces a phase diagram mimicking the $T$-$p$ phase diagram, underscored by the double-dome superconductivity and the presence of single-$Q$ stripelike CDW beyond the minima of $T_c$ [28-30] (Fig. 1a-c). Intriguingly, the pressure-induced and Sn-doping-induced charge stripe orders have similar periodicities; the former is a long-range order [24], whereas the latter is short-ranged (with an anisotropic correlation length $\xi_{CDW}$ ~ 6-18 nm at $x$ = 0.15 [29]) and gradually becomes fluctuating upon increase of $x$ [30,31]. The nonmonotonic pressure/Sn-doping dependence of superconductivity and its correspondence to the transformations of CDW highlight the intimate link between the two orders, strongly implying that their involved mutual relation is not simply competitive.

In this work, we utilize complementary probes of electrical transport, torque magnetometry, specific heat and scanning tunneling microscopy (STM) to explore the evolution of superconductivity upon Sn doping in $CsV_3Sb_{5-x}Sn_x$ up to $x \approx 0.30$ (where the charge stripes are likely of the fluctuating type). A peculiar heterogeneous superconducting state possessing (i) a distribution of local $T_c$, (ii) a spatially inhomogeneous $\Delta$ and (iii) a large population of unpaired electrons is identified, emerging uniquely from the (short-range) single-$Q$ CDW phase. These unusual signatures are irrelevant to the chemical disorder and residual strain effect from Sn doping, but are most likely to stem from the strong influence of pair-breaking scattering with the presence of short-range charge stripe [32]; such intense nonmagnetic pair breaking is in stark contrast to the expectation for conventional Bardeen-Cooper-Schrieffer (BCS) superconductors [33], instead being a hallmark of sign-reversal superconductivity. A phenomenological resemblance between this state and the overdoped high-$T_c$ cuprates [34-37] underpins the realization of unconventional superconductivity in $CsV_3Sb_{5-x}Sn_x$ in the intermediate doping range where the static stripelike CDW prevails.

## Results

### The electronic phase diagram of $CsV_3Sb_{5-x}Sn_x$

Single crystals of $CsV_3Sb_{5-x}Sn_x$ ($x$ = 0, 0.06, 0.07, 0.09, 0.12 and 0.30) were grown via flux method (Methods and Supplementary Note 1). A $T$-$x$ phase diagram (Fig. 1b and c) is constructed based on a combination of electrical resistivity ($\rho$), magnetic susceptibility ($\chi$) and specific heat ($C$) measurements. With the inference from previous results [29-31], three phase regions can be identified: (I) a long-range triple-$Q$ CDW phase up to $x$ = 0.06 (the turquoise shaded region in Fig. 1b), characterized by the in-plane 2×2 superlattice (Fig. 1a, left), (II) a single-$Q$ CDW phase featured by short-range [29], static stripelike CDW with periodicity $\lambda \sim 3a_0$ [30] ($a_0$ is the in-plane lattice parameter, see Fig. 1a, right), commencing at $x = 0.07$ (the orange shaded region in Fig. 1b), which crossovers to (III) a phase regime harboring very short-ranged single-$Q$ charge correlation, developed at $x \geq 0.12$, as illustrated by a fade-out of the orange shades in Fig. 1b. The CDW-transformation transition between phases I and II, occurring at $x_c = 0.07$, is concomitant with a sudden drop in $T_c$: the zero-resistivity onset ($T_{c0}$) decreases from 3.3 K for $x = 0.06$ to 1.2 K for $x = 0.07$ (Fig. 1d and g), underlining the strong influence of charge order on superconducting pairing. The CDW transition in phase II is ostensibly absent in thermodynamics [28, 29], in accordance with the weak structural impact of the short-range stripelike order; nonetheless, $T_{CDW}$ can be determined from an anomaly in $d\rho/dT$ (Supplementary Fig. 3b) and the unidirectional band folding due to the stripelike CDW is clearly manifested in the quasiparticle interference (QPI) in this phase [30]. With further increase of $x$, the correlation length of stripelike CDW becomes even shorter, resulting in fluctuating-type charge order [31] in phase III, wherein both the CDW transition in $d\rho/dT$ and the QPI features of the stripe-induced electronic band reconstruction are unresolvable (Supplementary Note 2). As such, we infer that the electronic structure in phase III is almost identical to the undistorted kagome lattice. The overall evolution of superconductivity upon Sn doping is nonmonotonic, with pronounced suppression of $T_{c0}$ occurring only in the narrow phase region II ($x = 0.07$-$0.09$) where static short-ranged charge stripe order that suffices to cause band reconstruction (Fig. 1b and c); a higher $T_{c0} = 2.8$ K is recovered at the onset of phase III at $x = 0.12$.

**Anomalous superconducting state in stripe-ordered $CsV_3Sb_{5-x}Sn_x$**

A comparison of $\rho(T)$, $\chi(T)$ and $C(T)$, summarized in Figs. 1d-o (see Supplementary Fig. 3 for data of undoped $CsV_3Sb_5$ and the $x = 0.30$ sample), further divides the six samples into two groups with distinct superconducting state properties. The first group includes members in phase regimes I ($x = 0, 0.06$) and III ($x = 0.12, 0.30$). Their superconducting states behave ordinarily: $T_{c0}$ coincides with the onsets of both the Meissner diamagnetism ($T_d$) and the specific heat anomaly ($T_c^s$), signaling the occurrence of a single transition at this critical temperature (Fig. 1d-f, m-o, Supplementary Fig. 3c-h). On the contrary, superconducting states in the stripe-ordered compounds in phase II ($x = 0.07, 0.09$) show remarkably differentiated characteristic temperatures for the three physical quantities (Fig. 1c). Concretely, for $x =$ 0.07 (0.09), diamagnetism appears at a relatively high $T_d \approx 2.4$ K (3.5 K) and develops a drastic enhancement at $T_c^s = 1.7$ K (2.1 K) where the transition in $C(T)$ occurs, yet $\rho$ becomes zero only at $T_{c0}$ = 1.2 K (2.0 K), as displayed in Fig. 1g-i (Fig. 1j-l). In addition, the resistive transitions in these two compounds are considerably broadened compared with the first group (Supplementary Fig. 3a); a nonohmic voltage-current ($V$-$I$) relation, manifested as a current-dependent $\rho(T)$ profile, is observed at the transition and extends to $\simeq$ 1 K above $T_{c0}$ (Fig. 1g and j).

The unusual superconducting behavior uniquely develops in phase region II cannot be assigned to the phase separation prompted by chemical inhomogeneity (e.g., clustering of Sn dopants) or residual strain fields, considering its immediate disappearance at $x = 0.06$ and $x = 0.12$. This inference is further confirmed by the STM results discussed below. Filamentary superconductivity [38] is also irrelevant: The formation of superconducting filaments typically renders $T_{c0}$ higher than the onsets of thermodynamic transitions ($T_d$ and $T_c^s$). Intriguingly, there are a few phenomenological parallels between these nonmagnetic, weakly-correlated superconductors and the high-$T_c$ cuprates that employ similar signatures: (1) The large disparity between $T_{c0}$ and $T_d$ (Fig. 1c), indicative of appreciable precursor superconducting diamagnetism ($\chi_d$) [35,39,40]; (2) the rounded onset of resistive transition resulted from the excess conductivity (paraconductivity) $\Delta\sigma$ introduced by superconducting precursor [41-43] and (3) the nonohmic resistivity drop, which is reminiscent of the "fragile" superconductivity [10,44] emerging due to an intricate compromise between superconductivity and CDW order [11].

**Percolative superconducting precursor**

Two possible origins of the superconducting precursors, i.e., fluctuating Cooper pairs [39,41,45] and percolation process [39,40,42,43], have been scrutinized in previous studies (mostly in high-$T_c$ superconductors). The former signifies fluctuations of the superconducting order parameter, happening in both its amplitude (Gaussian fluctuation) and phase [41,45,46]; the latter considers a spatial distribution of the superconducting order parameter and, consequently, $T_c$ [32,35,42]. In the percolation scenario, the development of superconductivity upon cooling is an incremental process, i.e., the growth and eventual proliferation of superconducting clusters embedded in the normal metal medium [42,43]. The two contributions can be distinguished by their $T$ evolutions. Gaussian-type fluctuations engender an approximately power-law $T$ dependence of $\Delta\sigma$ and $\chi_d$ (the Aslamazov-Larkin expressions, Supplementary Note 4) within the framework of Ginzburg-Landau (GL) theory [45], whereas the phase fluctuations are amenable to analysis in the Halperin-Nelson formalism which considers vortex-like excitations [41,46] (Supplementary Note 4). By contrast, the percolation process, commonly described using the effective-medium theory (EMT), is characterized by the exponential decay of $\Delta\sigma$ and $\chi_d$ with increasing $T$ [39,40,42,43]. We employed these approaches for superconducting fluctuations and the percolation scenario to our data for $CsV_3Sb_{5-x}Sn_x$ ($x$ = 0.07 and 0.09), as denoted by the fitting curves and simulations presented in Fig. 2a-d; the superconducting precursor herein unambiguously stems from the percolation process between $T \simeq (1.0\text{-}1.2)T_c^s$ and $\simeq 3.5$ K. In this temperature range, both $\Delta\sigma$ and $\chi_d$ decrease in a roughly exponential fashion with increasing $T$, a trend satisfyingly captured by the EMT approach (Methods, more detail in Supplementary Note 5) assuming a significant distribution width of $T_c$ ($\Xi_1 \simeq 1.2$ K and 0.6-0.7 K for $x$ = 0.07 and 0.09, respectively, Supplementary Note 5 and Supplementary Fig. 5c and d). Conversely, neither the Aslamazov-Larkin model (orange curves) nor the Halperin-Nelson model (wine curves) succeeds in reproducing the amplitude and $T$ dependence of the precursor superconducting effects (Fig. 2a-d). Hence, we conclude that the fluctuation effect is negligible in $CsV_3Sb_{5-x}Sn_x$ compared with the influence of the local $T_c$ distribution.

The presence of self-organized puddles wherein the superconductivity is more persistent than their surroundings is further confirmed by contrasting the profiles of magnetoresistance (MR) and magnetization $M(H)$. In most superconductors, the hysteresis loop of diamagnetism arising from shielding currents only occurs within the vortex solid phase, in which the resistivity is essentially zero at small excitations [47]; the irreversibility field $H_{irr}$ thus approximately coincides with the onset of nonzero $\rho(H)$. For $CsV_3Sb_{5-x}Sn_x$ with $x$ = 0.09 (Fig. 2e and f) and 0.07 (Supplementary Fig. 5e and f), however, the magnetization hysteresis closes at a position where MR almost reaches its normal-state values. The concurrence of irreversible diamagnetism and heavily dissipative transport strongly implies a picture of heterogeneity that disconnected superconducting grains scatter inside a metallic environment. As such, the nonohmic resistivity drop (Fig. 1g and j) reflects current-induced breaking of superconducting inter-grain links [48].

Our two-component EMT modeling (Methods), though being oversimplified, provide a good depiction of the paraconductivity as shown in Fig. 2a and c (note that the EMT analysis does not account for the characteristic size of the superconducting grains, Supplementary Note 5). Based on this, we propose a phenomenological understanding (Fig. 2g-i) of how superconductivity emerges in the stripe-ordered $CsV_3Sb_{5-x}Sn_x$: Small-sized superconducting grains appear at $T \gtrsim T_d$ (Fig. 2g) and grow in a percolative manner with decreasing $T$; at $T \simeq T_c^s$, proximity effect (depicted by a second component of distribution with scale $\Xi_2$ in the EMT analysis, see Methods and Supplementary Fig. 5a) drives a large portion of the surrounding medium superconducting (Fig. 2h), giving rise to the drastic enhancement of $\chi_d$ and onset of bulk transition in $C(T)$; a superconducting path across the sample (Fig. 2i) is eventually established at $T_{c0}$. It should be noticed that before the $\Xi_2$ term kicks in, the $\Xi_1$ distribution only contributes to about 0.1% of the total volume fraction (Supplementary Fig. 5b), hence the high-$T_c$ "tail" may evade the detection of many experimental probes.

**Visualization of inhomogeneous superconductivity**

Spectroscopic-imaging STM probes the spatial characteristics of electronic orders, with which the proposed heterogeneous superconductivity as well as its potential origin can be revealed. To this end, we performed dilution-refrigerator-based STM studies for $x = 0.09$ with a lattice temperature of 20 mK. Figure 3a shows the topography of a 70 nm × 10 nm antimony surface, where the atomic position of individual Sn dopant, substituted for the in-plane Sb atom, is clearly identified (Fig. 3b). The statistically homogeneous distribution of Sn dopants is remarkable: we did not find any propensity for them to form clusters; their concentration was appreciably uniform across seven measured samples (Supplementary Note 1). The normal-state electronic density of states (DOS), reflected by integrating the differential conductance (d$I$/d$V$) between energy windows of ±6 (Fig. 3c) and ±50 meV (Fig. 3d), shows no significant spatial inhomogeneity (also see Supplementary Fig. 2). However, the superconducting characteristics, provided by the gap amplitude $\Delta(\mathbf{r})$ and the gap depth $H(\mathbf{r})$ (defined Fig. 3j, see also Methods and Supplementary Note 6), exhibit pronounced real-space variations (Fig. 3e and f). The gradual variation observed for both $\Delta(\mathbf{r})$ and $H(\mathbf{r})$ marks a clear departure from the homogenous spatial distributions of Sn atoms and the normal state DOS (Fig. 3c, d and k). Statistical analysis of $\Delta(\mathbf{r})$ yields a mean gap value of $\bar{\Delta} =$ 323 μeV with a standard deviation of $\sigma =$ 16.6 μeV (Fig. 3g). The resulting $\sigma/\bar{\Delta} \approx 0.051$ is relatively smaller than those reported in cuprates [49,50] and iron pnictides [51], but similar to that of highly disordered superconducting thin films close to the superconductor-insulator transition [52,53]. Another spectral survey in a 40 nm × 40 nm field of view shows similar behavior (Supplementary Fig. 8); an autocorrelation analysis of the gap map over this larger range yields a decay length $L_\Delta \approx$ 58 nm (Supplementary Fig. 7), which is consistent with the expectation for the spatial variation of superconducting order parameter (Supplementary Note 7) according to the in-plane superconducting coherence length $\xi_{ab}$ ($\simeq$ 28-41 nm) attained from vortex imaging and analysis of the upper critical fields (Supplementary Note 3 and Supplementary Fig. 7) .

Notably, local spectra with a larger gap consistently exhibit higher coherence peaks and a deeper zero-bias suppression (Fig. 3i and j), implying a strong positive correlation ($\alpha_c \approx 0.8$) between $\Delta(\mathbf{r})$ and $H(\mathbf{r})$. This correlation explicitly suggests that the gap filling and smearing out of coherence peaks are due to pair-breaking scattering [54], which in turn leads to a reduction of $\Delta$ through $\Delta = V\sum_k\langle c_{k\uparrow}^{+}c_{-k\downarrow}^{+}\rangle$ as in the framework of the Bogoliubov–de Gennes theory. Consistently, local spectral fitting using the Dynes formula $N(E) = Re[(E - i\Gamma)/\sqrt{(E - i\Gamma)^2 - \Delta(\theta)^2}]$ reveals that the extracted pair-breaking parameter $\Gamma(\mathbf{r})$ is anti-correlated with $\Delta(\mathbf{r})$ (Supplementary Fig. 8h-k), further supporting that the gap inhomogeneity is closely associated with spatially inhomogeneous pair breaking. Overall, the observation of microscopic inhomogeneity at the lowest $T$ complements to the percolative superconducting transition (stemming from a distribution of $T_c$), consistently establishing a full picture of spatially heterogeneous superconductivity in these samples.

What could be the potential origin of such an inhomogeneity? First, the influence of local strains and surface Cs concentrations can be precluded (Supplementary Notes 9 and 10). Moreover, the cross-correlation between the position of Sn dopants and $\Delta(\mathbf{r})$ is almost zero (Fig. 3h and Supplementary Fig. 8 for more datasets), suggesting that the inhomogeneity is not inherited from the distribution of dopants. This is further verified by a systematic comparison of the gap distribution across different Sn doping levels (Supplementary Note 11): The normalized gap variation $\sigma/\bar{\Delta}$ remains small for $x$ = 0.06, 0.12 and 0.30 with values of 0.012, 0.010 and 0.020, respectively, whereas it reaches 0.051 in the stripe-ordered $x$ = 0.09 sample (Supplementary Table S2). The $x$ = 0.30 sample indeed exhibits noticeable spatial variations in its normal-state DOS, yet its superconducting gap remains rather homogeneous (Supplementary Fig. 13). Given that remarkable inhomogeneity of $\Delta$ is absent in both phase regions I and III in the $T$-$x$ phase diagram displayed in Fig. 1 (for pristine $CsV_3Sb_5$ $\sigma/\bar{\Delta}$ ≈ 0.018, [55]), we conclude that it is most likely linked to the occurrence of the unidirectional band folding from short-ranged charge-stripe order in phase II (Supplementary Fig. 15), as further discussed below.

**Normal fluids in the superconducting state**

Besides the spatially varying superconducting gap, superconductivity in stripe-ordered samples ($x$ = 0.07 and 0.09) in the phase region II is also distinguished by the presence of copious low-energy quasiparticle excitations at temperatures far below $T_c$. The first implication comes from specific heat study. In Figure 4a we present the electronic specific heat $C_e(T)$ of $CsV_3Sb_{4.91}Sn_{0.09}$ obtained after subtraction of the phonon contribution and nuclear Schottky term (Methods and Supplementary Fig. 17). Below $T_c^s$, $C_e/T$ develops an extremely broadened anomaly with a heavily suppressed jump height ($\Delta C_e/\gamma_n T_c$ ≈ 0.43 compared to the BCS value of 1.43), reflecting a substantial distribution of local $T_c$. The low-$T$ (< 0.7 K) profile can be fitted to $C_e/T = \gamma_0 + a'T^{-5/2}\exp\left[-\Delta(0)/k_B T\right]$ [56] with $\gamma_0$ ≈ 11.7 mJ mol$^{-1}$ K$^{-1}$ and $\Delta(0)$ = 0.27 meV; here $a'$ is a constant and $k_B$ is the Boltzmann constant. Such behavior provides two important insights: First, the specific heat of superconducting electrons takes the form suggestive of a fully gapped pairing state [detailed analysis of $C_e(T)$ is presented in Supplementary Note 13, where we verify the validity of two-band and (modified) single-band BCS $\alpha$-model for $x \leq$ 0.06 and $x \geq$ 0.07, respectively]. Second, a large residual linear term $\gamma_0 T$ with $\gamma_0/\gamma_n$ ≈ 38% ($\gamma_n$ ≈ 31.0 mJ mol$^{-1}$ K$^{-1}$ is the normal-state Sommerfeld coefficient, Fig. 4a) is necessarily introduced to trace the leveling off $C_e/T$ at the lowest $T$ that guarantees the entropy balance condition (Supplementary Note 13 and Supplementary Fig. 18); it presumably arises from gapless quasiparticles at the chemical potential that do not superconduct down to at least 0.1 K --- straightforwardly, it means more than one third of conduction electrons stay unpaired at $T/T_c$ ≈ 0.05, constituting a prominent branch of normal fluid in the superconducting state. Indeed, the presence of an abundant normal fluid is another unique feature of superconductivity in the charge-stripe phase region II as demonstrated in Fig. 4b: $\gamma_0/\gamma_n$ is remarkably enhanced for $x$ = 0.07 and 0.09 (strikingly reaching ~ 63% in the former), whilst stays relatively low for $x$ = 0, 0.06, 0.12 and 0.30 (these residual terms may not be intrinsic; see Supplementary Note 13).

The concurrence of a full superconducting gap and profuse quasiparticle DOS at the Fermi level ($E_F$) can originate from either phase separation (with a large fraction of the sample being nonsuperconducting), or a strong pair-breaking effect triggered by a specific scattering process [32,57]; the latter, under the condition of weak, nonmagnetic disorder [33], has been established as a hallmark of unconventional superconductivity possessing sign-reversal order parameters [34,36,37,58,59]. Our STM results unequivocally nail down the pair-breaking origin of the normal fluid. Figure 4c shows a direct comparison of the superconducting spectra for $x$ = 0 [60], 0.09 and 0.3, acquired at temperatures of $0.015T_{c0}$, $0.01T_{c0}$ and $0.09T_{c0}$, respectively. Only the composition $x$ = 0.09 in phase region II exhibits a prominent residual DOS (approximately 30% of the normal-state DOS) at zero energy down to lowest $T$; at $x = 0.3$, a large, U-shaped gap with negligible in-gap DOS emerges. This trend is in excellent agreement with the specific heat data.

Further elucidation of the momentum-space origin of the residual in-gap DOS comes from quasiparticle interference imaging. As sketched in Fig. 4d, the salient scattering channels of the normal state include intra-$p$-orbital scattering ($\mathbf{q_0}$), scattering of the quasi-1D segments of the kagome $d$-band ($\mathbf{q_1}$), and scattering between the arcs of triangular pockets ($\mathbf{q_2}$). The one-of-a-kind scattering element in phase region II is the unidirectional, quasi-1D stripe features (labeled as $\mathbf{q'}$ and $\mathbf{q''}$) that span almost the entire Brillouin zone (Fig. 4e); these features may rise from the scattering between the original quasi-1D bands and their folded replicas owing to the formation of single-$Q$ stripelike CDW (Supplementary Fig. 15) [30]. Upon entering the superconducting gap, most of the scattering events fade away owing to the depletion of quasiparticles by gap opening; the prominent feature at $\mathbf{q_1}$, however, persists down to zero-energy even at 20 mK. These unidirectional signals unveil the momentum origin of the residual DOS at $E_F$; that is, these unpaired quasiparticles mainly reside on the quasi-1D bands which are involved in the band folding corresponding to the stripelike CDW (Fig. 4e and h; Supplementary Fig. 15). Moreover, the same stripe-related electronic band reconstruction is also connected to the real-space gap inhomogeneity; our spatial-resolved analyses reveal that the gap distribution $\Delta(\mathbf{r})$ (while is anti-correlated with $\Gamma$, Supplementary Note 8) correlates positively with the stripe-induced folded-band QPI intensities, i.e., regions with larger $\Delta$ exhibit stronger features at $\mathbf{q'}$ and $\mathbf{q''}$ (Supplementary Note 12). This immediately rules out a straightforward competition between the stripelike CDW and superconductivity; instead, it implies that the heterogeneous superconductivity is tied to both charge-stripe-related electronic structure and spatial dependent pair-breaking effect. A most likely scenario is thus the short-range stripelike CDW imposes a strong pair-breaking scattering effect upon the superconducting state, consequently engenders the spatially varying order parameter and sizeable normal fluid DOS [32], as have been established in the cases of $d$-wave superconductors [36,37].

**Discussion**

A distinct heterogeneous superconducting state that develops peculiarly in the charge-stripe phase regime (II) of $CsV_3Sb_{5-x}Sn_x$ (Fig. 1b and c) has been uncovered by our multimodal investigations. It cannot be simply attributed to the inhomogeneity of dopant distribution or local strain fields, nor a simple competition between the charge order and superconductivity. This state, unforeseen in a three-dimensional (3D) weakly-correlated system and has been overlooked thus far in the study of $AV_3Sb_5$ series, adds vigorously to the rich emergent physics in kagome lattice superconductors [61,62]. We emphasize that our results can hardly be reconciled with the sign-invariant $s$-wave pairing for the following reasons. (i) Sn substitution only introduces nonmagnetic scattering centers [30], thus, Anderson's theorem [33] should be complied with. (ii) Accordingly, BCS-type superconductivity is insensitive to impurity scattering up to a substantial strength of disorder potential [33,58], which suffices to bring the system close to the mobility edge [63]; whilst spatially inhomogeneous superconductivity naturally sets in [63,64] at such high levels of disorder, there is compelling evidence that our 3D crystals are far from the localization threshold on the metallic side (Supplementary Note 14). (iii) STM results indicate that the pair breaking is more likely caused by scattering off the disorder potential associated with the short-range CDW (Supplementary Note 15), as evidenced by the remnant quasiparticle DOS pertains to the Fermi sheets participating in the charge-stripe-induced unidirectional band folding (Fig. 4h); in contrast, conventional $s$-wave pairing is known to be robust against this sort of perturbation [6,65].

In essence, all our major discoveries, i.e., (1) the percolative superconducting precursor, (2) spatially-varying superconducting order parameter on a length scale of the order of $\xi_{ab}$ and (3) the residual normal fluid concentrated on specific Fermi surfaces, are consistent with the phenomenology that a sign-changing unconventional superconducting phase is exposed to strong pair-breaking effect which has been widely investigated on the background high-$T_c$ cuprates [32,66,67]: scattering between momentum-space regions with opposite signs of superconducting order parameter is particularly effective in destroying the coherent superconducting order, engendering a spontaneous tendency toward heterogeneity. The resulted superconducting state exhibits inherent spatially nonuniform signatures (that can be unrelated to the distribution of disorder) and incorporates a remarkable density of gapless, unpaired quasiparticles [32], in agreement with both our experimental observations and earlier results in overdoped cuprates [34-37]. As such, we have managed to provide strong evidence for the occurrence of sign-reversal superconductivity in the kagome Cs135 family ($CsV_3Sb_5$ and its doped derivatives). We also note that the pair-breaking scattering in $CsV_3Sb_{5-x}Sn_x$ is predominantly attributed to the short-range stripelike CDW whose correlation length is putatively comparable to $\xi_{ab}$ [67] (Supplementary Note 15), instead of the point-like impurities as the case in cuprates [32,37]. The evolution of the vortex-core spectra is also consistent with this interpretation. The zero-energy conductance peak observed in the vortex cores at $x = 0$ [16] becomes broadened and disappears as the short-range stripe-like CDW develops in phase region II, leaving a featureless, normal-state-like spectrum (Supplementary Fig. 7). By contrast, finite-energy conductance peaks reappear at $x = 0.3$, where the stripe-like CDW is weak or no longer discernible (Supplementary Fig. 13).

Thanks to the unique quantum interference between the electron wavefunctions of three sublattices in the geometrically frustrated kagome lattice, diverse theoretical proposals of sign-reversal pairing states have been raised for kagome superconductors; these includes ($p$+$ip$)-wave (Fig. 5a), ($d$+$id$)-wave (Fig. 5b), $f$-wave (Fig. 5c) and $s^{+-}$ states (Fig. 5d) [68-71] as well as exotic PDW order [72]. Despite that the exact pairing symmetry cannot be conclusively determined at this stage, we suggest that the singlet $d$+$id$ (chiral $d$) [68,69,71] and $s^{+-}$ [70,71] pairing states are more promising candidates, both of which can give rise to the unidirectional pair-breaking scattering observed in the STM measurements (at present, the QPI analysis cannot give a more definitive identification of the pairing symmetry, see Supplementary Note 17). On the other hand, the nodal $f$-wave pairing (Fig. 5c) can be generally ruled out as there is no evidence for gap nodes in our specific heat (Supplementary Note 13) and STM data (Supplementary Note 16); the $p$+$ip$ state (Fig. 5a) is also less likely due to the absence of strong pair breaking effect of Sn dopants that is promised for spin-triplet pairing (Supplementary Note 17). Moreover, the unusual superconducting state in $CsV_3Sb_{5-x}Sn_x$ ($x = 0.09$) can hardly be associated with the PDW state, since the spatial modulation of the superconducting order parameter lacks the periodicity featuring a PDW order (Fig. 5e-h). Several important implications of our results deserve to be stressed: Heterogeneous unconventional superconductivity emerges only in the charge-stripe phase regime where the unidirectional band folding can be identified (Fig. 1), which highlights the strikingly intricate correspondence between CDW and superconductivity in the Cs135 family [73,74]. The effective pair breaking promoted by an extend CDW-related disorder potential may indicate distinct responses to disorder for unconventional kagome superconductors [75]. The persistent gapless feature of the Fermi sheet participating in the unidirectional folding of charge stripes (Fig. 4h) provides crucial information about the possible location of sign reversal of the superconducting order parameter (Supplementary Note 17).

Our work thus identifies charge-stripe-ordered $CsV_3Sb_{5-x}Sn_x$ as a privileged platform for the exploration of emergent unconventional superconductivity and its interplay with coexisting electronic orders. Future phase-sensitive probes and theoretical efforts may help to pinpoint the exact symmetry of superconducting order parameter and the microscopic mechanisms for the occurrence of heterogeneous superconductivity.

## Methods

### Single crystal growth and characterizations

Single crystals of $CsV_3Sb_{5-x}Sn_x$ were synthesized via a self-flux growth method from a mixture of $Cs_2Sb_3$, $VSb_2$ and SnSb. The mixture was sealed in an evacuated quartz ampoule and heated at 1000 °C for 30 h, followed by cooling at a rate of 2 °C/h down to 700 °C. The platelike as-grown $CsV_3Sb_{5-x}Sn_x$ single crystals with typical sizes of 5×5×0.1 $mm^3$ are stable in air. X-ray diffraction (XRD) measurement demonstrated their [00*l*] orientation; in this work, we define the *c* axis along the crystallographic [001] direction and the *ab* plane perpendicular to it. Chemical compositions were determined by energy-dispersive X-ray spectroscopy (EDS) and have been further verified by STM topographic images (Supplementary Note 1).

### Transport measurements

Electrical resistivity measurements were performed in a Quantum Design Dynacool-14 T physical property measurement system (PPMS) equipped with a dilution refrigerator. Single crystals of $CsV_3Sb_{5-x}Sn_x$ were cut into rectangular bar shapes. Electrical contacts were made by attaching platinum wires to the (001) surface of the sample in a conventional four-lead configuration using silver paste (DuPont 4929N). For the x = 0.07 and 0.09 samples, the zero-field paraconductivity $\Delta\sigma(T)$ is defined as: $\Delta\sigma(T) = \sigma(T)-\sigma_n(T) = 1/\rho(T) - 1/\rho_n(T)$, where the normal-state resistivity $\rho_n(T)$ was obtained by fitting to the form $\rho_n(T) = \rho_0+A_2T^2+A_5T^5$ ($\rho_0$, $A_2$ and $A_5$ are fitting parameters).

### Magnetization measurements and torque magnetometry

The dc magnetization was measured down to 0.4 K using a Quantum Design magnetic property measurement system (MPMS) equipped with a helium-3 ($^3He$) insert. The diamagnetic responses of samples were tracked in both zero-field-cool (ZFC) and field-cool (FC) processes, under a constant external field $H$ = 5 Oe applied parallel to the *c* axis. The magnetic susceptibility $\chi$ was calculated considering the demagnetizatng factors given by the sample geometries.

Measurement of the magnetic torque $\tau(H)$ were performed using the method of piezoelectric torque magnetometry. A small piece of single crystal was attached to the tip of a piezoresistive cantilevers (SCL Sensortech PRSA-L300), which was mounted to a $^3He$ probe and loaded into a Quantum Design PPMS. In practice, the resistance of SCL sensor (essentially reflecting the voltage drop across a built-in Wheatstone bridge) was measured and converted to the magnetic torque: $\tau \propto R(H) - R(H = 0)$. The transverse magnetization $M_\perp$ was derived as $M_\perp = \tau/\mu_0 H$.

### Effective-medium model for the percolation process

The putative percolation process above the bulk $T_c$ can be properly described within the effective-medium theory (EMT) [75,76], which has been frequently exploited in the analysis of spatially inhomogeneous superconductors (e.g. high-$T_c$ cuprates [42,43] and oxide heterointerfaces [48,76]) in a certain temperature range above $T_c$. Within this picture, the systems comprise a mostly random network of superconducting and metallic regions; accordingly, $T_{c0}$ marks the temperature at which the fraction $P$ of superconducting clusters reaches a threshold $P_\pi$ that renders a fully-connected superconducting path across the sample. Here, we chose a 3D nearest-neighbor site percolation model [43,77,78] for EMT analysis. Such a model treats the sample as an ensemble of well-connected small cubic grains, each having a local $T_c$; with decreasing $T$, the fraction of superconducting grains grows, eventually percolates at $T = T_\pi$ where $P$ adopts a critical value $P_\pi \approx 0.31$ [41,42]. The resistivity of an individual grain is set to a universal value $R_n$ above its local $T_c$ and a vanishingly small $R_0 = 10^{-4}R_n$ below it. Thereby, the spatial inhomogeneity stems only from the distribution of local $T_c$ that is depicted by a function $g(T)$. The $T$ dependence of superconducting fraction $P$ is thus:

$$P(T) = \int_T^\infty g(T')dT' \qquad (1)$$

To delineate the drastic enhancement of $P$ when $T$ approaches $T_c^S$ (presumably due to proximity effect [33]), we opted for a two-component form of $g(T)$: $g = ug_1(T)+(1-u)g_2(T)$; $u$ is a weighing factor, $g_1$, $g_2$ are two Gaussian distribution functions with the same mathematic expectation $T_0$ and half-width-at-half-maximum (HWHM) $\Xi_1$ and $\Xi_2$, respectively ($\Xi_1 > \Xi_2$, see Supplementary Fig. 5a). Above a critical temperature $T_\pi$ [$P(T_\pi) = P_\pi = 0.31$], the profile of dc conductivity $\sigma(T) = 1/\rho(T)$ follows a trend given by the EMT:

$$\frac{\sigma-(1-P+P/R_0)}{1-P+P/R_0+(1/P_\pi-1)\sigma}P+\frac{\sigma-\dfrac{1/R_0}{P+(1-P)/R_0}}{P+\dfrac{1-P}{R_0}+\left(\dfrac{1}{P_\pi}-1\right)\sigma}(1-P)=0 \qquad (2)$$

As demonstrated by the fitting curves in Figs. 2a and c, this EMT modeling provides a reasonable description of the experimental paraconductivity data for the $x = 0.07$ and 0.09 samples (in a temperature range of ~1.4-3.6 K and ~2.1-3.5 K, respectively). In particular, the exponential attenuation upon warming is reproduced by fits to Eq. (2). We suggest that such a percolation scenario, in spite of its oversimplification, captures some of the main characters of emerging superconductivity precursor above bulk $T_c$ in these samples. It should be noticed that our samples lack the inherit normal-state inhomogeneity (Supplementary Note 1 and Supplementary Fig. 2) that prevails in cuprates due to heterogeneous charge localization [79]; the percolative model instead reflects an unusual interplay between stripelike CDW and superconductivity in $CsV_3Sb_{5-x}Sn_x$.

**Measurement and analysis of specific heat**

Specific heat of $CsV_3Sb_{5-x}Sn_x$ was measured down to 0.1 K in a Dynacool-14T PPMS equipped with a dilution refrigerator and $^3$He system. During the measurements, the temperature rise in each relaxation cycle was kept at $\Delta T/T = 3\%$. In our samples, $C(T)$ incorporates contributions from electrons ($C_e$), phonons ($C_{ph}$) and the Schottky anomaly of nuclear spins ($C_{Sch}$). The electronic contribution $C_e$ was obtained by subtracting $C_{ph}$ and $C_{Sch}$ from the raw data. At low $T$, the phonon contribution takes the form $C_{ph} = \beta T^3 + \eta T^5$; we assumed that $\beta$ and $\eta$ are independent of $H$ and employed the fit $C(T) = \gamma_n T + \beta T^3 + \eta T^5$ [79,80] to data between 0.5 and 4 K under $\mu_0 H = 3$ T along $c$ (at which the superconductivity is fully quenched in all samples, thus $C_e = \gamma_n T$), see Supplementary Fig. 17. The value of Sommerfeld coefficient $\gamma_n$ was verified by fitting the zero-field normal state $C(T)$ to the same formula (Supplementary Fig. 17), endorsing that $\gamma_n$ does not vary under $H$ at $T > T_c$. The Schottky term $C_{Sch}$ is only resolvable below approximately 0.5 K; we fit it using a two-level model: $C_{Sch} = Dx^2e^x/(1+e^x)^2$ ($D$ is a constant, $x = \Delta E/k_BT$, $\Delta E$ is the energy separation between two levels) [80]. It yields reasonable fitting results in a temperature range from 0.2 to 0.4 K (Fig. 4a, Supplementary Fig. 17).

**STM measurements**

The majority of STM experiments were performed with a custom-built, dilution refrigerator-based systems with a working temperature of 20 mK at the Synergetic Extreme Condition User Facility A7 experimental station of the Institute of Physics, Chinese Academy of Sciences. Single crystals were cleaved *in situ* at 4.2 K under ultrahigh vacuum, then cooled to 20 mK at the STM head. PtIr tips were calibrated on polycrystalline Au prior to the measurements. Spectroscopic data were acquired using a standard lock-in technique at a frequency of 827.183 Hz with a modulation voltage of 10-50 μV. With cryogenic temperature cleaving, we were able to obtain relatively large regions with sparse Cs atoms on the surface; clean Sb surfaces were subsequently obtained by sweeping away loosely bonded Cs atoms with an STM tip (Fig. 3 and Supplementary Fig. 8). Clustering of Cs atoms on the surface might alter the local superconducting gap, but only within a spatial range of ~5 nm (Supplementary Note 10).

*Signatures of the charge stripes*. For $x = 0.07$ and 0.09, the existence of short-range charge stripe is evident from the sudden appearance of $\mathbf{q}'$ and $\mathbf{q}''$ in the QPI pattern (Fig. 4e): they are located at approximately 1/3 and 2/3 of $(3/2)^{1/2}\mathbf{Q}_{\mathbf{Bragg}}$ (along Γ-K), and their energy dispersions are in good agreement with the quasi-1D vanadium $d_{xz/yz}$ orbitals [36]; thus, a straightforward explanation is that the emergence of charge-stripe order folds the quasi-1D kagome *d*-bands with a vector of roughly 1/3 along one of the three lattice directions [36]; the two new channels arise from the scattering between their original and replica bands. In fact, short-range single-$Q$ $3a_0$ modulations can be directly observed in the *dI/dV* maps once the surface reconstruction of Sb atoms is suppressed (Supplementary Fig. 15).

*Superconducting gap map*. The spectra acquired at each pixel were first smoothed via a Gaussian filter and then interpolated with second-degree polynomials. The derivative of each spectrum was subsequently calculated using the central difference method, and the zero-crossings were identified as potential coherence peaks. The superconducting gap size was then defined as $\Delta = (\Delta^+ + \Delta^-)/2$.

*QPI analysis*. Momentum-space information of the superconducting gap function is reflected in the intensity variations in the Bogoliubov QPI maps. When the energy sweeps to inside the gap, the intensity of the intra-*p*-orbital scattering ($\mathbf{q}_0$) and the intra-*d*-orbital scattering ($\mathbf{q}_2$) is strongly suppressed (Fig. 4 and Supplementary Fig. 20). The QPI images shown in Fig. 4 were 2-fold symmetrized to improve signal-to-noise. To seek possible phase-sensitive evidence for sign-changing pairing states, we performed *H*-dependent and phase-referenced QPI analyses [81-84] (Supplementary Fig. 22); during the data analysis we used $g(\mathbf{q}, E)$ maps instead of the Fourier transform of the conductance-ratio maps $Z(\mathbf{r}, E) = g(\mathbf{r}, +E)/g(\mathbf{r}, -E)$, because the $Z(\mathbf{q}, E)$ signals in our experiments are rather weak at low energies inside the superconducting gap (Supplementary Fig. 22i-l) and hence the *Z*-map analysis becomes less reliable. The absence of a clear phase-sensitive QPI signature does not necessarily rule out the sign-changing pairing proposed here, because the dominant observed scattering is intra-pocket and could not connect regions with opposite gap signs (see details in Supplementary Note 17).

*Impurity states*. While no clear in-gap state is observed near the Sn dopants, we found that some instinct defects at the vanadium site are strong pair breakers and induce pronounced in-gap bound states near $E_F$. Corresponding local oscillations give rise to an enhanced intensity at $\mathbf{q}_2$ in the Fourier transformations. These impurity states are spatially confined to the defect sites, in sharp contrast to the scattering events at $\mathbf{q}_1$ that is more extended in real space (Supplementary Note 15).

**Data availability**

The primary data underlying all the plots within the main text, as well as the supplementary resistivity, magnetization, magnetic torque and specific heat data that support the findings of this study, have been deposited in the Figshare database under DOI 10.6084/m9.figshare.34332690, see ref. [85]. Other source data generated in this study are available from the corresponding authors upon request.

**Acknowledgements**
We thank Zi-Xiang Li, Kun Jiang, Yi Zhou, Hong Yao, Qianghua Wang, Ziqiang Wang, Haiwen Liu and Andreas Kreisel for insightful discussions. A portion of experimental work was carried out at the Synergetic Extreme Condition User Facility (SECUF, https://cstr.cn/31123.02.SECUF).

**Funding**
This work was financially supported by the National Natural Science Foundation of China (Grants Nos. 12488201, 12274390 and 52261135638), the National Key Projects for Research and Development of China (Grant No. 2022YFA1602602), the Quantum Science and Technology-National Science and Technology Major Project (Grant No. 2021ZD0302802), the Basic Research Program of the Chinese Academy of Sciences Based on Major Scientific Infrastructures (No. JZHKYPT-2021-08) and the Fundamental Research Funds for the Central Universities No. WK2030000111.

**Author contributions**
Z.X., Zhenyu Wang and X.C. supervised the project and designed the experiments. Hongyu Li, H.R. and T.W. prepared the samples. Hongyu Li and Z.X. performed the electrical transport, magnetic torque and specific heat measurements and analyzed the data with help from X.Z., Houpu Li and Z.G. S.Y. and Zhenyu Wang conducted the STM measurements and analyzed the data with assistance from Zhuying Wang, Y.Z., X.Y. and R.W. Hongyu Li, Z.X., S.Y. and Zhenyu Wang wrote the paper with input from all authors.

**Competing interests**
The authors declare no competing interests.

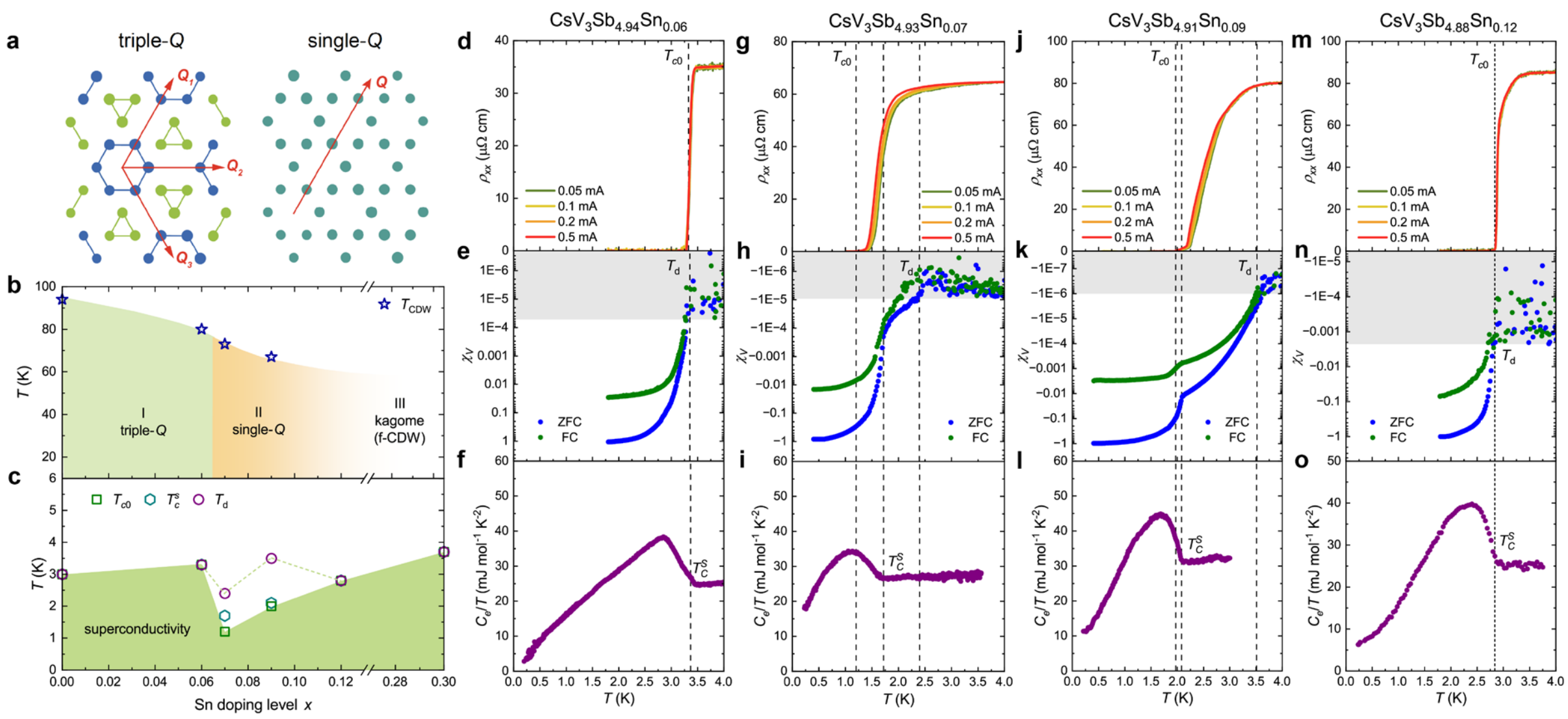

**Fig. 1 | Phase diagram of $CsV_3Sb_{5-x}Sn_x$ and the characteristic temperatures of superconductivity.** **a,** Charge-density-wave (CDW) patterns in the kagome vanadium network: (left) triple-$Q$ 2×2 CDW and (right) the stripelike single-$Q$ CDW [30]. Arrows denote the CDW wavevectors. **b, c,** electronic phase diagram of $CsV_3Sb_{5-x}Sn_x$ showing the evolution of the CDW (**b**) and superconductivity (**c**) upon increase of Sn doping level $x$. Different colours in **b** discriminate the CDW phase regions with stars representing the transition temperature $T_{CDW}$ determined from resistivity (Supplementary Fig. 3). Note that the single-$Q$ CDW in **b** is a short-range order; the corresponding phase regime (II, orange-shaded) crossover to a nearly-undistorted kagome phase (III) with fluctuating single-$Q$ charge correlation, as denoted by the colour gradient. Circles, hexagons and squares in **c** represent the onset temperatures of diamagnetism ($T_d$), specific heat anomaly ($T_c^S$) and zero resistivity ($T_{c0}$), respectively. **d-o,** In-plane resistivity $\rho_{xx}$ (**d, g, j, m**), volume magnetic susceptibility $\chi_v$ (**e, h, k, n**) and electronic specific heat $C_e/T$ (**f, i, l, o,** see Methods for the extraction of $C_e$) plotted as a function of temperature ($T$) for samples with $x$ = 0.06 (**d-f,** $T \leq 4$ K), 0.07 (**g-i,** $T \leq 4$ K), 0.09 (**j-l,** $T \leq 4$ K) and 0.12 (**m-o,** $T \leq 4$ K). $\chi_v$ was measured with a magnetic field of 5 Oe applied along the $c$ axis and was corrected for the demagnetizing effects; both zero-field-cooled (ZFC, olive) and field-cooled (FC, blue) data are shown. Vertical dashed lines indicate the characteristic temperatures $T_{c0}$, $T_c^S$ and $T_d$ for each sample.

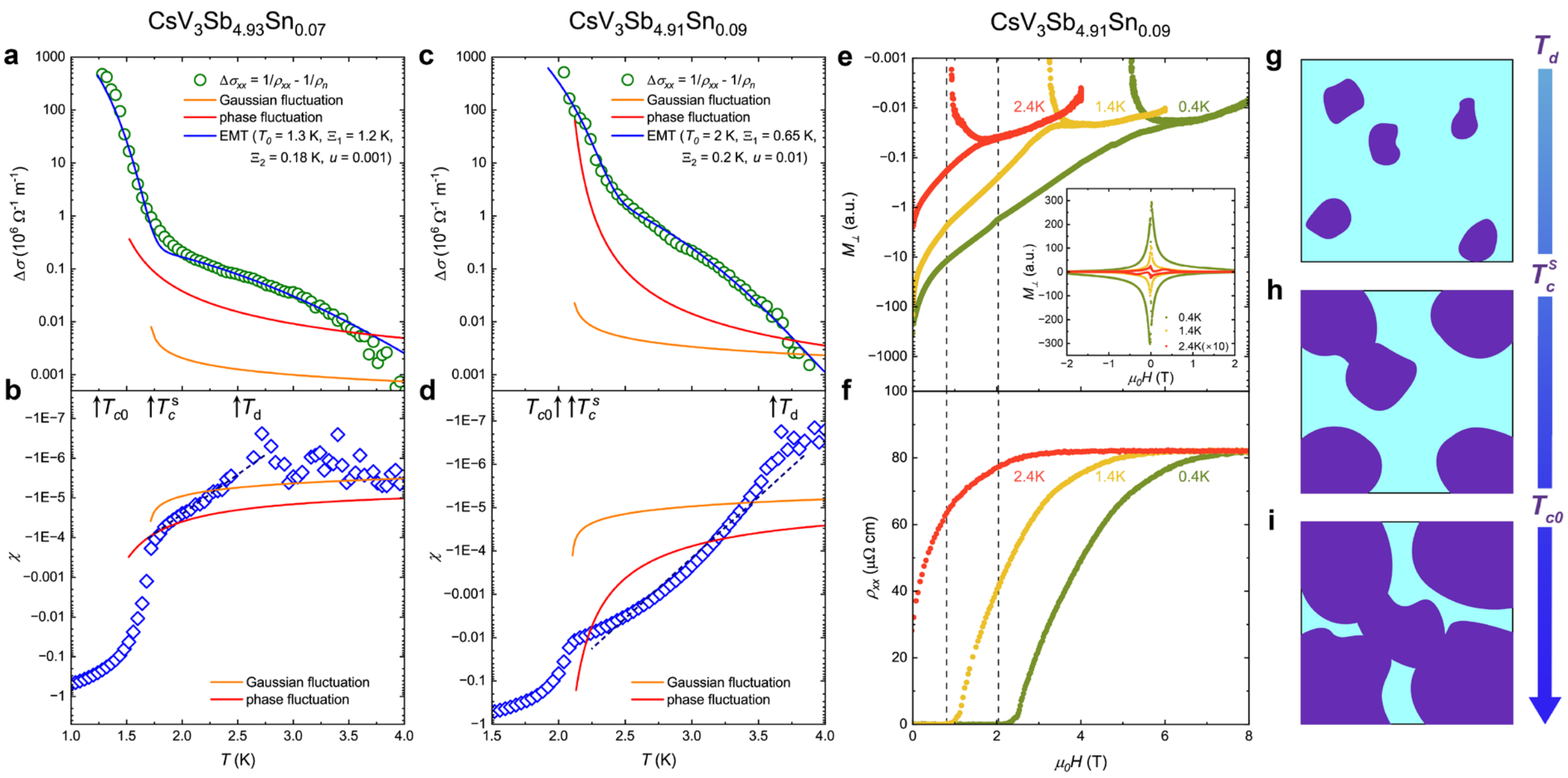

**Fig. 2 | Percolative superconducting precursors in charge-stripe-ordered $CsV_3Sb_{5-x}Sn_x$. a-d,** In-plane paraconductivity $\Delta\sigma_{xx}$ (**a** and **c**) and ZFC susceptibility $\chi$ (**b** and **d,** same data as presented in Figs. 1**h** and **k**, respectively), measured in $CsV_3Sb_{4.93}Sn_{0.07}$ (**a** and **b**) and $CsV_3Sb_{4.91}Sn_{0.09}$ (**c** and **d**). Data are displayed as hollow symbols in logarithmic scales. Orange and red solid lines are simulations based on the 3D Gaussian-type fluctuations and phase fluctuations, respectively (Supplementary Notes 3 and 4). Blue lines in **a** and **c** represent the fits of $\Delta\sigma_{xx}$ to the EMT model (with a two-component distribution of $T_c$, see Methods and Supplementary Note 5). Dashed lines in **b** and **d** guide the eyes. Black arrows indicate characteristic temperatures $T_{c0}$, $T_c^S$ and $T_d$. **e, f,** A contrast between the transverse magnetization $M_\perp$ (**e**) derived from torque magnetometry (Methods) and $\rho_{xx}$ (**f**) measured in a $CsV_3Sb_{4.91}Sn_{0.09}$ sample under in-plane $H$ at different temperatures (olive, $T = 0.4$ K; orange, 1.4 K; red, 2.4 K). Inset of **e** displays the full hysteresis loops of $M_\perp$. Vertical dashed lines mark the onset fields of zero resistivity. **g-i,** Schematic representations of the percolation process which is partially captured by the EMT approach. With decreasing $T$, gradual emergence of superconducting grains (dark purple) from the metallic surroundings (light cyan) gives rise to detectable signals of diamagnetism below $T_d$ (**g**); the superconducting grains are substantially enlarged due to proximity effect when $T$ approaches $T_c^S$ (**h**) and a percolation path across the sample appears at $T = T_{c0}$ (**i**).

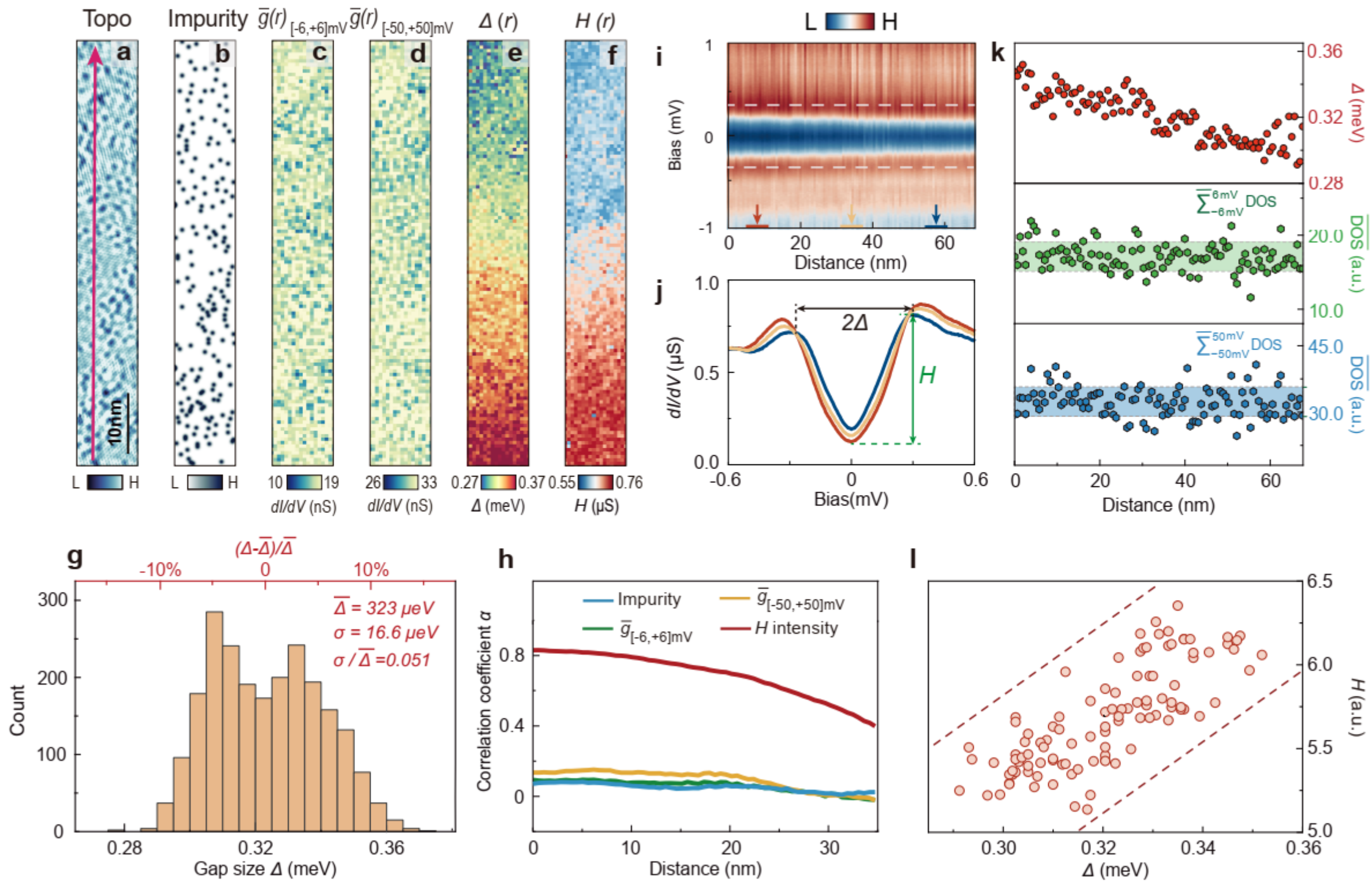

**Fig. 3 | Spatial inhomogeneity of superconductivity in $CsV_3Sb_{4.91}Sn_{0.09}$. a-f,** Topologic and spectroscopic characteristics in a 70 nm × 10 nm field of view, including topographic image (**a**), dopant distribution (**b**), integrated DOS between ± 6 meV (**c**) and ± 50 meV (**d**), superconducting gap map (**e**) and gap-depth map (**f**). **g**, Histogram showing the distribution of the gap amplitude in **e**. **h**, Cross-correlation coefficients between the gap map *versus* impurity distribution (blue), integrated DOSs of ±6 meV (green) and ±50 meV (orange), and the gap-depth map (red). **i,** $dI/dV$ linecut along the purple arrow in **a**, showing the spatial variation in the superconducting gap. **j**, Representative spectra averaged over three different regions marked in **i**. **k**, gap amplitude (upper panel), integrated DOS between ± 6 meV (middle panel) and ± 50 meV (bottom panel) extracted from the linecut of **i**, in which the shaded regions indicate the standard deviation. **l**, gap depth ($H$) versus gap amplitude ($\Delta$) obtained from the line-profiled spectra, showing a positive correlation. STM setups: **a.** $V_s$ = -90 mV, $I_t$ = 0.5 nA; **c, d**, $V_s$ = -90 mV, $I_t$ = 1 nA, $V_m$ = 2 mV; **e, f,** $V_s$ = -1 mV, $I_t$ = 0.6 nA, $V_m$ = 20 µV; **i, j,** $V_s$ = -1 mV, $I_t$ = 0.6 nA, $V_m$ = 20 µV.

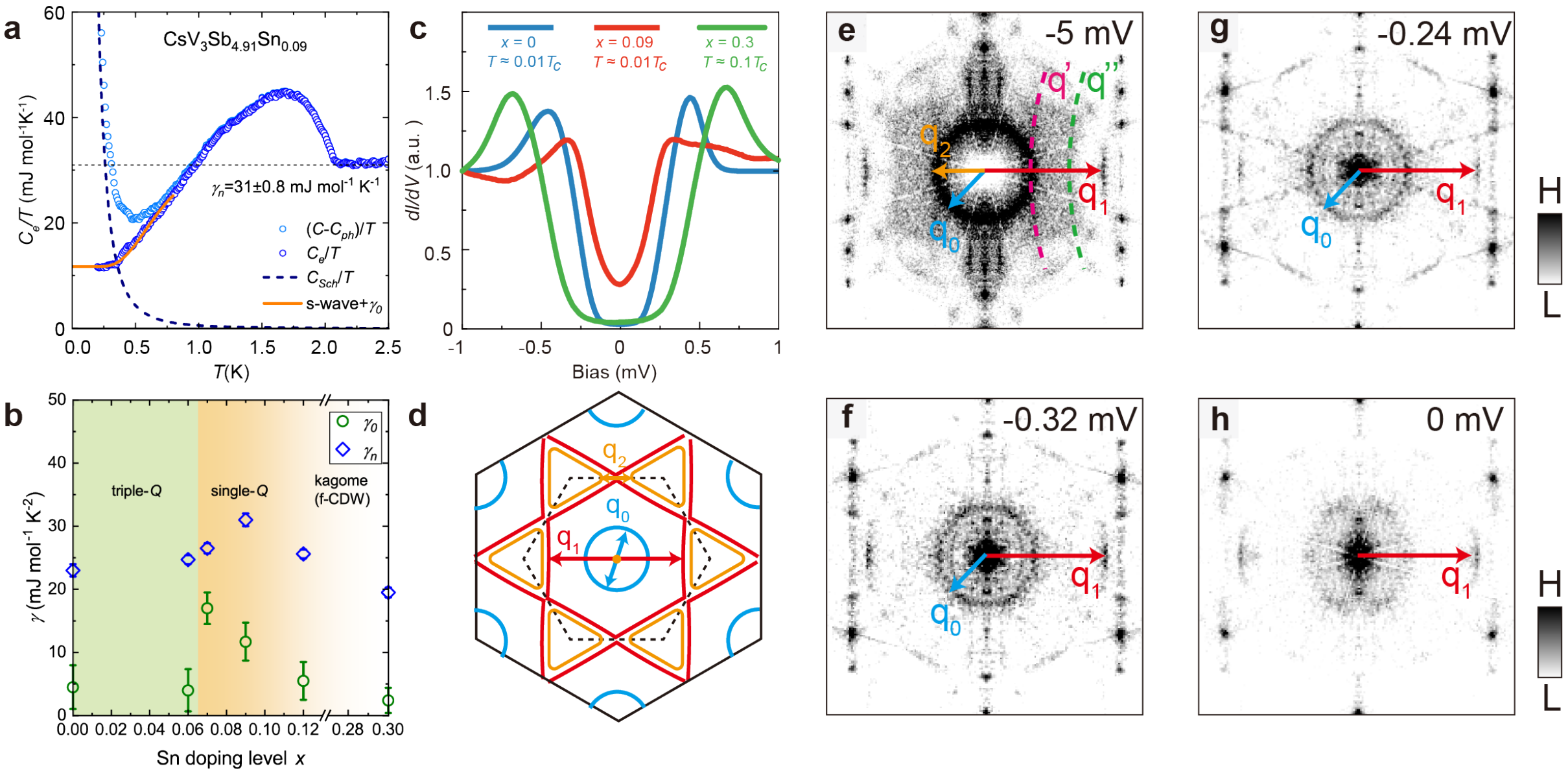

**Fig. 4 | Residual normal fluid in $CsV_3Sb_{4.91}Sn_{0.09}$. a,** The electronic specific heat $C_e(T)/T$ of $CsV_3Sb_{4.91}Sn_{0.09}$ (blue circles) with the Schottky anomaly (navy dashed line) subtracted from data of $C(T)$-$C_{ph}(T)$ (cyan symbols, see Methods for details). Orange line indicates a BCS *s*-wave fit with a residual linear term $\gamma_0$: $C_e/T = \gamma_0 + a'T^{-5/2}\exp[-\Delta(0)/k_BT]$. Horizontal dashed line indicates the normal-state Sommerfeld coefficient $\gamma_n$. **b,** $\gamma_n$ (blue diamonds) and $\gamma_0$ (olive circles) plotted against Sn doping level $x$. Background colours differentiate phase regimes. **c,** Superconducting spectra for $x = 0$ [58] (blue), $x = 0.09$ (red) and $x = 0.3$ (olive). The experimental temperature is shown in units of $T_{c0}$. **d,** Schematic illustration of the Fermi surface in $CsV_3Sb_{5-x}Sn_x$, with colour-coded arrows highlighting the dominant scattering vectors. The dashed dark lines denote the first Brillouin zone. **e-h,** Twofold-symmetrized quasiparticle interference patterns at several energies measured in $CsV_3Sb_{4.91}Sn_{0.09}$: **e,** -5 meV, away from the superconducting gap; **f, g,** -0.32 mV and -0.24 mV, within the superconducting gap; **h,** 0 meV, deep inside the gap (see Method and Supplementary Note 15). The quasi-1D scattering event labelled as $\mathbf{q_1}$ persists to zero energy even at the lowest temperature. STM setups: **c**. $V_s = 1.2$mV, $I_t = 0.15$nA, $V_m = 30$ μV for $x = 0$; $V_s$ =-1mV, $I_t = 0.6$nA, $V_m = 10$μV for $x = 0.09$; $V_s = 2$mV, $I_t = 1$nA, $V_m = 40$μV. **e**, $V_s = -5$mV, $I_t = 1$nA, $V_m = 2$mV; **f-h**. $V_s = -2$mV, $I_t = 0.5$nA, $V_m = 50$μV.

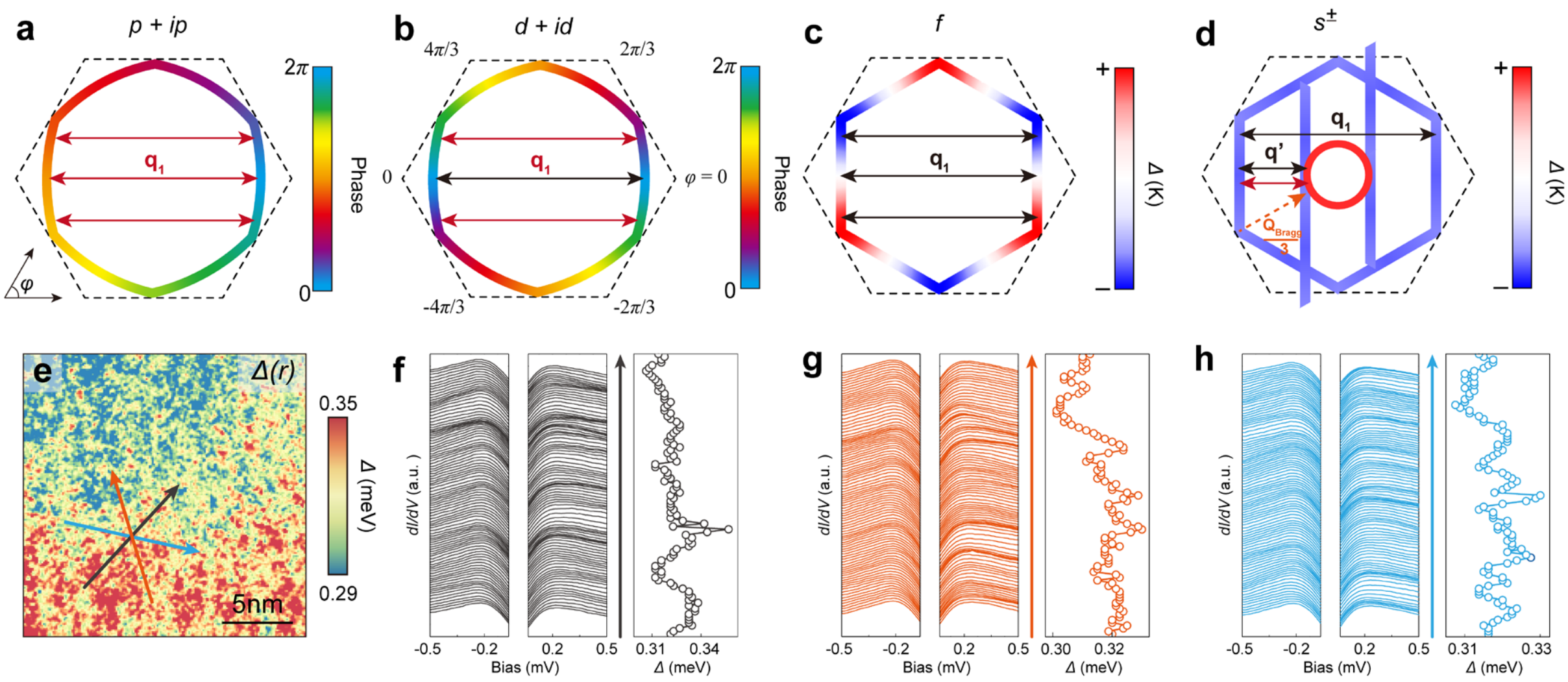

**Fig. 5 | Possible sign-reversal pairing states and the absence of PDW state. a-d,** Illustrations of four representative sign-reversal pairing states: **a,** chiral *p*-wave (*p*+*ip*); **b,** chiral *d*-wave (*d*+*id*); **c,** *f*-wave; **d,** $s^{+-}$ wave. Colours in **a** and **b** indicate the phase of superconducting order parameter, which varies as a function of the polar angle $\varphi$ (inset of **a**). In **c** and **d**, red and blue colours denote the positive and negative sign of superconducting order parameter, respectively, with their darkness implying the gap amplitude. The dashed hexagons mark the boundary of the first Brillouin zone; coloured hexagon and circle rings represent the Fermi surface contributed by the vanadium (kagome) *d*-orbitals and Sb *p*-orbitals, respectively. Black arrows denote scattering processes that preserve the superconducting order, whereas red arrows indicate scattering events potentially cause breaking of Cooper pairs. The two quasi-1D bands (blue) in **d** are folded *d*-electron bands induced by the charge-stripe order. **e**, Gap map with higher spatial resolution of a 20nm × 20 nm region of $x$ = 0.09, showing no signature of PDW state. **f-h**, Left panel: d$I$/d$V$ linecuts along three different lattice directions (arrows in **e**, black, ochre and cyan for **f**, **g** and **h**, respectively); right panel: corresponding gap variation, both show absence of discernible periodic modulation.

**Supplementary Information for**

# Heterogeneous superconductivity emerging from a charge-stripe state in kagome lattices

Hongyu Li[1], Shuikang Yu[2], Huachen Rao[1], Zhuying Wang[2], Xingchen Zhou[2], Houpu Li[2], Yunmei Zhang[2], Zhao Gong[1], Xin Yu[3], Rui Wu[3], Tao Wu[1,4], Zhenyu Wang[2,4], Ziji Xiang[1,4], Xianhui Chen[1,2,4]

[1] *Hefei National Research Center for Physical Sciences at the Microscale, University of Science and Technology of China, Hefei, 230026, China*

[2] *Department of Physics, University of Science and Technology of China, Hefei, 230026, China*

[3] *Beijing National Laboratory for Condensed Matter Physics and Institute of Physics, Chinese Academy of Sciences, Beijing 100190, China*

[4] *Hefei National Laboratory, University of Science and Technology of China, Hefei, 230088, China*

**This PDF File includes:**

**Supplementary Note 1:**

**Determination of Sn doping level *x* and local dopant concentrations**

The Sn doping level $x$ was determined by two methods in this work, i.e., the energy-dispersive X-ray spectroscopy (EDS) measurements on single crystals and a direct counting of the number of Sn dopants within the field of view of STM. The EDS results for samples with five nominal doping levels are presented in Supplementary Table 1; the actual compositions are close to the nominal values with a spatial inhomogeneity less than 10%. For the STM-based estimation, we note that there are two types of Sb sites in the crystalline structure of $CsV_3Sb_5$: Sb1 sites in the hexagonal center of the V-kagome lattice and Sb2 site that form two honeycomb layers above and below each kagome layer (Supplementary Fig. 1a). The Sn dopant (substitutions to the Sb atoms) appears as a dark depression in the STM topographies with its center located right in the middle of six topmost Sb atoms (Supplementary Fig. 1b), indicating that the Sn dopants favour to substitute the Sb atoms in the $V_3Sb$ kagome plane (the Sb1 site), in agreement with previous results [1]. By counting the number of these defects in large fields of view (typically 50-nm-squares), the concentrations can be determined with high accuracy. Supplementary Fig. 1c shows the local dopant concentration counted from topographies obtained on seven samples with the nominal composition $x$ = 0.09; a weak spatial variation of $x$ among these samples is consistent with the EDS results (Supplementary Table 1), whereas the average concentration is slightly below 0.09 ($x \approx 0.076$). Nevertheless, we use the nominal compositions across the manuscript since such a small deviation does not alter our conclusions.

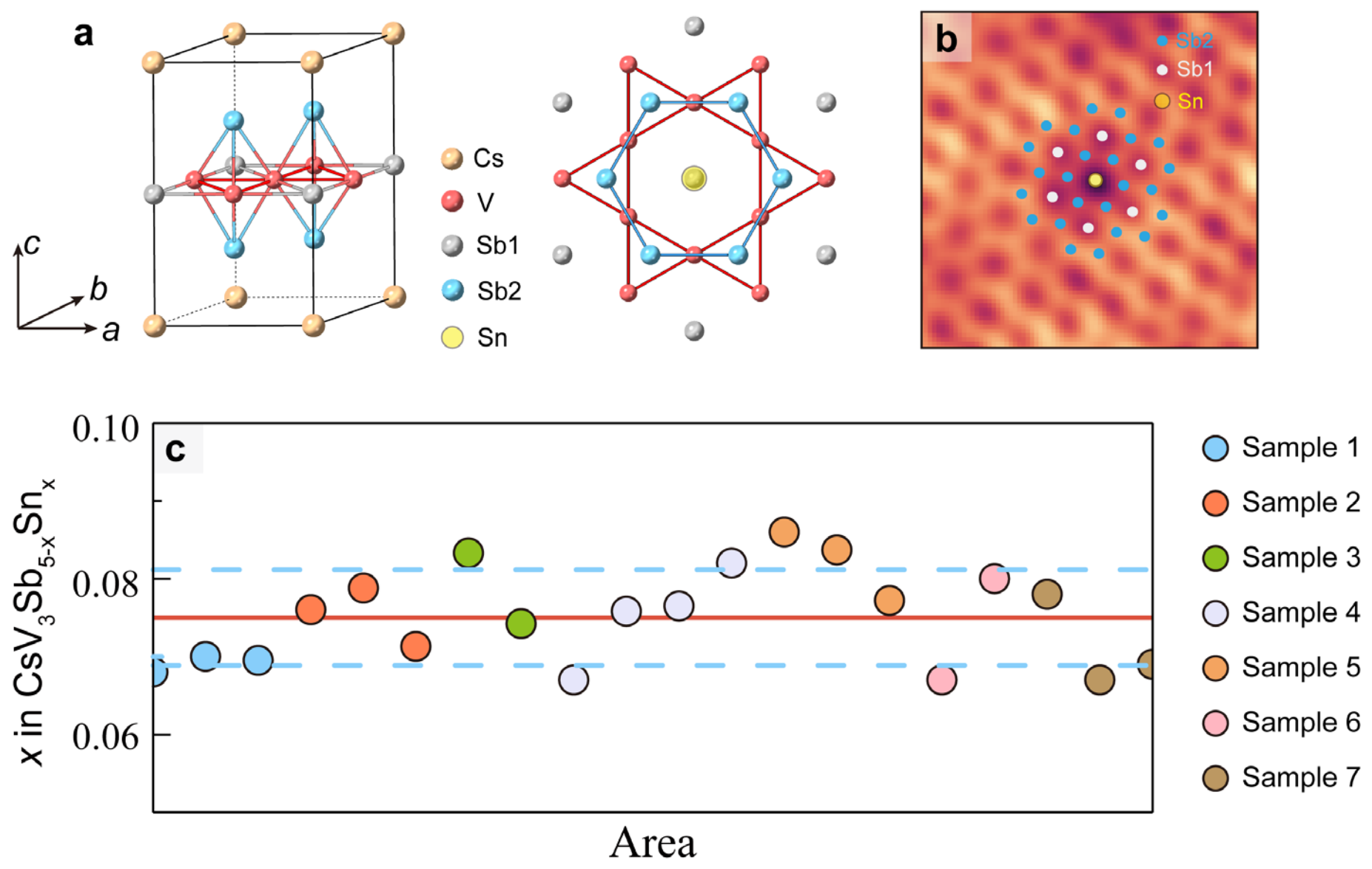

**Supplementary Fig. 1. Local Sn concentration determined from the STM images measured for $CsV_3Sb_{4.91}Sn_{0.09}$. a,** A schematic showing the crystal structure of $CsV_3Sb_5$ and the position of an Sn dopant. **b**, STM topographic image showing the Sb2 honeycomb lattice and one doped Sn atom at the Sb1 site. **c,** Local Sn concentrations determined by counting the Sn atoms in different regions of seven $CsV_3Sb_{4.91}Sn_{0.09}$ samples. The orange solid line indicates the mean value, while the blue dashed line indicates the standard deviation, revealing minor variations.

**Supplementary Table 1. EDS results of the doping level $x$ of $CsV_3Sb_{5-x}Sn_x$** with nominal $x = 0.06$, 0.07, 0.09, 0.12 and 0.30. For each sample, five areas were measured, and $x$ was determined from the average of areas 1-5.

| **Nominal *x*** | **0.06** | **0.07** | **0.09** | **0.12** | **0.30** |
|---|---|---|---|---|---|
| area 1 | 0.0541 | 0.0687 | 0.0959 | 0.1152 | 0.311 |
| area 2 | 0.0631 | 0.0617 | 0.0852 | 0.1206 | 0.318 |
| area 3 | 0.0627 | 0.0799 | 0.0849 | 0.1269 | 0.287 |
| area 4 | 0.0655 | 0.0745 | 0.0916 | 0.1260 | 0.312 |
| area 5 | 0.0567 | 0.0717 | 0.0853 | 0.1215 | 0.287 |
| **Average** | **0.0604** | **0.0713** | **0.0886** | **0.1220** | **0.301** |

We further note that for a given surface region, the Sn concentration and the one-of-a-kind QPI features for the charge-stripe order have been checked before studying the superconducting state. As displayed in Supplementary Figure 2, the normal-state tunneling spectra of $CsV_3Sb_{4.91}Sn_{0.09}$ are uniform without notable spatial variation on the length scale of tens of nanometers. We conclude that the normal state can be regarded as spatially homogeneous in all our samples (including those displaying remarkable inhomogeneity in the superconducting state).

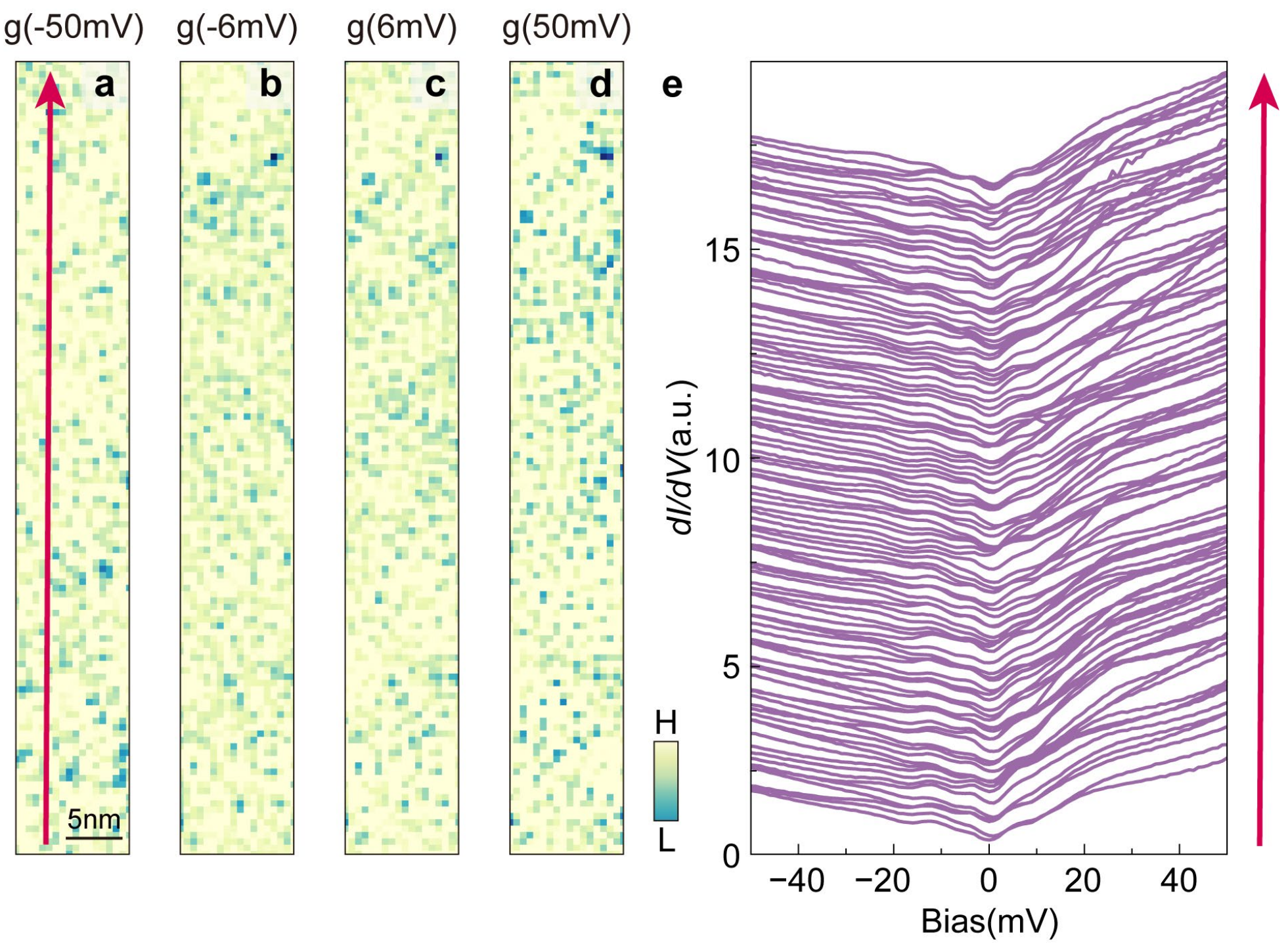

**Supplementary Fig. 2. Normal-state d*I*/d*V* maps and linecut for $CsV_3Sb_{4.91}Sn_{0.09}$. a-d**, d*I*/d*V* maps taken at bias voltages of -50 mV **(a)**, -6 mV **(b)**, 6 mV **(c)** and 50 mV **(d)** in the same 10 × 70 nm$^2$ field of view shown in main text Fig. 3. **e**, d*I*/d*V* profile along the purple arrow shown in **a**.

**Supplementary Note 2:**

**CDW transitions in $CsV_3Sb_{5-x}Sn_x$ observed by means of resistivity**

The transition entering the triple-$Q$ CDW phase manifests itself as a slope change in the temperature-dependent resistivity $\rho_{xx}(T)$, as seen on the $x = 0$ and 0.06 curves in Supplementary Fig. 3a. Such feature vanishes once the short-range single-$Q$ stripelike CDW emerges upon Sn doping at $x = 0.07$ [1]; nevertheless, a very broad anomaly can still be seen in the temperature derivative of resistivity, $d\rho_{xx}(T)/dT$, the onset of which is marked by arrows in Supplementary Fig. 3b. We identify this temperature as $T_{CDW}$ for $x = 0.07$ and 0.09, as displayed in main text Fig. 1. At higher $x$ ($\geq 0.12$), the anomaly in $d\rho_{xx}(T)/dT$ cannot be resolved, implying that the stripelike CDW becomes increasingly short-ranged and may eventually develop a fluctuating nature [2]. This is underpinned by the disappearance of band-folding signatures induced by the single-$Q$ CDW at $x \geq 0.12$ revealed by the STM measurements (see Supplementary Fig. 15 and Supplementary Note 12). We stress that in the present study, all experimental fingerprints of stripelike CDW (and its impact on superconductivity) are significantly weakened at $x \geq 0.12$; in this high-doping regime, the electronic structure closely resembles a pristine kagome lattice with the absence of superlattice reconstruction.

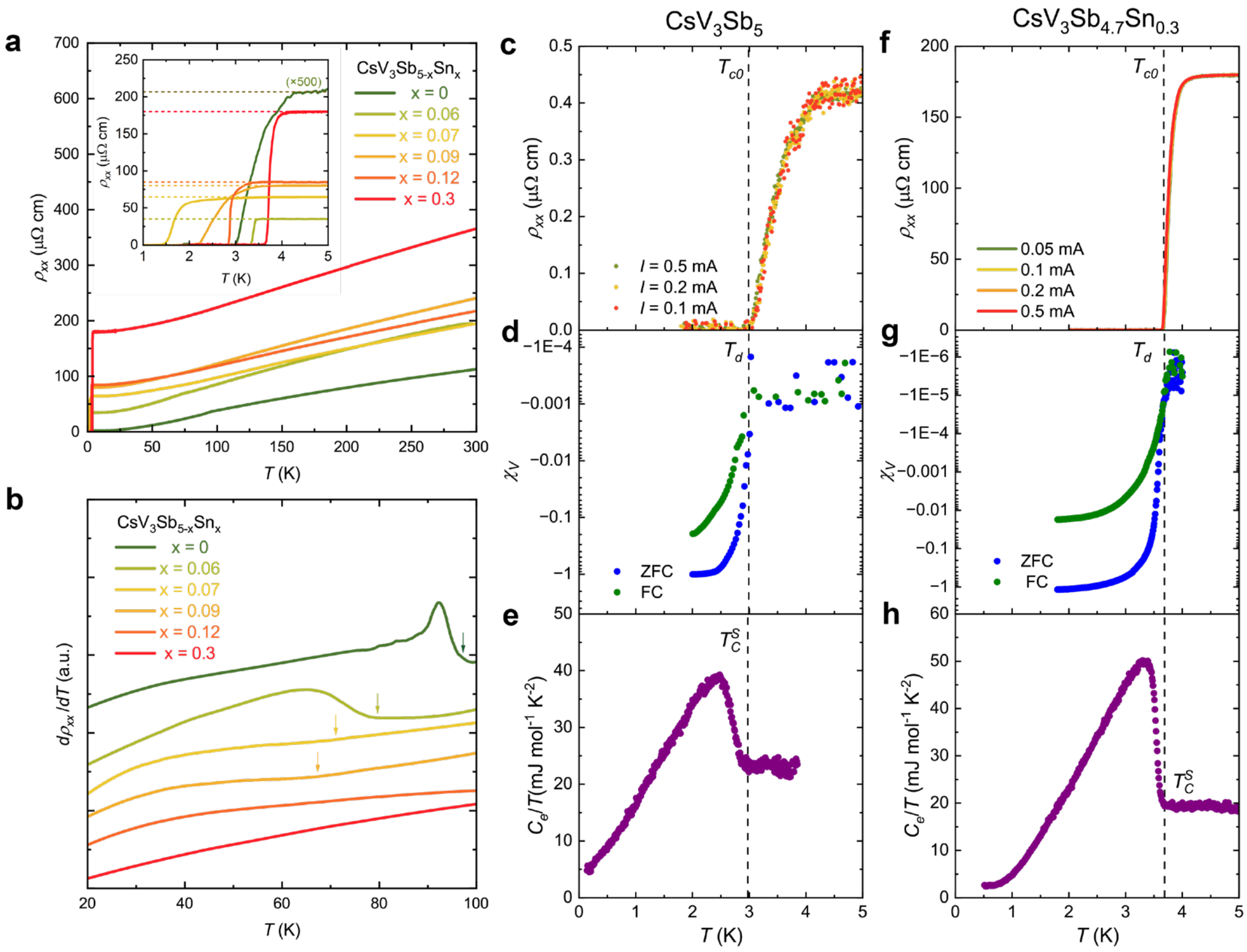

**Supplementary Fig. 3. *T*-dependent resistivity of $CsV_3Sb_{5-x}Sn_x$ and the characteristic temperatures of superconductivity for $x = 0$ and 0.30. a,** $\rho_{xx}$ measured from 0.4 K to 300 K in samples with different $x$. Inset displays a zoomed view of the superconducting transitions. Note that despite strongly enhanced residual resistivity (dotted lines in the inset) in the Sn-doped samples, metallic behavior persists up to $x = 0.30$ (red curve). **b,** Temperature derivative of longitudinal resistivity, $d\rho_{xx}/dT$, as a function of $T$ for the six compositions shown in **a**. Arrows indicate the onset of the anomaly manifesting the CDW transition. At $x = 0.12$, this transition becomes invisible. **c-e,** In-plane resistivity $\rho_{xx}$ (**c**), volume magnetic susceptibility $\chi_v$ (**d**) and electronic specific heat $C_e/T$ (**e**) measured in undoped $CsV_3Sb_5$. **f-h,** same as **c-e** but for a sample of $x = 0.30$. The characteristic temperatures of superconducting states, i.e., $T_{c0}$, $T_d$ and $T_C^S$ are identical for $x = 0$ and $x = 0.30$ (vertical dashed lines). $C_e/T$ data in (**h**) may contain a small Schottky contribution at the lowest $T$.

**Supplementary Note 3:**

**Upper critical fields and superconducting coherence lengths**

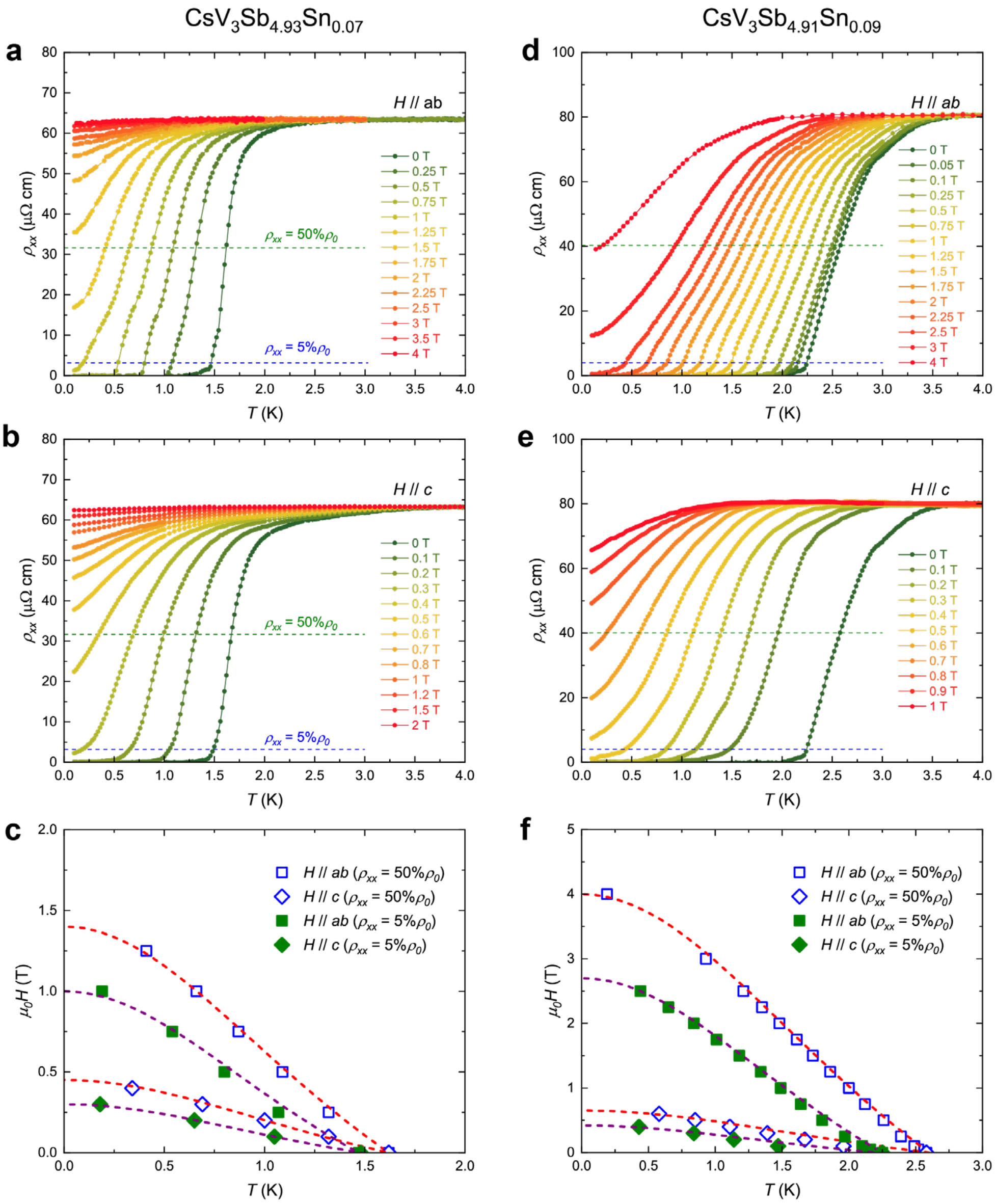

**Supplementary Fig. 4. Upper critical fields of $CsV_3Sb_{5-x}Sn_x$ ($x$ = 0.07 and 0.09). a, b, d, e,** $\rho_{xx}(T)$ measured under varying magnetic fields in samples with $x = 0.07$ (**a** and **b**) and $x = 0.09$ (**d** and **e**), with $H$ applied in the $ab$ plane (**a** and **d**) and along the $c$ axis (**b** and **e**), respectively. All data were measured with an applied current $I = 0.1$ mA. Dashed horizontal lines denote the 50% (green) and 5% (blue) criteria (with regard to the normal-state $\rho_{xx}$) of determination of $H_{c2}$. **c, f,** The upper critical field $H_{c2}(T)$ for $x = 0.07$ (**c**) and 0.09 (**f**). Symbols are values obtained using the 50% and 5% criteria. Dashed lines are trajectories predicted by the GL models.

The zero-temperature values of in-plane and out-of-plane coherence lengths ($\xi_{ab}$ and $\xi_c$, respectively) can be derived from the measurement results of upper critical fields $H_{c2}^{ab}$ and $H_{c2}^{c}$. According to the GL relations for an anisotropic three-dimensional (3D) superconductor [3], $\mu_0 H_{c2}^{c} = \Phi_0/(2\pi\xi_{ab}^2)$, $\mu_0 H_{c2}^{ab} = \Phi_0/(2\pi\xi_{ab}\xi_c)$ ($\Phi_0 = 2.07\times10^{-15}$ Wb is the superconducting magnetic flux quantum). We determined $H_{c2}^{ab}$ and $H_{c2}^{c}$ to be the $H$-fields at which the resistivity reaches 50% of the normal-state value at each temperature (Supplementary Fig. 4; a 5% criterion was also used for comparison). The $H_{c2}(T)$ data are generally tracked by the GL fits, $\mu_0 H_{c2}(T) = \mu_0 H_{c2}(0)[(1-t^2)/(1+t^2)]$, where $t = T/T_c$ (Supplementary Fig. 4); the zero-$T$ limit was subsequently approached by extrapolating the fits to $T$ = 0. Under the 50% criterion, $\xi_{ab}$ ($\xi_c$) is estimated to be 27.0 (8.7) and 22.5 (3.8) nm for samples with $x$ = 0.07 and $x$ = 0.09, respectively. $\xi_c$ far exceeds the distance $d$ between adjacent kagome planes ($d \sim 0.9$ nm [4]), corroborating the validity of our anisotropic 3D GL modeling. The values of $\xi_{ab}$ determined using the 5% criterion are 33.1 nm for $x$ = 0.07 and 28.7 nm for $x$ = 0.09. Furthermore, a parallel estimation of $\xi_{ab} \approx 41.1$ nm can be extracted from an exponential fit to the zero-energy density of states (DOS) distribution around a vortex core in STM measurements, see Supplementary Note 7 and Supplementary Fig. 7d-f.

**Supplementary Note 4:**
**The contribution of superconducting fluctuations**

Aside from the local $T_c$ inhomogeneity, the resistivity rounding above the superconducting transition (main text Fig. 1g and j, Supplementary Fig. 3a, inset) can potentially result from strong superconducting fluctuations [5]. Here we consider two types of fluctuations of the superconducting order parameter: (i) The Gaussian amplitude fluctuations [6] and (ii) the phase fluctuations [7]. The influence of (i) is usually analyzed in the GL scheme, with the corrections to the normal-state kinetic characteristics described by the well-known Aslamazov-Larkin (AL) theory [6]. For an anisotropic 3D superconductor at temperatures not far above $T_c$, the AL expressions for fluctuating diamagnetism and paraconductivity in the weak field limit are [8-10]:

$$\chi_{\mathrm{AL}} \simeq \frac{2\pi^2}{3}\frac{k_B T_c}{\Phi_0^2}\frac{\xi_{ab}^2}{\xi_c}\left(\frac{T-T_c}{T_c}\right)^{-\frac{1}{2}} \quad \text{(S1)}$$

$$\Delta\sigma_{\mathrm{AL}} \simeq \frac{e^2}{32\hbar}\frac{1}{\xi_c}\left(\frac{T-T_c}{T_c}\right)^{-\frac{1}{2}} \quad \text{(S2)}$$

The AL simulations displayed in main text Fig. 2 were constructed with $T_c$ defined as $T_c^{\mathrm{s}}$ and the coherence lengths ($\xi_{ab}$, $\xi_c$) using the estimations in Supplementary Note 3 (the 50% criterion).

The effect of phase fluctuations can be significant in two-dimensional (2D) superconducting systems, wherein long-range phase coherence is established at a Berezinskii-Kosterlitz-Thouless (BKT) transition occurring at $T_{\mathrm{BKT}}$ below the mean-field $T_c^{\mathrm{MF}}$ [7,8,11]. In bulk superconductors, phase fluctuation only becomes important when the superfluid density is low [8], or the system is close to a crossover towards Bose-Einstein condensation [10]. Considering the moderate anisotropy of $H_{c2}$, $\Gamma = H_{c2}^{ab} / H_{c2}^{c} \sim$ 3-6, inherent phase fluctuations are unexpected in $CsV_3Sb_{5-x}Sn_x$. Nevertheless, we also estimated the possible contributions from this effect, whose fluctuating diamagnetism and paraconductivity are given by the Halperin-Nelson formulas [7,11] in the weak-field limit:

$$\chi_{\mathrm{HN}} \simeq \frac{k_B T}{2\Phi_0^2}\frac{\xi_{ab}^2}{d}\sinh^2\left(\frac{b}{\sqrt{\varepsilon}}\right) \quad \text{(S3)}$$

$$\frac{\Delta\sigma_{\mathrm{HN}}}{\sigma_n} \simeq \left(\frac{2}{A}\right)^2 \sinh^2\left(\frac{b}{\sqrt{\varepsilon}}\right) \quad \text{(S4)}$$

where $A$ is a constant with the order of unity, $d$ is the thickness of superconducting layer, $\varepsilon = (T - T_{BKT})/T_{BKT}$, $b \simeq 2(T_c^{MF}/T_{BKT} - 1)^{1/2}$. In practice, we took $A = 22.2$, $d = 0.9$ nm (spacing between adjacent kagome planes), $T_{BKT} = T_{c0}$, $T_c^{MF} = T_d$. (Again, we emphasize that this modeling is completely formal and may not have a clear connection with the underlying physics in our samples.)

Simulations presented in main text Fig. 2a-d based on both the AL [Eqs. (S1) and (S2)] and the HN [Eqs. (S3) and (S4)] models fail to trace the evolution of $\chi_d$ and $\Delta\sigma$ with increasing $T$ --- yielding neither the correct amplitudes nor the exponential manner of decay. In high-$T_c$ cuprates where the spatial inhomogeneity has been extensively studied [12-15], only the $T$-dependence of the superconducting precursor deviates from Eqs. (S1)-(S2) with a faster suppression upon warming. Whether this is due to a percolation process [13,14] or a short-wavelength cutoff [8,16,17] is debatable. In our case, however, the stark disagreement in the amplitudes effectively invalidates any interpretations referring to fluctuation. Therefore, we propose that the profound superconducting precursor effect in stripe-ordered $CsV_3Sb_{5-x}Sn_x$ is not caused by strong superconducting fluctuations.

**Supplementary Note 5:**
**Additional information about the EMT model**

As shown in main text Fig. 2, the two-component EMT model (Methods and Supplementary Fig. 5a) yields fitting parameters for $CsV_3Sb_{4.93}Sn_{0.07}$ and $CsV_3Sb_{4.91}Sn_{0.09}$: $T_0 = 1.3$ K, $\Xi_1 = 1.2$ K, $\Xi_2 = 0.18$ K, $u = 0.001$ and $T_0 = 2.0$ K, $\Xi_1 = 0.65$ K, $\Xi_2 = 0.2$ K, $u = 0.01$, respectively. Following Eq. (1) in Methods, these values yield $T_\pi = 1.5$ K for $x = 0.07$ and 2.1 K for $x = 0.09$. The relatively wide distribution of $T_c$, characterized by the parameter $\Xi_1$, implies a very small fraction of the sample that hosts a considerably higher $T_c$ (and presumably larger gap) than the average: As shown in Supplementary Fig. 5b, the high-$T$ "tail" of $T_c$ distribution above the cut-off by the second, narrower distribution (with $\Xi_2 < \Xi_1$) only contributes to about 0.1% of the total volume fraction. Obviously, such a small portion is unlikely to be captured by STM measurements. We mention that $T_0$ does not correspond to any of the observed characteristic temperatures ($T_d$, $T_c^s$ or $T_{c0}$) and $T_\pi$ is close to (but slightly higher than) $T_{c0}$; $\Xi$, while it may reflect the real $T_c$ distribution to some extent, is not a specific parameter as well (Supplementary Fig. 5c and d) [14]. This notwithstanding, the EMT analysis describes the superconducting precursors in the charge-striped $CsV_3Sb_{5-x}Sn_x$ quite satisfactorily. The proximity effect simulated by the inclusion of $g_2$ generally reproduces the abruptly boosted superconducting fraction at $T \simeq T_c^s$ (main text Fig. 1h, i, k and l); it engenders an expansion of superconducting clusters by bringing their surrounding metallic regions superconducting (main text Fig. 2g).

Nevertheless, our model ignores the inter-grain Josephson coupling, which may also become preponderant close to the bulk $T_c$. Hence, the fits become less reliable in the vicinity of $T_{c0}$. Note that the resistive tail above $T_{c0}$ (main text Fig. 1g and j) hints at a possibility that the network of clusters manifests a one-dimensional-type spatial correlation [18], in accord with the stripelike textures of charge order in the normal state.

We emphasize that the EMT does not impose strict constraints on the characteristic length scale of gap variations. The form of EMT fitting function [Eqs. (1) and (2) in Methods] does not include information about the size of superconducting patches, or grains/clusters in the 3D case. In principle, we would expect that the superconducting gap varies over a length scale set by the superconducting coherence length $\xi$ [19], as discussed in detail in Supplementary Note 7. Since the EMT analysis was performed for millimeter-sized samples, we do not see any inconsistency over these relevant length scales.

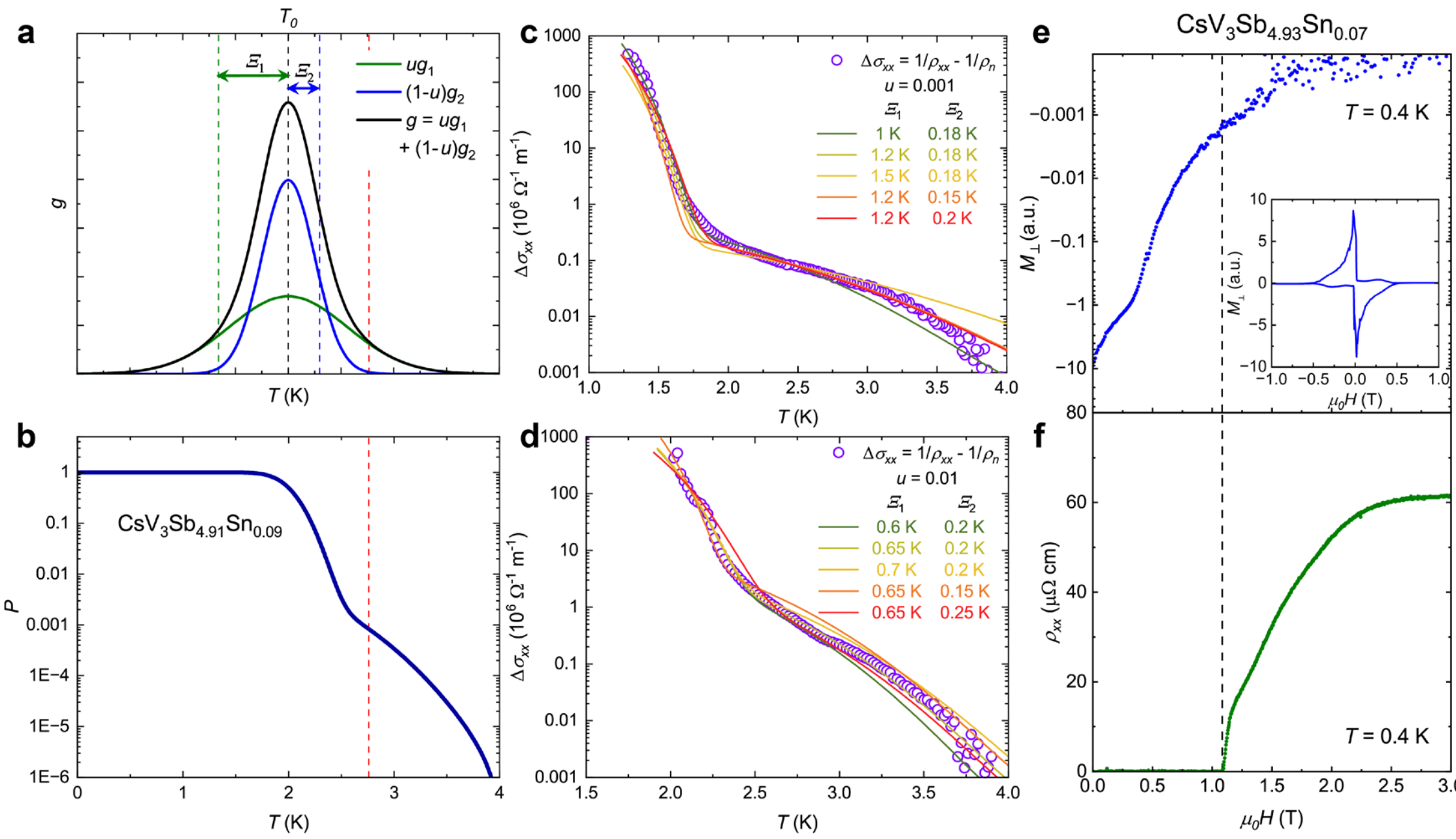

**Supplementary Fig. 5. The EMT analysis of the paraconductivity in $CsV_3Sb_{5-x}Sn_x$ ($x$ = 0.07 and 0.09). a,** The two-component distribution $g(T)$ of local superconducting transition temperature $T_c$: $g(T) = ug_1(T) + (1\text{-}u)g_2(T)$, with $u$ being a weighing factor; $g_1(T)$ (olive) and $g_2(T)$ (blue) are two Gaussians with the same mathematical expectation $T_0$ and half-width-at-half-maximum (HWHM) $\Xi_1$ and $\Xi_2$, respectively ($\Xi_1 > \Xi_2$ as illustrated by the vertical dashed lines). See Methods for more detail. **b,** The volume fraction of superconducting clusters $P$ at a temperature $T$ given by integrating $g(T')$ from $T' = T$ to $+\infty$. Here we adopted the fitting parameters that are applied to the $x$ = 0.09 sample (main text Fig. 2c): $T_0$ = 2.0 K, $\Xi_1$ = 0.65 K, $\Xi_2$ = 0.2 K and $u$ = 0.01. Red vertical dashed lines in **a** and **b** indicate the temperature at which the distribution $g_2$ (presumably stems from the proximity effect and inter-grain Josephson coupling) becomes unimportant. **c, d,** Contrasting the simulations of EMT model utilizing varying parameters $\Xi_1$ and $\Xi_2$ with the data of paraconductivity for $x$ = 0.07 (**c**) and $x$ = 0.09 (**d**). Here $u$ is fixed at 0.001 in **c** and 0.01 in **d**, referring to the superconducting volume fraction estimated from diamagnetic susceptibility (main text Fig. 1). **e, f,** A contrast of the transverse magnetization $M_\perp(H)$ (**e**) derived from torque magnetometry (Methods) and $\rho_{xx}(H)$ (**f**) measured in a $CsV_3Sb_{4.93}Sn_{0.07}$ sample at $T$ = 0.4 K. $H$ is applied within the $ab$ plane. Inset of **e** displays the full hysteresis loops of $M_\perp$. Vertical dashed line marks the onset field of zero resistivity.

**Supplementary Note 6:**

**Determination of superconducting gap size from STM spectra**

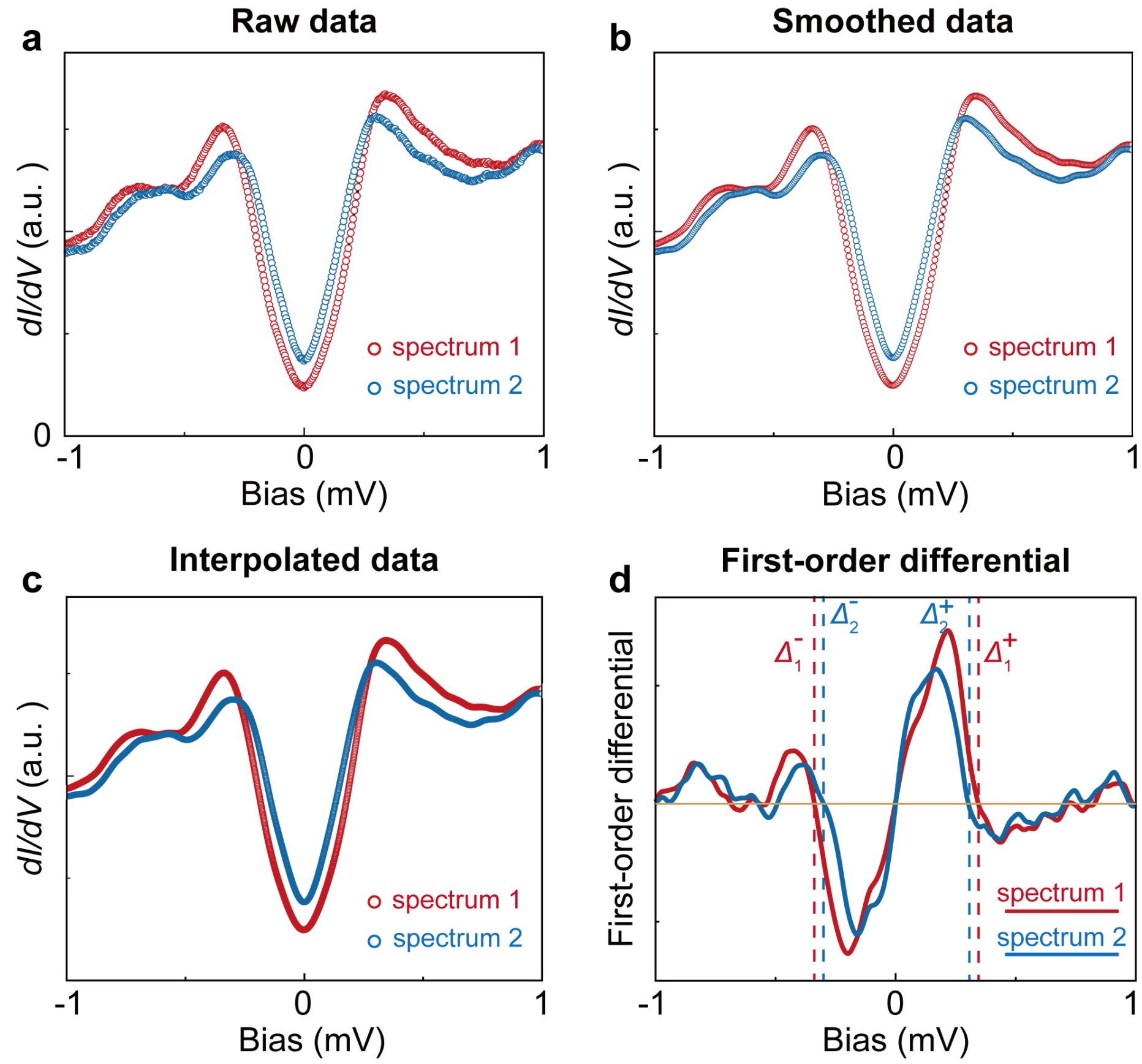

**Supplementary Fig. 6. Example of superconducting gap extraction from tunneling spectra**. **a,** Two raw tunneling spectra with different superconducting gap sizes. **b,** The same spectra after Gaussian smoothing. **c,** Interpolated spectra obtained using second-degree polynomials. **d,** First-order differential of the interpolated spectra calculated by the central-difference method. The zero-crossings marked by the dashed lines identify the coherence-peak positions at $\Delta^-$ and $\Delta^+$, from which the superconducting gap is defined as $\Delta = (\Delta^+ + \Delta^-)/2$.

**Supplementary Note 7:**
**Characteristic length of the spatial inhomogeneity of superconductivity at $x = 0.09$**

According to our STM data taken for $x = 0.09$, the characteristic length scale of gap variation can be determined as $L_\Delta \approx 58$ nm from the autocorrelation coefficients of the gap map (Supplementary Fig. 7a-c). Compared with $\xi_{ab} \approx 41.1$ nm extracted from DOS distribution around a vortex core (Supplementary Fig. 7d-f), we have $L_\Delta \sim (1.5\text{-}2)\xi_{ab}$; this is perfectly consistent with the expectation for a superconducting state with spatially inhomogeneous superconducting order parameter. Previous studies of the local order parameter in strongly disordered superconducting films have congruently revealed self-organized superconducting islands/regions with a size of a few times the coherence length [20-23]. Our observations in $CsV_3Sb_{4.91}Sn_{0.09}$ strongly imply that the superconducting gap $\Delta$ varies over a spatial distance set by (but not necessarily precisely equal to) $\xi_{ab}$. The close correspondence between these two length scales $L_\Delta$ and $\xi_{ab}$ verifies that the variation of $\Delta$ (over a distance of a few tens of nanometers) is not an arbitrary long-wavelength artifact, but rather reflects self-organized formation of inhomogeneous superconductivity induced by pair-breaking scattering stemming from short-ranged disorder potential (see further descriptions in Supplementary Note 15).

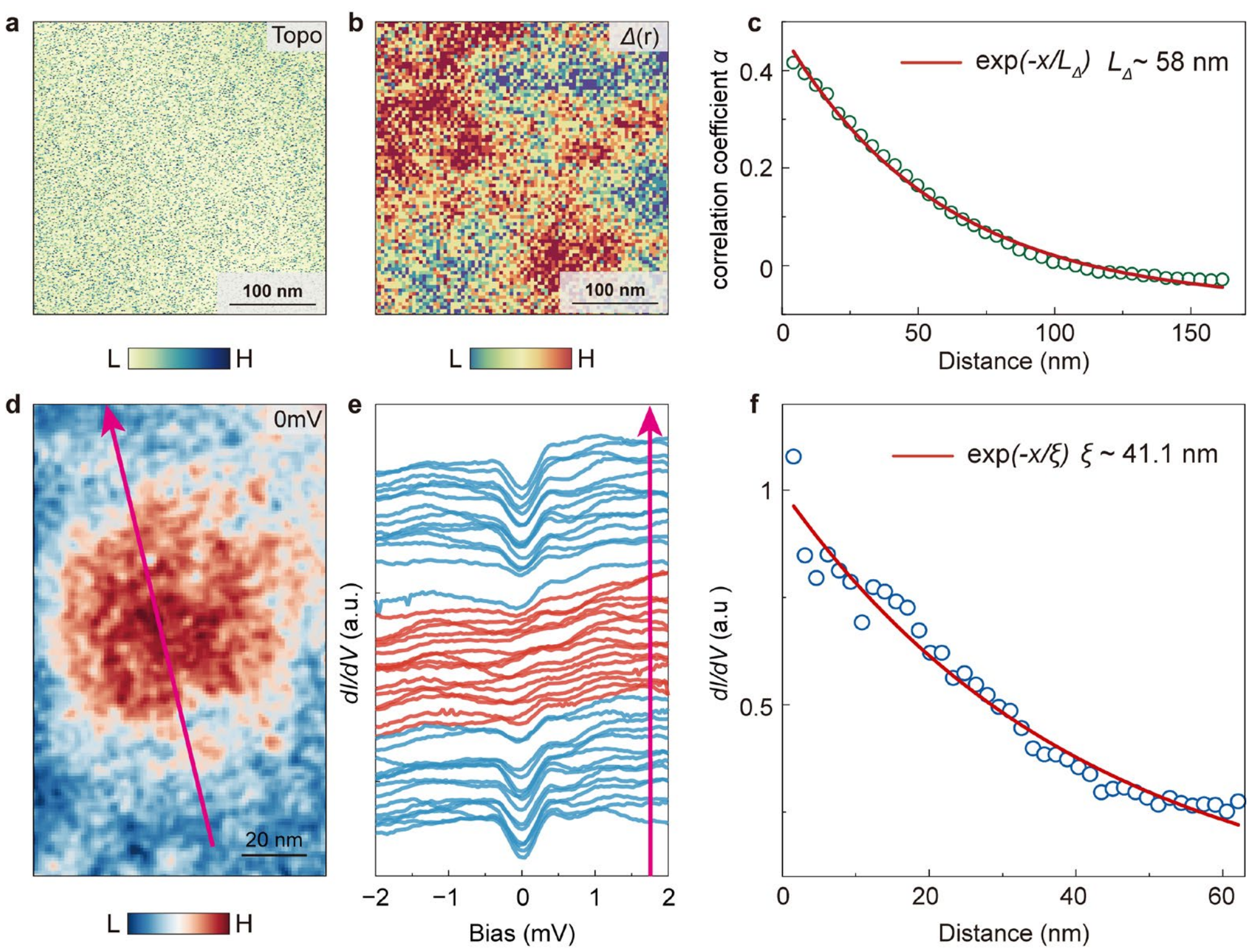

**Supplementary Fig. 7. Two interconnected length scales for a spatially inhomogeneous superconducting state.** **a**, Topographic image of a 500 nm × 500 nm Sb surface with sparse Cs atoms. **b**, Corresponding superconducting gap map. **c**, Azimuthally averaged autocorrelation coefficients of the gap map (green circles), exhibiting an exponential decay characterized by a length scale $L_\Delta \approx 58$ nm (orange line). **d**, Zero-bias *dI/dV* conductance map with an out-of-plane magnetic field of 0.15 T, showing a superconducting vortex. **e**, *dI/dV* line-profile along the pink arrow in **d**. The flat DOS inside the gap without well-defined Caroli-de Gennes-Matricon vortex states indicates the presence of strong scattering in these crystals. **f**, Exponential fit to the zero-energy DOS distribution around the vortex (azimuthally averaged), yielding a coherence length of $\xi \approx 41.1$ nm.

We further point out that the most likely origin of this pair-breaking disorder potential is the short-ranged charge stripes in Sn-doped $CsV_3Sb_5$, as emphasized in the discussion of the manuscript. According to data from other groups, the correlation length of the short-ranged charge-stripe order is 6–18 nm in $CsV_3Sb_{5-x}Sn_x$ with $x = 0.15$ [24]. In our samples, the correlation length of the short-ranged charge-stripe can be longer than this (and becomes close to $\xi_{ab}$), considering that the CDW correlation length appears to decrease with increasing $x$ [25]. It has been proposed in several influential theoretical works on cuprate superconductors [26-28] that when the Cooper pair coherence length is comparable to the correlation length of the disorder potential, the relevant pair-breaking causes the pair-field amplitude to be spatially heterogeneous. Consequently, the short-ranged charge stripe order with a correlation length of a few tens of nanometers promisingly serves as the source of pair-breaking scattering in our samples, given that the superconducting order parameter changes its sign in momentum space.

**Supplementary Note 8:**

**Spatially inhomogeneous pair breaking revealed by local tunneling spectra in $x = 0.09$ sample**

To verify the spatially inhomogeneous superconductivity in $CsV_3Sb_{4.91}Sn_{0.09}$, we performed spectroscopic-imaging STM measurements in a 40 nm × 40 nm field of view. The topographic image and the corresponding Sn dopant distribution are shown in Supplementary Fig. 8a and b. Similar to the data in main text Fig. 3, Sn dopants are distributed rather uniformly without evident clustering. In contrast, the superconducting gap map $\Delta$(r) exhibits a pronounced spatial variation over a length scale of several tens of nanometers (Supplementary Fig. 8c). The histogram of $\Delta$(r) gives an average gap value $\bar{\Delta}$ = 303 μeV with a standard deviation $\sigma$ = 16.4 μeV, corresponding to $\sigma/\bar{\Delta}$ = 0.054 (Supplementary Fig. 8e). This value is close to that obtained in main text Fig. 3, confirming that the superconducting gap inhomogeneity is reproducible in different regions of the same composition.

Moreover, similar to the field of view in main text Fig. 3, local spectra with a larger gap exhibit sharper coherence peaks and a deeper zero-bias suppression, implying a positive correlation between $\Delta$(r) and $H$(r) in this independent region (Supplementary Fig. 8d, f and g). This correlation suggests that the spatial variation of $\Delta$ is accompanied by gap filling and coherence-peak smearing, as expected for an inhomogeneous pair-breaking effect.

To examine whether the spectral evolution can be attributed to spatially varying pair breaking, we fitted the local tunneling spectra using the Dynes formula, $N(E) = Re[(E - i\Gamma)/\sqrt{((E - i\Gamma)^2 - \Delta(\theta)^2)}]$, where $\Gamma$ denotes the pair-breaking parameter. The superconducting gap was taken as a nodeless form with sixfold in-plane anisotropy, $\Delta(\theta) = \Delta_0 + \Delta_6 \cos(6\theta)$. A representative fit is shown in Supplementary Fig. 8h. By applying the same fitting procedure pixel by pixel to the dI/dV map, we obtained the spatial distribution of $\Gamma(r)$, as displayed in Supplementary Fig. 8i.

We then compared $\Gamma(r)$ with the superconducting gap map $\Delta(r)$. $\Gamma(r)$ is spatially inhomogeneous and anti-correlated with $\Delta(r)$: regions with smaller superconducting gaps exhibit larger $\Gamma$, whereas larger-gap regions show weaker pair breaking. This trend is summarized by the $\varDelta$-$\varGamma$ density plot in Supplementary Fig. 8j and is further confirmed by the negative cross-correlation between $\varDelta(r)$ and $\varGamma(r)$ in Supplementary Fig. 8k. These results indicate that the gap filling and suppression of coherence peaks cannot be described by a simple reduction of the gap with spatially constant broadening, but instead point to spatially inhomogeneous pair breaking as a likely origin of the observed superconducting gap inhomogeneity.

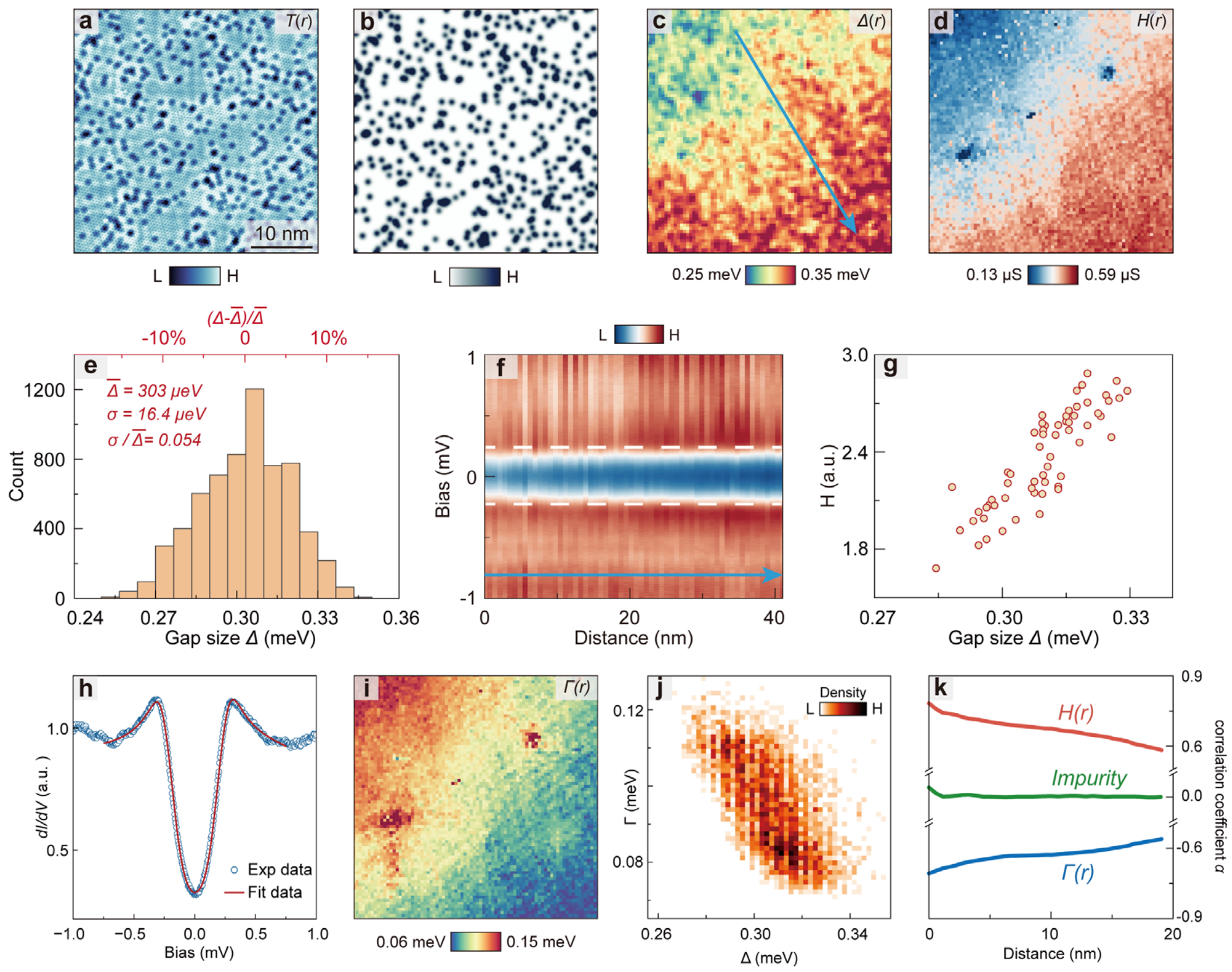

**Supplementary Fig. 8. Spatially inhomogeneous pair breaking in $CsV_3Sb_{4.91}Sn_{0.09}$. a-d,** Topographic image (**a**), dopant distribution (**b**), gap map (**c**) and gap-depth map (**d**) of a 40 nm × 40 nm field of view. **e**, Histogram showing the distribution of the gap amplitude in **c**. **f**, d*I*/d*V* linecut along the arrow in **c**, showing the spatial variation in the superconducting gap. The dashed white line serves as a reference. **g**, Gap depth ($H$) versus gap size ($\varDelta$) along the linecut, revealing a positive correlation. **h**, Representative local tunneling spectrum and fit using a nodeless superconducting gap with sixfold rotational symmetry. **i**, Spatial distribution of the fitted pair-breaking parameter $\varGamma(\mathbf{r})$ in the same region of **c**. **j**, Density plot of $\varGamma$ versus $\varDelta$, showing a clear anti-correlation between the fitted pair-breaking parameter and the local superconducting gap. **k**, Azimuthally averaged cross-correlation coefficients between the gap amplitude (**c**) versus the gap depth (**d**, red curve), the impurity dopants (**b**, green curve) and the pair-breaking parameter (**i**, blue curve).

**Supplementary Note 9:**
**Absence of inhomogeneous strain effects in the $x = 0.09$ sample**

It is well-established that the residual strain fields in a superconductor can effectively influence its superconducting properties [29]. If an inhomogeneous strain effect exists, it is likely to equally affect samples with all doping levels $x$ from 0 to 0.30; there must be specific reasons why the heterogeneous superconducting state only emerges within a narrow window of $x$ (most profoundly, $x = 0.07$ and 0.09), where the single-$Q$ stripelike CDW is also the most pronounced [25]. To nail down the influence of internal strain, we have conducted a series of experiments. These new data not only confirm our conclusions but also provide deeper insights into the unusual properties of stripe-ordered kagome lattices.

(i). We had measured the magnetoresistance (MR) and magnetic torque $\tau(H)$ in two more $CsV_3Sb_{5-x}Sn_x$ samples with $x = 0.09$ and the results are shown in Supplementary Fig. 9. Similar to the behavior displayed in main text Fig. 2e and f, the diamagnetic hysteresis loop in $\tau(H)$ of these two samples closes at a magnetic field $H_{irr}$ where the resistivity nearly recovers its normal-state value; $H_{irr}$ appears to be much higher than the vortex depinning field determined from the termination of the zero resistivity, which is quite rare in type-II superconductors. The consistency among observations in all $x = 0.09$ samples indicates that a coexistence of magnetic irreversibility and strongly dissipative transport is an inherent property of this composition (and is not caused by extrinsic mechanisms), verifying our identification of the heterogeneous superconducting state therein.

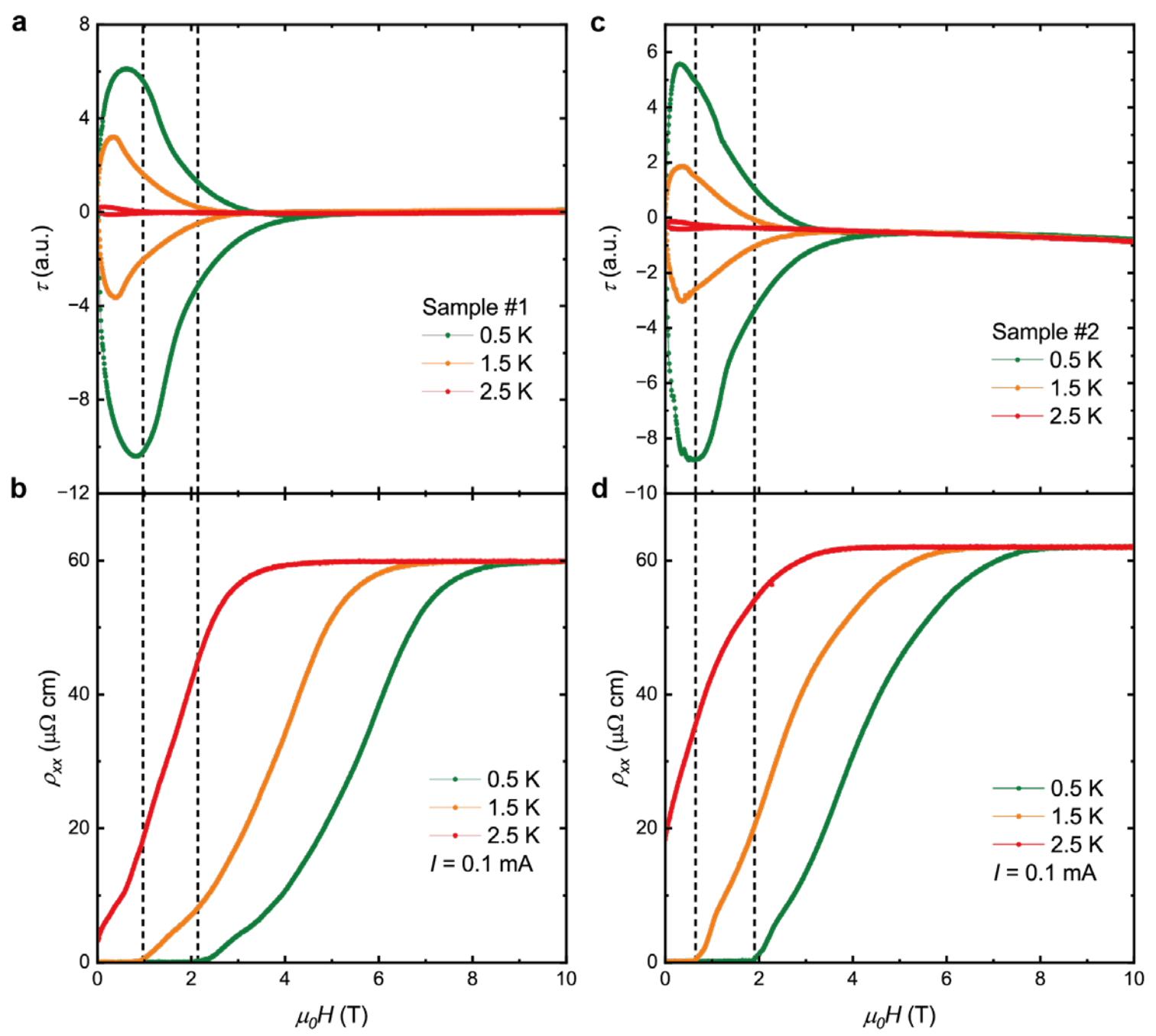

**Supplementary Fig. 9. The magnetoresistance (MR) and magnetic torque $\tau(H)$ measured in two additional $x = 0.09$ samples (#1 and #2).** A comparison of the torque magnetometry data (**a, c**) and $\rho_{xx}(H)$ (**b, d**) measured in $CsV_3Sb_{4.91}Sn_{0.09}$ samples #1 (**a** and **b**) and #2 (**c** and **d**) under in-plane $H$ at $T = 0.5$ K (olive), 1.5 K (orange) and 2.5 K (red). Dashed vertical lines denote the position of the onsets of the dissipative state (presumably the depinning field).

(ii). Temperature cycling has been used as an effective method to modify residual strains across the sample [29]. We have measured the in-plane resistivity $\rho_{xx}$ during repeated temperature cycling as plotted in Supplementary Fig. 10. In practice, an $x = 0.09$ sample was subjected to five successive thermal cycles within a temperature range of 1.8 to 5 K (as illustrated in Supplementary Fig. 10a); $\rho_{xx}(T)$ was collected during the cooling process of each cycle. We adopted a slow cooling rate of 3.3 mK/s (0.2 K/min) to effectively modify the strain inhomogeneity within the sample. As shown in Supplementary Fig. 10b and c, the $\rho_{xx}(T)$ data taken during the five cycles --- including the broadened superconducting transition --- fall exactly on top of each other, exhibiting no visible dependence on the history of temperature change. We thus infer that the influence of strain inhomogeneity on the transport behavior of $CsV_3Sb_{4.91}Sn_{0.09}$ is negligible.

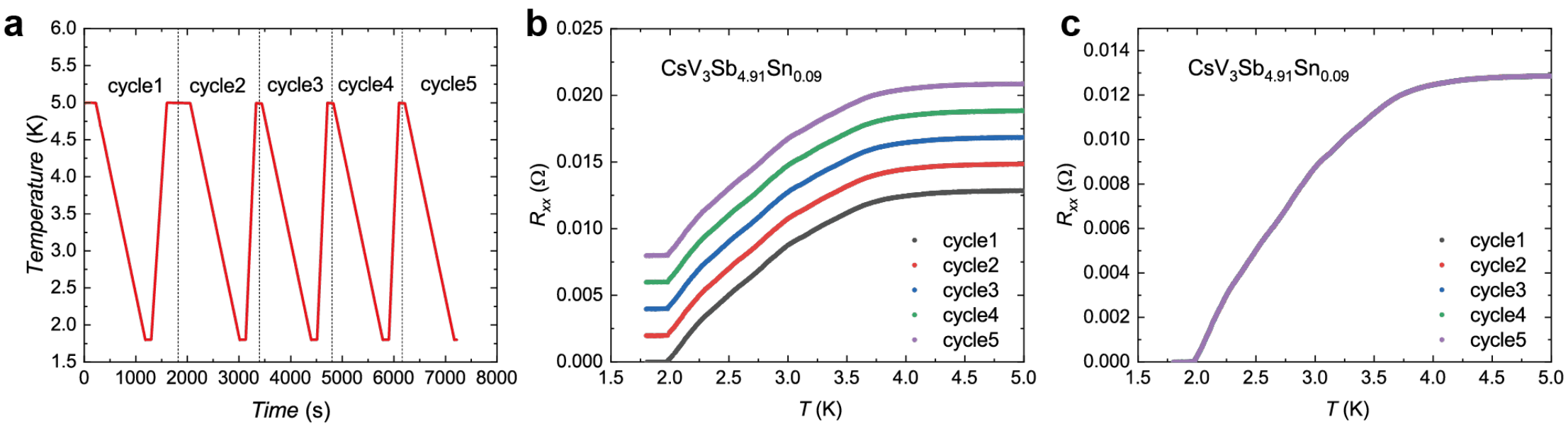

**Supplementary Fig. 10. Resistivity measurement during repeated thermal cycles across the superconducting transition. a,** A sketch showing a total of five thermal cycles over the temperature range from 1.8 to 5 K; $\rho_{xx}(T)$ was measured during the cooling process of each cycle. **b,** $\rho_{xx}(T)$ curves obtained for each cycle, which are shifted vertically for clarity. **c,** $\rho_{xx}(T)$ data without vertical offsets.

(iii). To further examine whether the observed superconducting inhomogeneity could originate from local strain, we analyzed the strain fields in two surface regions on an $x = 0.09$ sample; within each region, pronounced spatial inhomogeneity of the superconducting gap can be observed. The topographies and corresponding superconducting gap maps for the two fields of view (with sizes of $40 \times 40$ and $60 \times 60$ nm$^2$, respectively) are displayed in Supplementary Fig. 11a, b and g, h, respectively. For each region, we extracted the strain components along the three Bragg directions, $U_{aa}$, $U_{bb}$, and $U_{cc}$, as shown in Supplementary Fig. 11d-f and j-l, respectively.

The most important implication of this study is the absence of correlation between the spatial distribution of the superconducting gap, $\Delta(\mathbf{r})$, and any of the extracted strain components. None of the strain maps show a global spatial pattern that matches the inhomogeneity of $\Delta(\mathbf{r})$ [see the distinct trends of spatial variation when contrasting the patterns in Supplementary Fig. 11d-f(j-l) with Supplementary Fig. 11b(h)]. This is further confirmed by the cross-correlation analysis between the strain fields and the gap maps. As shown in Supplementary Fig. 11c and i, the correlation coefficients remain close to zero for all three strain components. Therefore, the superconducting gap inhomogeneity observed in the $x = 0.09$ sample cannot be attributed to a local strain distribution since the latter fails to track the local superconducting gap.

Combining (i), (ii) and (iii), we can confidently conclude that the unique properties of $CsV_3Sb_{4.91}Sn_{0.09}$, in particular the heterogeneous nature of its superconducting state highlighted by a spatial distribution of the superconducting gap, are more likely intrinsic for this compound. We cannot find evidence in our transport, torque magnetometry and STM studies that randomly distributed strain fields across the sample contributed to our observations, to any extent.

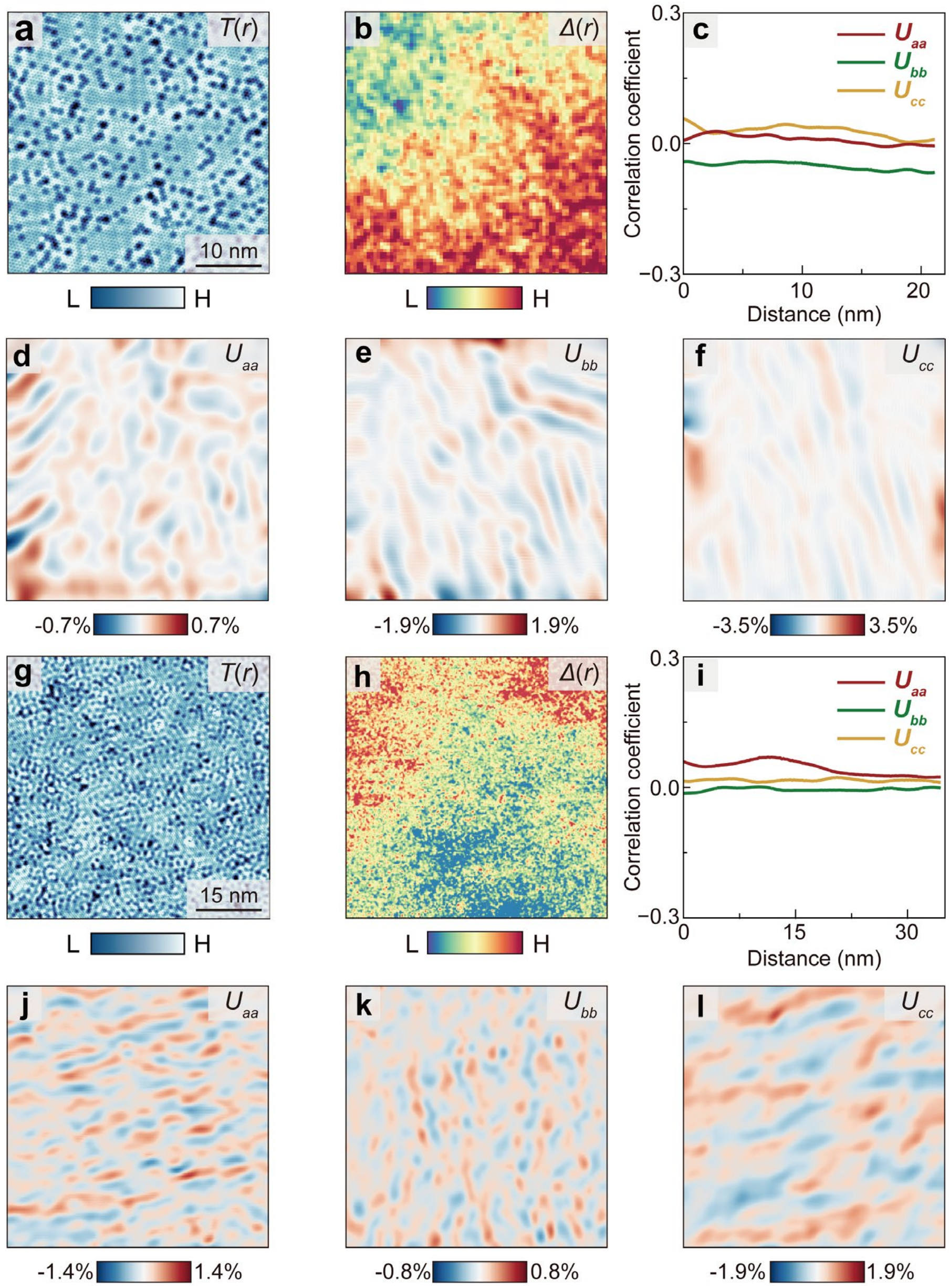

**Supplementary Fig. 11. Strain analysis based on STM measurements on a sample with *x* = 0.09. a, b,** Topographic image $T(\mathbf{r})$ (**a**) and superconducting gap map $\Delta(\mathbf{r})$ (**b**) in the first field of view with an area of 40 × 40 nm$^2$. **c,** The cross-correlation coefficients between $\Delta(\mathbf{r})$ and the strain components $U_{aa}$ (red), $U_{bb}$ (olive), and $U_{cc}$ (orange) along three Bragg directions. **d-f,** Strain maps extracted along the three Bragg directions, $U_{aa}$ (**d**), $U_{bb}$ (**e**), and $U_{cc}$ (**f**). **g-l,** Same as **a-f** but for data taken within a second field of view with an area of 60 × 60 nm$^2$. In both fields of view, the strain distributions do not show a global pattern matching the superconducting gap inhomogeneity; the correlation coefficients between the strain distribution and gap distribution are close to zero for all directions.

**Supplementary Note 10:**

**Influence of surface Cs clusters on the local superconducting gap**

Another trivial origin of the STM-resolved spatial distribution of the superconducting gap is the variation of the chemical environment at the surface. The local superconducting gap could change at boundaries between Cs-diluted and Cs-rich regions in Ta-doped $CsV_3Sb_5$ samples, probably due to surface electronic doping effects [30]. Therefore, one may doubt that the observed gap variation can originate from the Cs clusters in the surrounding area. In the following we present two key facts that exclude surface Cs clusters as a possible reason for the inhomogeneity in our case.

First, a large field of view that includes the area shown in Supplementary Fig. 8 is displayed in Supplementary Fig. 12a. The gradual variation in the superconducting gap from the up-left to the bottom-right (Supplementary Fig. 8c), at first glance, does not show any relationship with the surrounding Cs clusters. Second, we also found boundaries between Cs-diluted and Cs-rich regions in our Sn-doped samples, and, similar to the case reported in Ref. [30], the gap variation occurred abruptly at the boundary (Supplementary Fig. 12b-d). This abrupt change is in sharp contrast to the intrinsic gap inhomogeneity that is more continuous over the length scale of $\xi_{ab}$ as discussed in the main text. Moreover, near such a Cs boundary, a spectrum with a larger superconducting gap exhibits lower coherence peaks (Supplementary Fig. 12e), which is also different from the data shown in main text Fig. 3. We note that for the data shown in Supplementary Fig. 12b-e, a superconducting Nb tip was used with $\Delta_{tip} \approx 1.3$ meV.

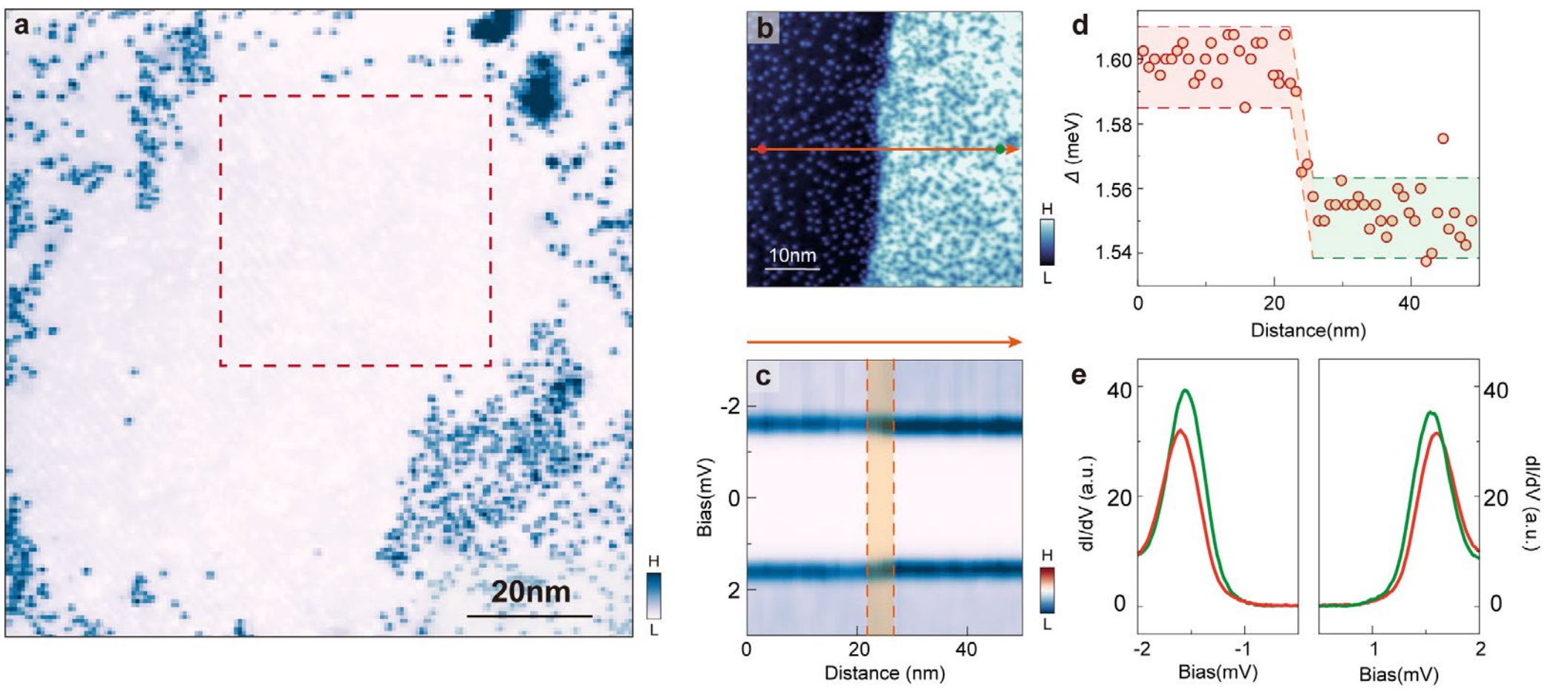

**Supplementary Fig. 12. Influence of the surface Cs cluster on the local superconducting gap**. **a,** STM topographic image in a large FOV which includes the region shown in Supplementary Fig. 8. **b**, Topographic image depicting a boundary between Cs-diluted and Cs-rich regions on the Sb surface. **c**, dI/dV linecut along the orange line crossing the boundary in panel **b**. **d**, Superconducting gap variation along the line. **e**, Representative spectra from the Cs-diluted (orange) and Cs-rich (green) regions; a superconducting Nb tip is used with $\Delta_{tip} \approx 1.3$ meV.

**Supplementary Note 11:**

**Doping dependence of superconducting gap inhomogeneity**

To further exclude the possibility that the superconducting gap inhomogeneity is simply caused by Sn substitution, we compared the spectroscopic properties of $CsV_3Sb_{5-x}Sn_x$ samples with $x = 0.06$ (long-range triple-$Q$ CDW), $x = 0.09$ (short-range single-$Q$ CDW), $x = 0.12$ (single-$Q$ CDW with further reduced correlation length) and $x = 0.30$ (likely fluctuating single-$Q$ CDW or very short-ranged stripelike charge correlation, nearly undistorted kagome lattice) [2,25]. The zero-bias conductance maps measured in the normal state show that the $x = 0.30$ sample has a much stronger spatial variation of the normal-state density of states than the $x = 0.09$ sample (Supplementary Fig. 13a and b). This is also reflected by the larger normalized standard deviation $\sigma/\mu$ of the differential conductance over a wide energy range in $x = 0.30$ (Supplementary Fig. 13c), indicating that the normal-state electronic environment is more inhomogeneous at higher Sn concentration.

This notwithstanding, the superconducting gap in $x = 0.30$ is rather homogeneous. Supplementary Fig. 13d and e show the topographic image and superconducting gap map measured in a 50 nm × 50 nm field of view. Despite the strong normal-state inhomogeneity, the distribution of $\Delta(r)$ is narrow, with an average gap value $\bar{\Delta}$ = 670 μeV, a standard deviation $\sigma$ = 13.7 μeV and $\sigma/\bar{\Delta}$ = 0.020 (Supplementary Fig. 13f), markedly smaller than $\sigma/\bar{\Delta}$ = 0.051 observed in the stripe-ordered $x = 0.09$ sample. The linecut spectra exhibit nearly unchanged coherence-peak positions across the field of view, further confirming the weak spatial variation of $\Delta$ in $x = 0.30$ (Supplementary Fig. 13g). The observation of Caroli-de Gennes-Matricon bound states inside the vortex core (Supplementary Fig. 13h and i) further indicates weaker quasiparticle scattering in $x = 0.30$, consistent with its more homogeneous superconducting gap.

More information about the close correlation between the charge-stripe order and spatially inhomogeneous superconductivity has been revealed by STM measurements on two additional compositions, $x = 0.06$ and 0.12, on both sides of $x = 0.09$. The $x = 0.06$ sample is positioned in the triple-$Q$ (2 × 2) CDW phase region (see the phase diagram in main text Fig. 1b), whereas the $x = 0.12$ sample shows the absence of CDW-related anomaly in charge transport (Supplementary Fig. 3b). Here we elucidate that the superconducting state is nearly homogeneous in both cases.

For $x = 0.06$, the gap map and linecut spectra show only weak spatial variation (Supplementary Fig. 14a-d). The gap histogram gives an average value $\bar{\Delta}$ = 425 μeV, a standard deviation $\sigma$ = 5 μeV, and $\sigma/\bar{\Delta}$ = 0.012 (Supplementary Fig. 14c), much smaller than $\sigma/\bar{\Delta}$ = 0.051 for the stripe-ordered $x = 0.09$ sample. Similarly, for $x = 0.12$, the gap map and linecut spectra also reveal marginal gap variation (Supplementary Fig. 14e-h), with $\bar{\Delta}$ = 390 μeV, $\sigma$ = 4 μeV, and $\sigma/\bar{\Delta}$ = 0.010, Supplementary Fig. 14g). The representative spectra in both compositions show fully opened superconducting gaps with negligible residual density of states.

The gap statistics for different doping levels are summarized in Supplementary Table 2. The normalized gap variation $\sigma/\bar{\Delta}$ is small for $x = 0.06$, 0.12 and 0.30, with values of 0.012, 0.010 and 0.020, respectively. By contrast, $\sigma/\bar{\Delta}$ reaches a more remarkable value of 0.051 at $x = 0.09$. Therefore, the pronounced superconducting gap inhomogeneity appears uniquely in this compound wherein the charge-stripe-induced band reconstruction can be clearly resolved in QPI (Supplementary Note 12 and Supplementary Fig. 15). This strongly suggests that the heterogeneous superconducting state is not a direct consequence of chemical disorder introduced by Sn substitution, but is instead linked to the charge-stripe order (or more precisely, its effectiveness in inducing band reconstruction).

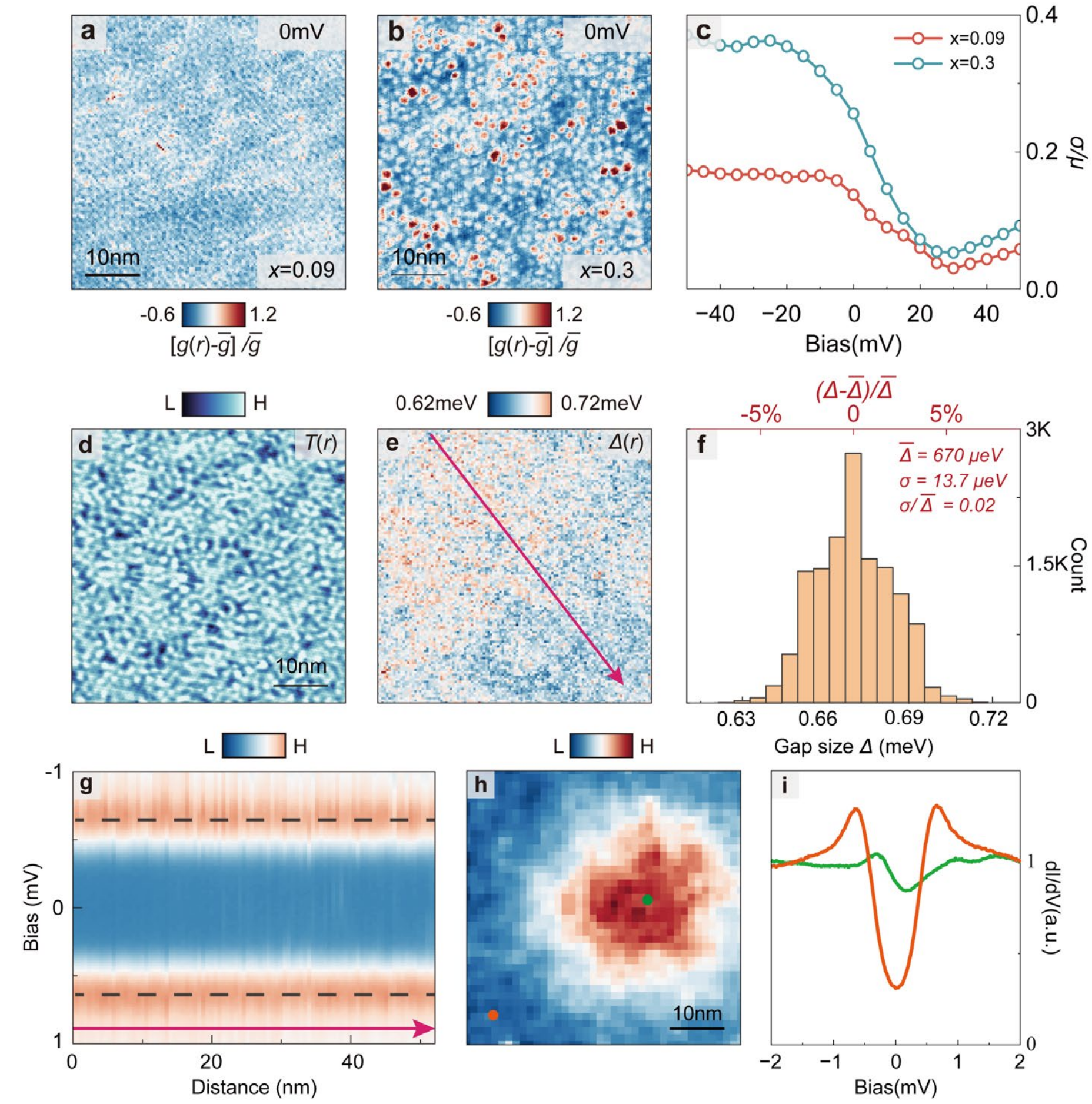

**Supplementary Fig. 13. Spatial characteristics of the normal and superconducting states in $CsV_3Sb_{4.7}Sn_{0.3}$. a, b,** Differential conductance maps at 0 mV for $x$ = 0.09 **(a)** and $x$ = 0.3 **(b)** measured at 4.2 K. The data were collected in 50 nm × 50 nm FOVs at 4.2 K for $x$ = 0.09 and $x$ = 0.3. The latter clearly exhibits a more inhomogeneous normal-state DOS. **c**, Standard deviation ($\sigma$) divided by the mean value ($\mu$) of the differential conductance values at different energies for $x$ = 0.09 and $x$ = 0.3. **d, e,** Topography and superconducting gap map in a 50 nm × 50 nm FOV for $x$ = 0.3. **f,** Histogram of the gap amplitude of **e. g,** *dI/dV* linecut along the purple arrow in **e. h,** Zero-bias *dI/dV* conductance map with an out-of-plane magnetic field of 0.5 T, showing a superconducting vortex. **i**, Spectra showing the bound state inside the superconducting vortex in **h** (green curve)**,** together with that far away from the vortex (orange curve). The presence of Caroli–de Gennes–Matricon vortex states in the $x$ = 0.3 sample suggests weaker scattering for this composition compared with $x$ = 0.09.

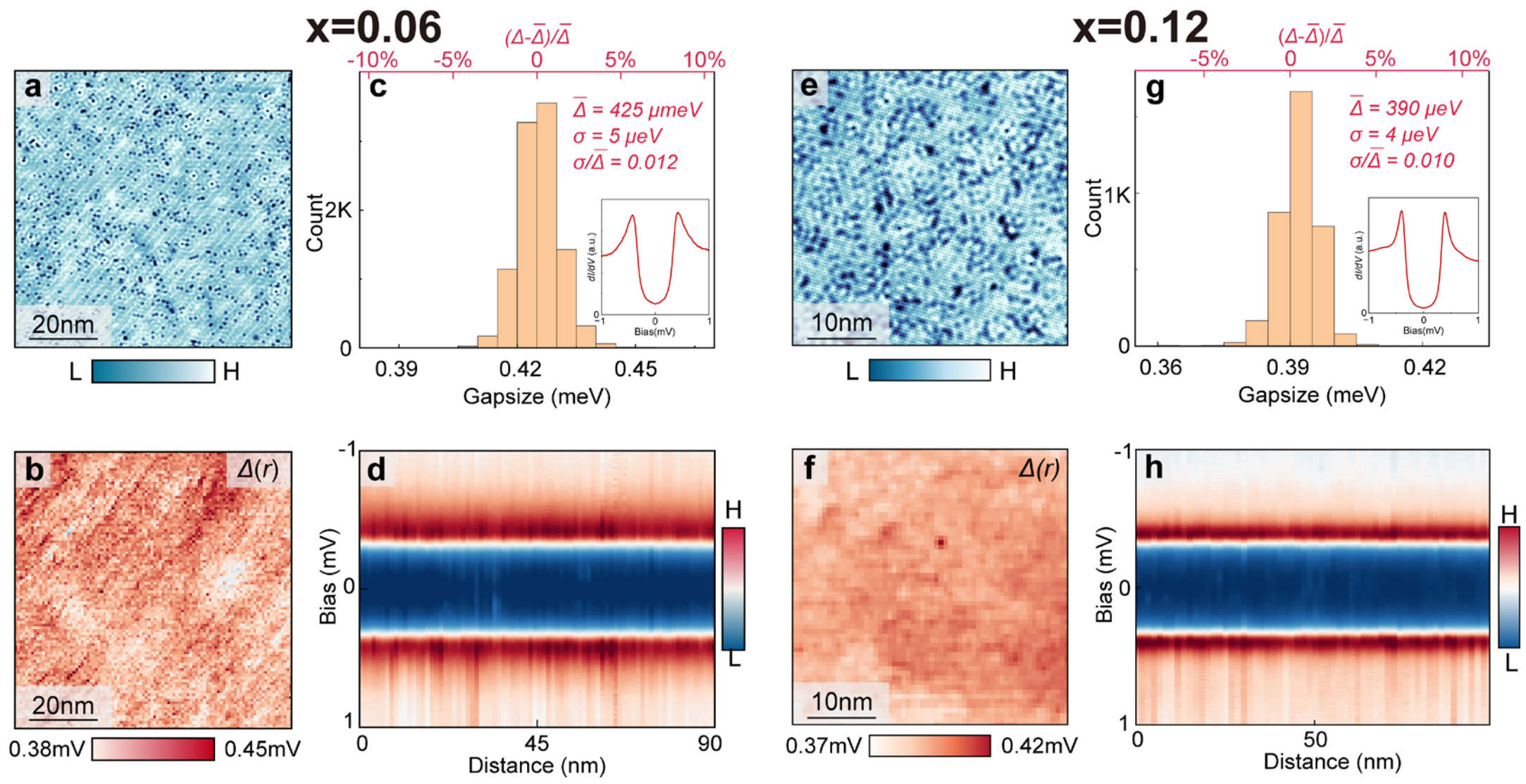

**Supplementary Fig. 14. Spatially homogeneous superconducting states in $CsV_3Sb_{5-x}Sn_x$ samples with $x$ = 0.06 and 0.12. a,** Topographic image measured on the $x$ = 0.06 sample. b, Superconducting gap map Δ($r$) in the same field of view as in **a**. **c**, Histogram of the gap values extracted from **b**. The distribution is narrow, with $\bar{\Delta}$ = 425 μeV, $\sigma$ = 5 μeV and $\sigma/\bar{\Delta}$ = 0.012. Inset: representative superconducting spectrum showing a full gap with negligible residual density of states. **d**, dI/dV linecut showing a nearly position-independent superconducting gap in $x$ = 0.06. **e-h**, Same analysis as in **a-d**, but measured on the $x$ = 0.12 sample. The gap map and linecut spectra reveal marginal spatial variation of the superconducting gap. The gap histogram gives $\bar{\Delta}$ = 390 μeV, $\sigma$ = 4 μeV and $\sigma/\bar{\Delta}$ = 0.010. The narrow gap distributions in both compositions indicate spatially homogeneous superconducting states outside the stripe-ordered $x$ = 0.09 regime.

**Supplementary Table 2.** The mean value $\bar{\Delta}$ of the spatially varying superconducting gap $\Delta(r)$, its standard deviation $\sigma$, and the ratio between them, $\sigma/\bar{\Delta}$, determined from STM measurements in $CsV_3Sb_{5-x}Sn_x$ with $x$ = 0.06, 0.09, 0.12 and 0.30.

| Sn doping level $x$ | 0.06 | 0.09 | 0.12 | 0.30 |
|---|---|---|---|---|
| $\bar{\Delta}$ (meV) | 0.425 | 0.323 | 0.390 | 0.670 |
| $\sigma$ (meV) | 0.005(0) | 0.016(6) | 0.004(0) | 0.013(7) |
| $\sigma/\bar{\Delta}$ | 0.012 | 0.051 | 0.010 | 0.020 |

**Supplementary Note 12:**

**Spatial correlation between stripe-induced folded-band QPI and superconducting gap in the x = 0.09 sample**

In our STM measurements, the stripe-related signature is mainly identified from the QPI patterns, where the charge-stripe order folds the quasi-one-dimensional kagome bands and gives rise to additional scattering vectors **q′** and **q″** (Supplementary Fig. 15a-c). Short-range real-space modulations with a periodicity of approximately $3a_0$ can also be observed in strained samples (Supplementary Fig. 15d). However, applying strain may introduce additional complexity, since superconductivity itself can also be tuned by external strain. Therefore, in the following we use the folded-band QPI features as practical fingerprints of stripe-related electronic reconstruction.

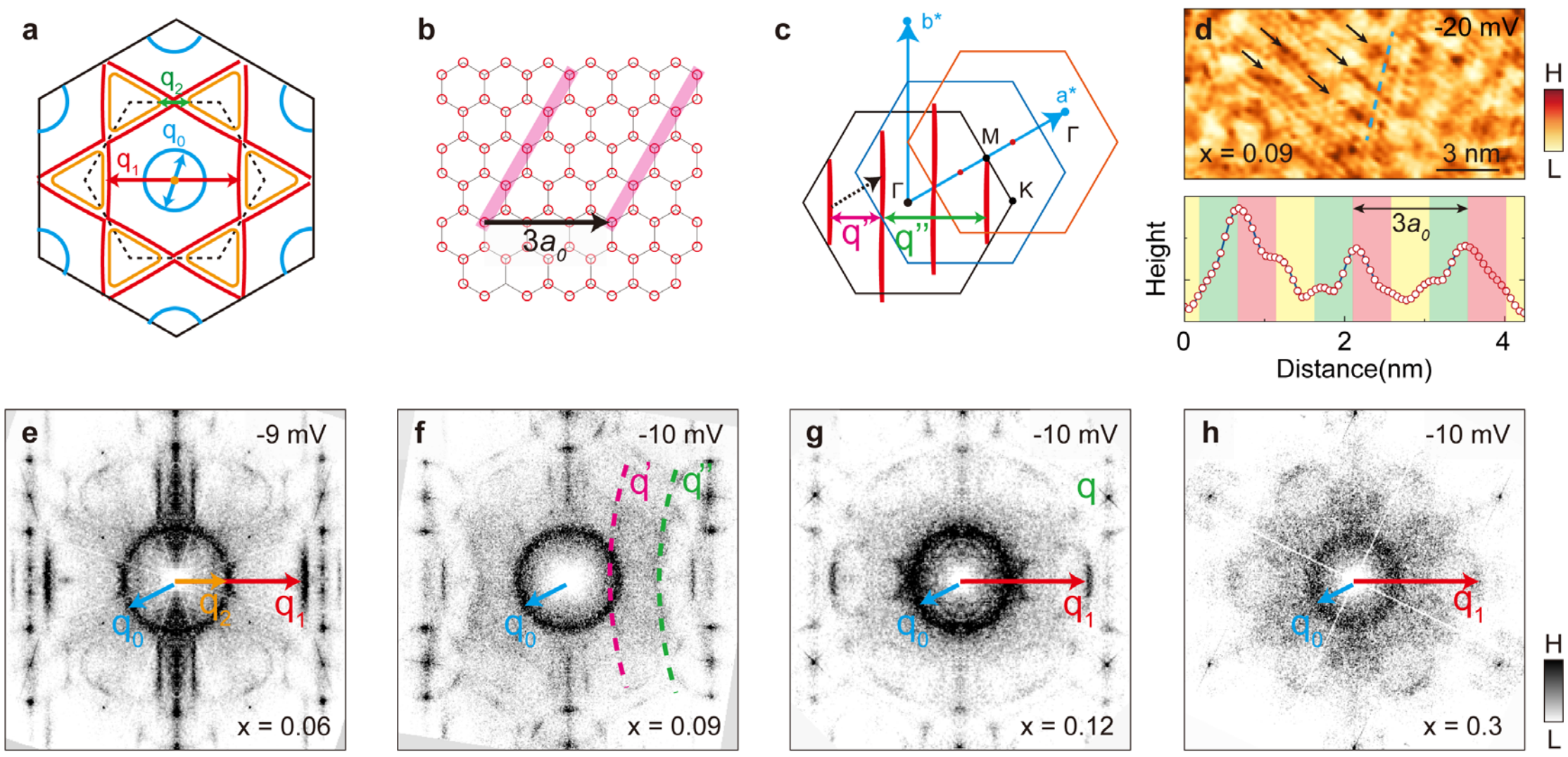

**Supplementary Fig. 15. Charge-stripe order revealed by QPI images. a,** Schematic illustration of the Fermi surface of $CsV_3Sb_{5-x}Sn_x$. The color-coded arrows denote the main scattering vectors. **b,** Schematic of the stripe order, with a unidirectional periodicity of approximately $3a_0$. **c,** Band folding induced by the charge stripe order: the quasi-1D bands (shown in red) are folded along the Γ–M direction (dashed black arrow), giving rise to two new scattering vectors **q′** and **q″**. **d**, Real-space visualization of short-range stripe order in strained $x$ = 0.09 sample. Upper panel: $dI/dV$ map showing the short-range modulations (marked by black arrows); lower panel: line-profile along the dashed green line, revealing a periodic modulation of approximately $3a_0$. **e-h**, QPI images at high energies for $x$ = 0.06 **(e)***,* 0.09 **(f)**, 0.12 **(g)** and 0.3 **(h)**. The unique scattering wavevectors of **q′** and **q″** can only be observed in the $x$ = 0.09 samples.

In Supplementary Fig. 15e-h we compare the normal-state QPI patterns for four representative compositions, $x$ = 0.06, 0.09, 0.12 and 0.30. The stripe-induced **q′** and **q″** features are clearly resolved only in the $x$ = 0.09 sample, whereas they are absent in the other three compositions. Note that transport measurements also show the absence of CDW-related anomaly in $x$ = 0.12 and 0.30 (Supplementary Note 2). This indicates a profound electronic band reconstruction induced by the short-range stripelike CDW occurring only for $x$ = 0.09 among these four compositions; its disappearance at $x$ = 0.12 is likely attributed to the reduced correlation length of the charge-stripe order at higher $x$. (In the phase diagram displayed in main text Fig. 1b, we thus anticipate that pronounced impact of the single-$Q$ CDW happens only within a narrow window above $x = 0.07$, as highlighted by the background colour.)

To investigate the relationship between the charge stripe order and the local superconducting gap $\Delta(r)$, we compared the stripe-induced QPI intensity with $\Delta(r)$ in the $x = 0.09$ sample. we performed spatial-resolved Fourier-transformed (FT) analysis on two independent datasets for x = 0.09. For each dataset, both QPI and gap map $\Delta$(r) were measured in the same field of view. Then, for each dataset, we selected two equal-area regions (with nearly identical local Sn concentration) exhibiting distinct local gap amplitudes, labeled as A1 and A2 for dataset #1 (Supplementary Fig. 16a, b), and B1 and B2 for dataset #2 (Supplementary Fig. 16g, h), respectively. Here A1 and B1 (A2 and B2) correspond to the relatively larger-gap (smaller-gap) regions. The spatial-resolved FT-QPI maps for all these four regions were obtained and shown in Supplementary Fig. 16c, d and i, j. This allows us to check whether the characteristic QPI signals for the charge stripe order (**q′** and **q″**) are stronger in regions with larger superconducting gap.

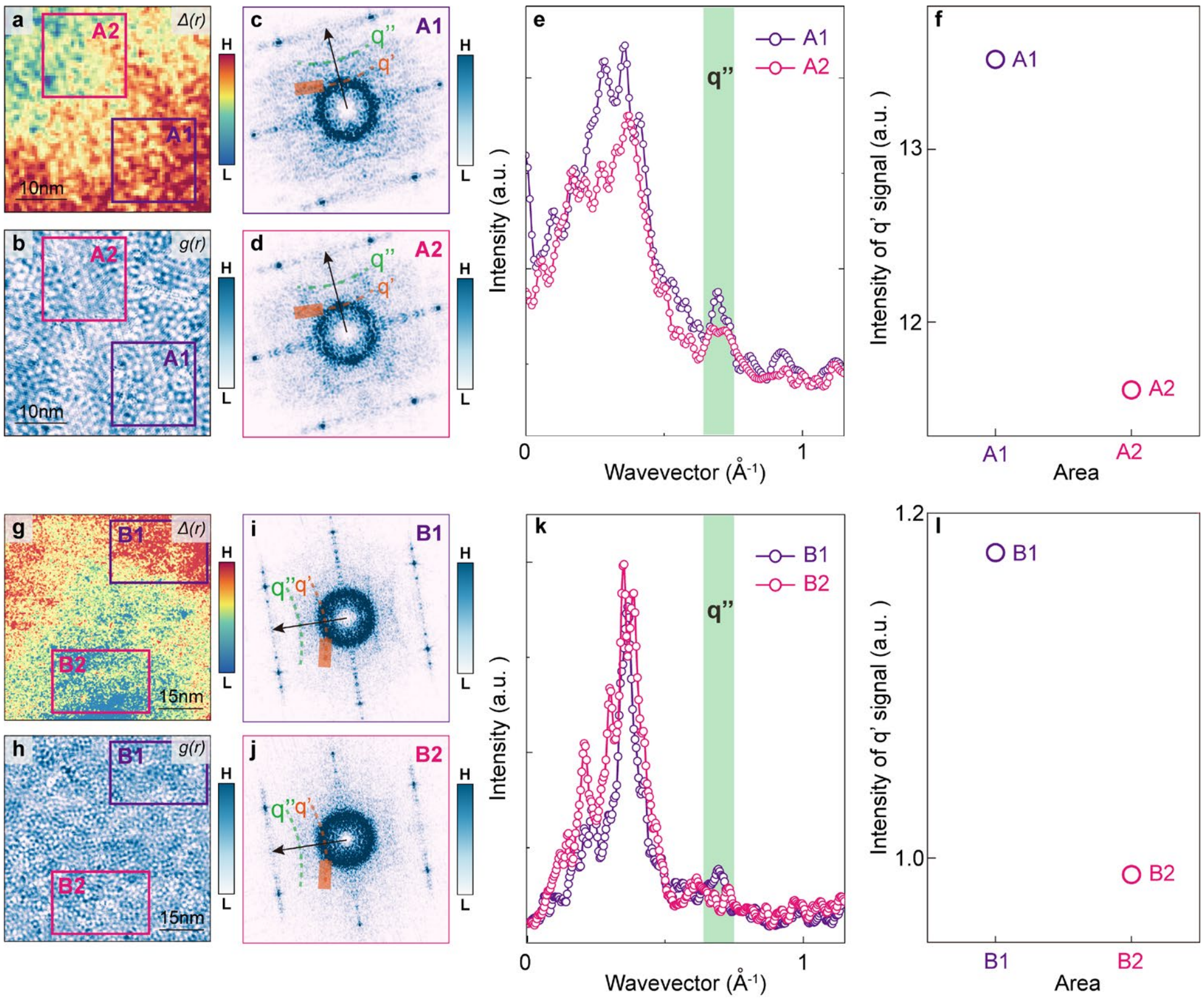

**Supplementary Fig. 16. Spatial correlation between stripe-induced folded-band QPI and superconducting gap for $CsV_3Sb_{4.91}Sn_{0.09}$. a, b,** Superconducting gap map $\Delta$(r) (**a**) and differential-conductance map $g$(r) (**b**) measured in the first field of view. Two equal-area regions with relatively large and small $\Delta$ are marked as A1 and A2, respectively. **c, d,** FT-QPI maps obtained from A1 (**c**) and A2 (**d**). The stripe-induced folded-band QPI features **q′** and **q″** are highlighted by orange and green colors in **c**, respectively. **e,** Linecuts through **q″** along the direction marked by the black arrows in **c** and **d**. The green shaded region marks the **q″** feature. **f,** Averaged FT-QPI intensity within the momentum-space window (orange shaded regions in **c** and **d**) around $\boldsymbol{q'}$ for A1 and A2. **g–l,** Same analysis as in **a–f**, but for another field of view wherein the equal-area regions with relatively large and small $\Delta$ are marked as B1 and B2, respectively. In both fields of view, the larger-gap region A1(B1) shows higher intensity of stripe-induced folded-band QPI than A2(B2).

From Supplementary Fig. 16c, d and i, j, a visual inspection finds that the characteristic QPI features of $\mathbf{q}'$ and $\mathbf{q}''$ appear to be more pronounced in the larger-gap region A1 (B1) as compared to A2 (B2). This can be cross-checked by quantitatively comparing their FT intensity. We first look into the signal at $\mathbf{q}''$: the linecuts across $\mathbf{q}''$ are shown in Supplementary Fig. 16e and k (taken along the arrows in Supplementary Fig. 16c, d and i, j), and one can clearly see that its intensity is larger in A1 (B1) compared with that in A2 (B2). In these linecuts, $\mathbf{q}'$ is overlapped with the strong QPI signal of Sb p-orbital (the circular shape QPI), which obscures a direct comparison of the q′ intensities. Instead, we can plot the averaged intensity of $\mathbf{q}'$ within a **q**-space window away from the Sb-p-orbital QPI patterns as marked by the orange rectangles in Supplementary Fig. 16c, d, i and j. The results are shown in Supplementary Fig. 16f and l, and the intensity of $\mathbf{q}'$ is also larger in the regions with larger superconducting gap. The consistent results of $\mathbf{q}'$ and $\mathbf{q}''$ demonstrate that the regions with stronger stripe-induced band folding also exhibit larger superconducting gap. We note that in these regions the number of local Sn dopants (the number of scattering centers) is nearly identical (205 and 207), thus it is unlikely to be the reason for the observed different QPI intensity. This suggests a positive correlation between the local strength of charge stripe and the local amplitude of the superconducting order parameter, instead of a competition.

Hence, according to the FT-QPI signatures, our data mark a clear departure from the simple anti-correlation between the charge stripe order and the local superconducting gap; in contrast, our observations imply that these two quantities increase in tandem. This argues against the crude picture of spatial-dependent competition between the two orders (in which a stronger stripe-related reconstruction directly suppresses the superconductivity). Alternatively, results shown in Supplementary Fig. 16 suggest that the observed heterogeneous superconducting state is the ramification of unusual corporation of the charge stripe and superconducting pairing instabilities.

Combined with the analysis of the local pair-breaking parameter $\Gamma$ presented in Supplementary Note 8, this positive correlation can be understood phenomenologically. The larger-gap regions exhibit enhanced stripe-induced folded-band QPI signals but weaker pair breaking, as reflected by the smaller $\Gamma$. We propose that stronger $\mathbf{q}'/\mathbf{q}''$ QPI intensity may reflect stripe correlations that are locally more coherent in phase, which generate a more well-defined band-folding signal while producing weaker pair-breaking scattering. By contrast, regions with more disrupted or frustrated stripe correlations may give weaker folded-band QPI signals but stronger pair breaking, leading to an enhanced $\Gamma$ and a reduced local gap $\Delta$. We emphasize that this is a phenomenological interpretation; a microscopic understanding of how short-range charge stripes modulate superconducting pairing and pair breaking requires further theoretical work.

**Supplementary Note 13:**

**Analysis of the specific heat**

The electronic specific heat $C_e$ of $CsV_3Sb_{5-x}Sn_x$ ($x$ = 0, 0.06, 0.07, 0.09, 0.12 and 0.3) was obtained by subtracting $C_{ph}$ and $C_{Sch}$ from the raw data (see Methods), as plotted in Supplementary Fig. 17. We analyzed the $C_e(T)/T$ data in the superconducting state using the BCS $\alpha$-model [31,32] and the results are presented in Supplementary Fig. 18.

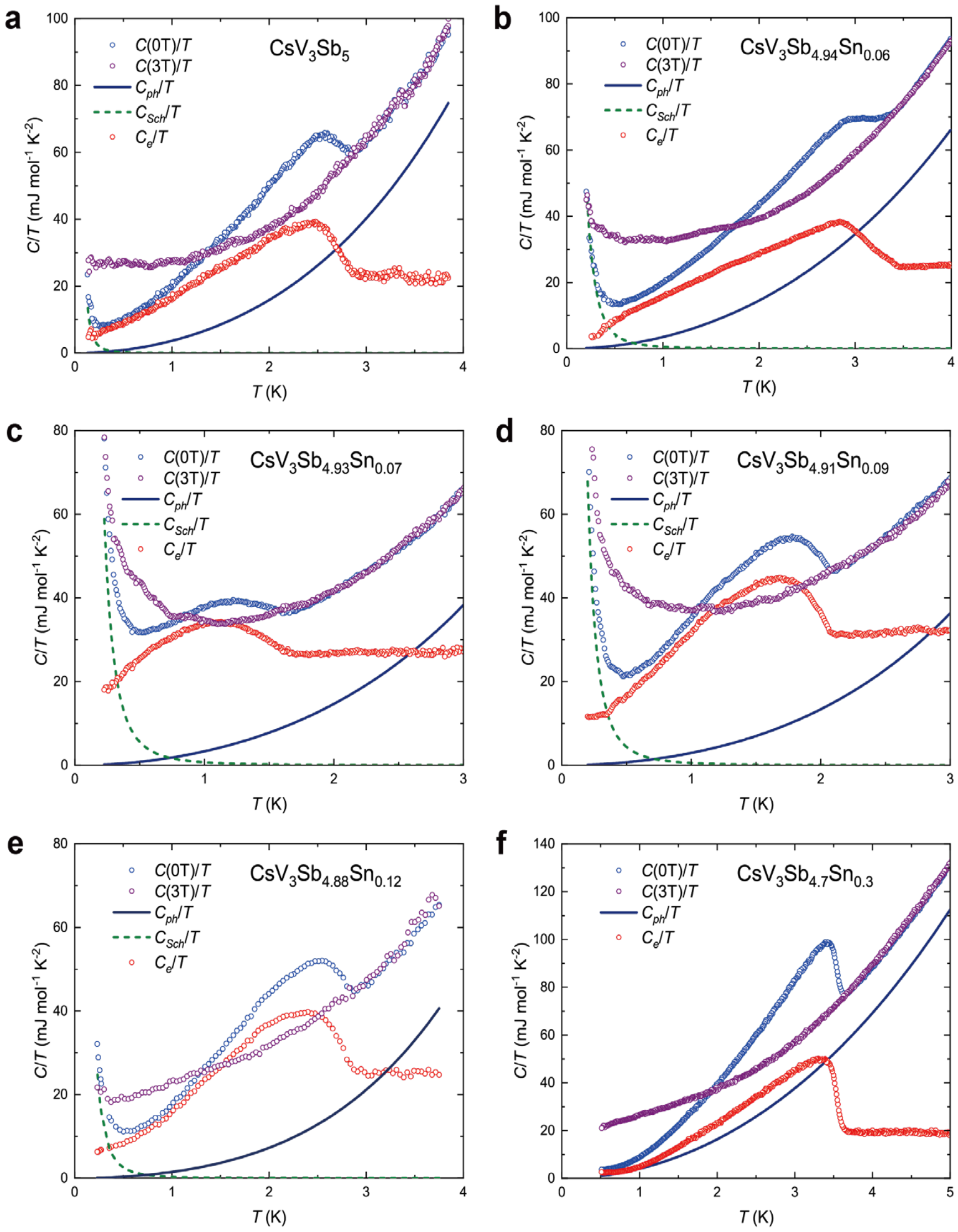

**Supplementary Fig. 17. Procedures for separating the electronic contribution $C_e$ from the raw data of $C(T)$. a-f,** Low-$T$ specific heat data measured in $CsV_3Sb_{5-x}Sn_x$ with $x$ = 0 (**a**), $x$ = 0.06 (**b**), $x$ = 0.07 (**c**), $x$ = 0.09 (**d**), $x$ = 0.12 (**e**) and $x$ = 0.3 (**f**). Blue and purple circles are $C/T$ measured under zero magnetic field and $\mu_0 H$ = 3 T ($H \parallel c$), respectively. Royal solid line and green dashed lines are fitting curves of the phononic term ($C_{ph}/T$) and the nuclear Schottky term ($C_{Sch}/T$) as described in Methods. Red symbols denote the electronic specific heat ($C_e/T$) attained from $C/T$ - $C_{ph}/T$ - $C_{Sch}/T$ (Methods).

For pristine $CsV_3Sb_5$, the data are incompatible with a single-gap $\alpha$-model, but can be successfully described by a two-band model with superconducting gaps $\Delta_L$ = 0.4 meV, $\Delta_S$ = 0.14 meV and a residual $\gamma_0$ = 4.5 mJ mol$^{-1}$K$^{-1}$ (Supplementary Fig. 18a). This result adds to the existing evidence for nodeless multigap superconductivity in $CsV_3Sb_5$ [33-37]. Moreover, the substantially reduced specific heat jump $\Delta C_e/\gamma_n T_c \approx 0.72$ (here $T_c$ was taken as the temperature where $C_e/T$ reaches its maximum) compared with the BCS value 1.43 (Supplementary Fig. 18g) can be associated with both the superconducting gap anisotropy [38] and the existence of multiple gaps [39,40]. The data measured in the $x = 0.06$ sample (which also possesses long-range triple-$Q$ CDW) appear to show qualitatively consistent two-gap behavior with the pristine sample (Supplementary Fig. 18b).

For the sample with $x = 0.07$, $C_e(T)$ ($T < 0.8$ K) can be fitted by the $\alpha$-model with one full gap ($\Delta = 0.17$ meV) with a large residual linear term $\gamma_0 \approx 17$ mJ mol$^{-1}$ K$^{-1}$ (Supplementary Fig. 18c). Similarly, $C_e(T)$ ($T < 1$ K) for $x = 0.09$ can be fitted by the single-gap $\alpha$-model assuming $\Delta = 0.27$ meV with $\gamma_0 \approx 11.7$ mJ mol$^{-1}$ K$^{-1}$ (Supplementary Fig. 18d). At $T \rightarrow T_c^S$, $C_e(T)$ deviates from this fit, with the specific heat jump profoundly suppressed from the model's prediction (main text Fig. 4a and Supplementary Fig. 18c, d). This phenomenon further corroborates a distribution of local $T_c$ that smears the sharp jump of $C_e$ at $T = T_c$, yielding a broad and stunted specific heat anomaly (whose height is further reduced by the large proportion of normal fluid, Supplementary Fig. 18h). As such, the gap value provided by the fit below 1 K should be taken as a spatial average over the sample.

In particular, we stress that the ostensible saturation of $C_e/T$ at a notable non-zero value ($\approx \gamma_0$) at the lowest $T$ for $x = 0.07$ and 0.09 sample can only be understood by proposing large residual unpaired DOS; the direct observation of these unpaired quasiparticles in STM data (main text Fig. 4e-h) serves as the most compelling support. Such phenomenon cannot be explained by the presence of a very small second superconducting gap $\Delta_S$. As displayed in Supplementary Fig. 18c and d, a two-gap $\alpha$-model assuming an extremely small $\Delta_S$ (blue solid lines) fails to reproduce the measured profile of $C_e/T$ in both $x = 0.07$ and $x = 0.09$, mostly manifested by the absence of the humplike anomaly heralding the low-$T$ drop of $C_e/T$ caused by $\Delta_S$. The occurrence of this anomaly has been ubiquitously observed in superconductors with small gaps, e.g., $Nb_3Sn$ [41], $Lu_2Fe_3Si_5$ [39], $KFe_2As_2$ [40] and FeSe [42]. Moreover, the presence of such a small $\Delta_S$ yielded by the fits, which is below 0.05 meV ($\Delta_S < 0.3k_BT_c$ for $x = 0.09$), indisputably evades our STM measurements down to 20 mK (main text Fig. 4c) in the **k**-space (Supplementary Note 16).

The analysis based on the entropy balance arguments, i.e., the entropy $S_{el}$ of the electronic system must recovers to the value equal to the normal-state entropy at a temperature above the second-order superconducting transition ($T \gtrsim T_c$) [43,44], provides another strong argument against the presence of a small $\Delta_S$. The entropy calculations presented in Supplementary Fig. 18g with the assumption of a constant residual DOS $\gamma_0$ (solid curves) generally fulfill the entropy balance condition. If $C_e/T$ abruptly drops to zero at $T < 0.2$ K owing to the full opening of the small gap $\Delta_S$ (dashed lines in Supplementary Fig. 18c and d), the entropy balance will be broken as demonstrated by the dashed curves in Supplementary Fig. 18g, which challenges the basic understanding of superconductivity.

When the charge-stripe order is almost collapsed at $x = 0.30$, $C_e(T)$ falls close to the single-gap $\alpha$-model fit with $\Delta = 0.52$ meV and a small residual term $\gamma_0 = 2.4$ mJ mol$^{-1}$ K$^{-1}$ (Supplementary Fig. 18f), indicating that a more isotropic full superconducting gap is established. Similar suppression of gap anisotropy upon chemical substitution has been observed in other doped $CsV_3Sb_5$ superconductors [30,45,46]. The value of $\Delta$ corresponds to $2\Delta/k_BT_c = 3.54$ close to the weak-coupling BCS value 3.53 (correspondingly, $\Delta C_e/\gamma_n T_c \approx 1.59$, Supplementary Fig. 18g); the empirical relation $C_e/\gamma_n T_c \propto \exp[-1.44(T_c/T)]$ holds for $CsV_3Sb_{4.70}Sn_{0.30}$ over a range $1.6 < T_c/T < 2.7$, also in accord with the prediction of the weak-coupling BCS theory [47]. Thereby, this compound appears to be a standard BCS superconductor. The data for $x = 0.06$ qualitatively matches that for $x = 0.30$ yet with a larger $\gamma_0$ (Supplementary Fig. 18e). It should be mentioned that in both the triple-$Q$ CDW ($x$ = 0, 0.06) and fluctuating CDW ($x$ = 0.12, 0.30) phases, $\gamma_0$ is constantly smaller than 6 mJ mol$^{-1}$ K$^{-1}$; such an amplitude barely surpasses the uncertainties in our specific heat analysis (main text Fig. 4b), hence obscuring the unambiguous determination of intrinsic nature of the residual term.

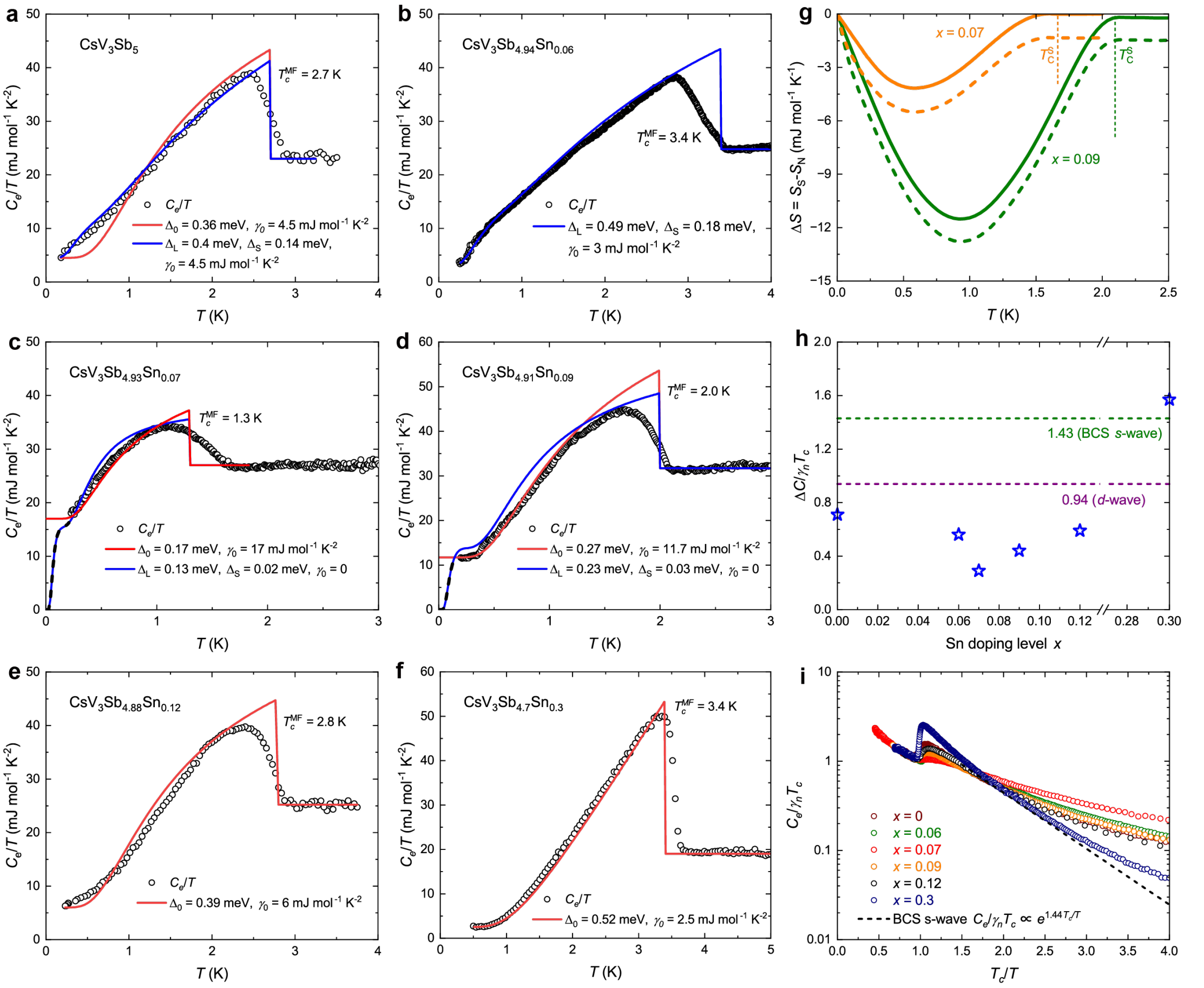

**Supplementary Fig. 18. Analysis of electronic specific heat $C_e$ of $CsV_3Sb_{5-x}Sn_x$. a-f,** Fitting the $C_e/T$ data to the BCS $\alpha$-model (solid lines) for $x$ = 0 (**a**), $x$ = 0.06 (**b**), $x$ = 0.07 (**c**), $x$ = 0.09 (**d**), $x$ = 0.12 (**e**) and $x$ = 0.3 (**f**). Red and blue full lines represent the single-gap and two-gap $\alpha$-models, respectively (all assuming full gaps and a residual term $\gamma_0$). **g,** The entropy difference $\Delta S = S_S - S_N$ between the electronic entropy of the superconducting state ($S_S$) and the normal state ($S_N$) for $x$ = 0.07 (orange) and $x$ = 0.09 (olive). The solid lines (dashed lines) represent the results of integrating $C_e/T$ with the assumption of saturation towards $\gamma_0$ (decrease towards zero) at $T$ < 0.2 K, as denoted by the red solid curves (gray dashed curves) in **c** and **d**. Only the former assumption satisfies the entropy balance condition that $\Delta S$ becomes zero at $T \approx T_c^S$. **h,** The specific heat jump $\Delta C/\gamma_n T_c$ ($\gamma_n$ is the Sommerfeld coefficient in the normal state, $T_c$ is determined from the maximum of $C_e/T$) plotted against the Sn doping level $x$. Note that the residual term $\gamma_0$ is not subtracted from $\gamma_n$. **i,** Tests of the exponential $T$ dependence of $C_e/\gamma_n T_c$ [47]. Black dashed line indicates the expected behavior in a BCS single-gap weak-coupling limit.

These results of specific heat analysis strongly imply nodeless superconductivity across the $CsV_3Sb_{5-x}Sn_x$ series, with the anisotropy and/or orbital-dependent feature of $\Delta$ weakened upon Sn doping. We find no evidence for nodal superconductivity. Remarkably, $C_e$ of the $x$ = 0.09 sample cannot be fitted by the form $C_e(T) \propto T^2$ describing a gap structure with line nodes (main text Fig.

4a). Even the superconducting state is prone to nonmagnetic pair breaking as implied by the enormous $\gamma_0$ and is thus supposedly sign changing [48-50], it is still more likely to be fully gapped rather than nodal (see also Supplementary Note 16 for further evidence). Nonetheless, the nodal pairing states (e.g., *f*-wave pairing in a kagome lattice [51,52]) cannot be completely ruled out because the contribution of nodal quasiparticles might be masked by the abundant low-energy excitations provoked by pair-breaking scattering [50,53].

**Supplementary Note 14:**
**Absence of localization effects in $CsV_3Sb_{5-x}Sn_x$**

It has been well established that conventional (BCS-type *s*-wave) superconductivity exhibits remarkable resilience against Anderson localization and persists even in strongly disordered systems in the absence of magnetic impurities: For spatial dimension $d$, appreciable disorder-induced suppression of the superconducting order parameter $\Delta$ (and strong inhomogeneity of $\Delta$) occurs when [54,55]:

$$\mathfrak{D}\Delta\xi_{loc}^{d} < 1, \qquad \text{(S5)}$$

here $\mathfrak{D}$ is the average DOS at the Fermi energy $E_F$ and $\xi_{loc}$ is the localization length. In two dimensions (2D), a perturbative estimate $\xi_{loc} \approx l\exp[(\pi/2)k_F l]$ [56] works across the transition from weak to strong disorder ($k_F$ is the Fermi wave vector, $l$ is the mean free path); $k_F l$ is frequently noted as the Ioffe-Regel parameter characterizing the strength of localization (the Anderson critical regime is entered when $k_F l \lesssim 1$ [57]). For disordered superconducting films ($d$ = 2), Eq. (S5) converts to a threshold for the occurrence of superconductor-insulator (S-I) transition: $\xi_{loc} \simeq \xi_{super}$, where $\xi_{super} = (\xi_{BCS} l)^{1/2}$ is the superconducting coherence length in the dirty limit, $\xi_{BCS} = \hbar v_F/(\pi\Delta)$ is the BCS coherence length ($v_F = \hbar k_F/m^*$ is the Fermi velocity, $m^*$ is the effective mass); the experimentally determined S-I transition happens at a larger localization scale with $\xi_{loc}$ varying from ~2 to ~30 $\xi_{super}$ [58,59] and a pronounced tendency of spatially nonuniform $\Delta$ develops in the vicinity of this threshold.

Now we demonstrate that our $CsV_3Sb_{5-x}Sn_x$ ($x$ = 0.07, 0.09) samples are far from these thresholds for strong impacts of disorder on superconductivity. In general, the Ioffe-Regel parameter $k_F l$ can be estimated from the residual resistivity extrapolated to $T$ = 0 [$\rho_0 = (ne^2\tau/m^*)^{-1}$ in a simple Drude model, $\tau$ is a characteristic relaxation time] (Supplementary Fig. 3a, inset); the carrier density $n$ can be extracted from the Hall coefficient $R_H = (ne)^{-1}$. Considering the multiband electronic structure of the Cs135 family [60], we employed a two-band model [61] with electron and hole carriers to fit the Hall resistivity $\rho_{xy}(H)$ and magnetoresistance $\rho_{xx}(H)$ of $CsV_3Sb_{4.91}Sn_{0.09}$ at $T$ = 5 K:

$$\rho_{xx} = \frac{1}{e}\frac{(n_h\mu_h + n_e\mu_e) + (n_h\mu_h + n_e\mu_e)\mu_h\mu_e B^2}{(n_h\mu_h + n_e\mu_e)^2 + (n_h - n_e)^2\mu_h^2\mu_e^2 B^2}, \qquad \text{(S6)}$$

$$\rho_{xy} = \frac{B}{e}\frac{(n_h\mu_h^2 - n_e\mu_e^2) + (n_h - n_e)\mu_h^2\mu_e^2 B^2}{(n_h\mu_h + n_e\mu_e)^2 + (n_h - n_e)^2\mu_h^2\mu_e^2 B^2}. \qquad \text{(S7)}$$

In Eqs. (S6) and (S7), the subscripts $h$ and $e$ represent hole and electron carriers, respectively, and $\mu = e\tau/m^*$ is the carrier mobility. Fitting the experimental data to these expressions, as shown in Supplementary Fig. 19b and c, we can obtain the following parameters for $CsV_3Sb_{4.91}Sn_{0.09}$ (at $T$ =

5 K): $n_e = 5.5\times10^{20}$ cm$^{-3}$, $n_h = 5.9\times10^{20}$ cm$^{-3}$, $\mu_e = 72$ cm$^2$ V$^{-1}$ s$^{-1}$ and $\mu_h = 65$ cm$^2$ V$^{-1}$ s$^{-1}$. Starting from quasi-2D cylinder-shaped Fermi surfaces ($n = k_F^2/2\pi c$) and three-dimensional (3D) spherical Fermi surfaces ($n = k_F^3/3\pi^2$), the derived results of $k_F l$ and $\xi_{loc}$ (only for the 2D case) are listed in Supplementary Table 3. The fact that $k_F l$ is on the order of 10 and hence the gigantic 2D $\xi_{loc} \gg \xi_{super}$ ($\xi_{ab} \simeq$ 20-40 nm, Supplementary Note 3) excludes the presence of strong localization effect in the $x$ = 0.09 sample: It is far on the metallic side of the above-mentioned thresholds for appreciable localization effects. Therefore, spatial inhomogeneity of $\Delta$ ought to vanish in the case of BCS $s$-wave superconductivity because of the weakness of disorder potential. The same is true for $x$ = 0.07, since $\rho_0$ of this sample is lower than that of $x$ = 0.09, suggesting better metallicity (Supplementary Fig. 3a; note that the heterogeneous signatures of superconductivity are strongest at $x$ = 0.07, see main text).

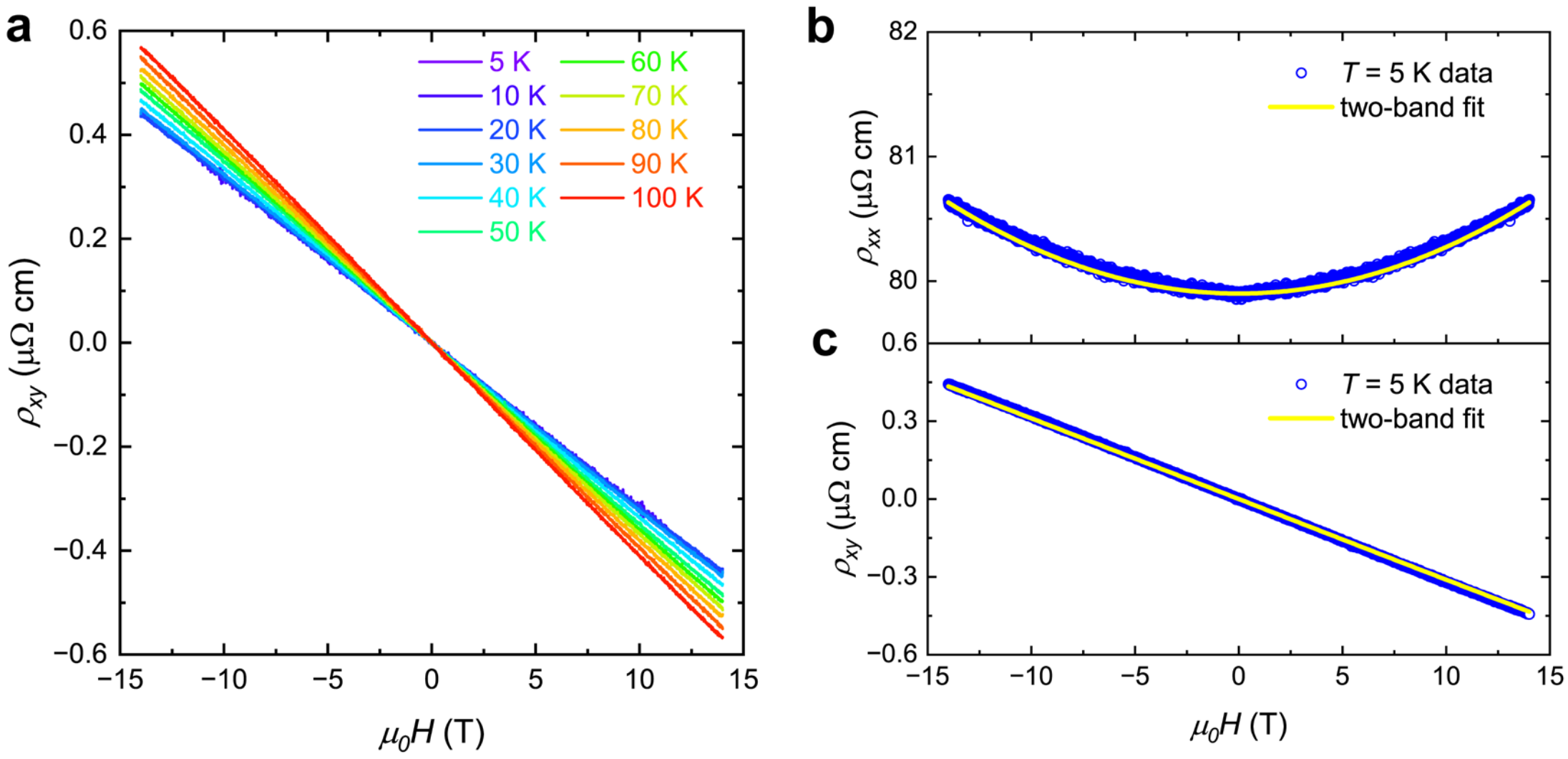

**Supplementary Fig. 19. Hall resistivity of $CsV_3Sb_{4.91}Sn_{0.09}$ and two-band model fits.**
**a,** Hall resistivity $\rho_{xy}$ plotted as a function of $H$ up to 14 T, measured in a $CsV_3Sb_{4.91}Sn_{0.09}$ sample at varying $T$ between 5 and 100 K. **b, c,** Two-band model fits of $\rho_{xx}(H)$ (**b**) and $\rho_{xy}(H)$ (**c**). Blue circles are data taken at $T$ = 5 K. Yellow lines are fits given by Eqs. (S6) and (S7).

We further elucidate that the 3D metal nature of $CsV_3Sb_{5-x}Sn_x$ imposes stricter constraints on the influence of disorder. In a mean-field picture, the criterion Eq. (S5) with $d$ = 3 (where $\xi_{loc}$ diverges at the mobility edge [56]) leads to the understanding that $\Delta$ stays almost unchanged (and spatially uniform) across the critical region at the metal-insulator transition [54]. As such, strong inhomogeneity of $\Delta$ as observed in this work can only emerge in a system on the insulating side of the mobility edge; in practice, distinct fluctuation effect stemming from disorder has been observed in a 3D superconductor which locates inside the Anderson critical region near the mobility edge (with $k_F l < 1$ and notably insulating normal-state resistivity) [57]. The absence of insulating $\rho_{xx}(T)$ in all $CsV_3Sb_{5-x}Sn_x$ samples (Supplementary Fig. 3), supplied by the lack of signatures of Altshuler-Aronov-type suppression of the DOS at $E_F$ [62] in the STM data (Supplementary Fig. 2), unambiguously implies that these samples are far beyond the mobility edge on the metallic side. We thus conclude that our observation of heterogeneous superconductivity cannot be understood within the BCS framework because the localization effect is too weak to yield these outcomes.

As a closing remark, we mention that there are theoretical proposals considering factors beyond the mean-field picture mentioned above (e.g., mesoscopic fluctuations, Coulomb interaction, etc.) [55,63]; these models predict that self-organized granularity of superconducting state can occur at a weaker disorder strength compared with the original mean-field results. Nevertheless, our samples are still too conductive with regard to the criterion given by these models (for instance, the residual conductivity of $CsV_3Sb_{4.91}Sn_{0.09}$ is three orders of magnitude larger than the critical conductance $g_{cF}$ derived from Eq.1 in Ref. [63], see contexts therein for details). Hence, our main conclusion of negligible localization effect stays unchanged.

**Supplementary Table 3. Physical parameters of $CsV_3Sb_{4.91}Sn_{0.09}$ derived from fitting of $\rho_{xx}(H)$ and $\rho_{xy}(H)$.** The Fermi vector ($k_F$), Fermi velocity ($v_F$), relaxation time ($\tau$), mean free path ($l$), Ioffe-Regel parameter ($k_Fl$) and localization length ($\xi_{loc}$, only for the quasi-2D case) of the hole-type and electron-type carriers in $CsV_3Sb_{4.91}Sn_{0.09}$, derived for quasi-2D and 3D Fermi surface structures (see texts). The effective mass $m^*$ was set to 1.0 $m_e$ ($m_e$ is the mass of free electron) in the calculation.

| | | $k_F$ (Å$^{-1}$) | $v_F$ (m/s) | $\tau$ (s) | $l$ (nm) | $k_Fl$ | $\xi_{loc}$ (nm) |
|---|---|---|---|---|---|---|---|
| 2D cylinder-shaped Fermi surface | *h* | 0.184 | $2.13\times10^5$ | $3.7\times10^{-14}$ | 7.88 | 14.5 | $6.1\times10^{10}$ |
| | *e* | 0.177 | $2.05\times10^5$ | $4{,}0\times10^{-14}$ | 8.2 | 14.5 | $6.3\times10^{10}$ |
| 3D spherical Fermi surfaces | *h* | 0.260 | $3.0\times10^5$ | $3.7\times10^{-14}$ | 11.1 | 28.6 | — |
| | *e* | 0.250 | $2.9\times10^5$ | $4.0\times10^{-14}$ | 11.6 | 29 | — |

**Supplementary Note 15:**

**Origin of pair-breaking scattering revealed by STM**

Nonmagnetic impurities with moderate scattering potential can break Cooper pairs and induce in-gap states in an unconventional superconductor with a sign-changing order parameter. In Sn-doped $CsV_3Sb_5$, we did not find notable impurity states inside the gap near the Sn dopants that should be nonmagnetic. However, this does not necessarily mean a conventional pairing in this kagome system, since point-like disorder may not serve as an effective phase-sensitive perturbation of superconductivity in kagome lattices owing to the sublattice character of the electronic states [64]. Alternatively, it is also possible that the scattering potential of Sn substitution is simply too weak to induce any in-gap states [65].

Moreover, we observed sparse vanadium-site defects in these crystals and found that they can induce clear in-gap states near $E_F$ in the superconducting state (Supplementary Fig. 20a, b). Their identity is not yet clear but most likely to be vanadium vacancies. If they are nonmagnetic or weakly magnetic, the pronounced in-gap states near zero energy might be a sign for unconventional pairing. Interestingly, we find that the real-space modulations of the in-gap state are unidirectional; this is because the site-symmetry of the defect on one sublattice is lower than the point group symmetry, thus implying the role of the sublattice degree of freedom. This anisotropy further carries over to the QPI images, resulting in a $C_2$-symmetric signal at $\mathbf{q}_2$ at the energy of the bound state (Supplementary Fig. 20c). Note that the direction of this signal at $\mathbf{q_2}$ depends on the atomic site of the sublattices for the defects, and is distinct from the direction of $\mathbf{q_1}$ feature that is associated with the charge stripe.

To elucidate the relationship between these impurity states (green circles in Supplementary Fig. 20a) and the scattering signals in BQPI, we perform Fourier filter on $\mathbf{q_2}$: the real-space features corresponding to $\mathbf{q_2}$ are strongly confined to the V-site defects (Supplementary Fig. 20d), and are exactly the same as the standing wave patterns near these defects in the raw d$I$/d$V$ map (marked by red arrows in Supplementary Fig. 20a), indicating a close correlation between the impurity states and the origin of $\mathbf{q_2}$ signals at $E_F$.

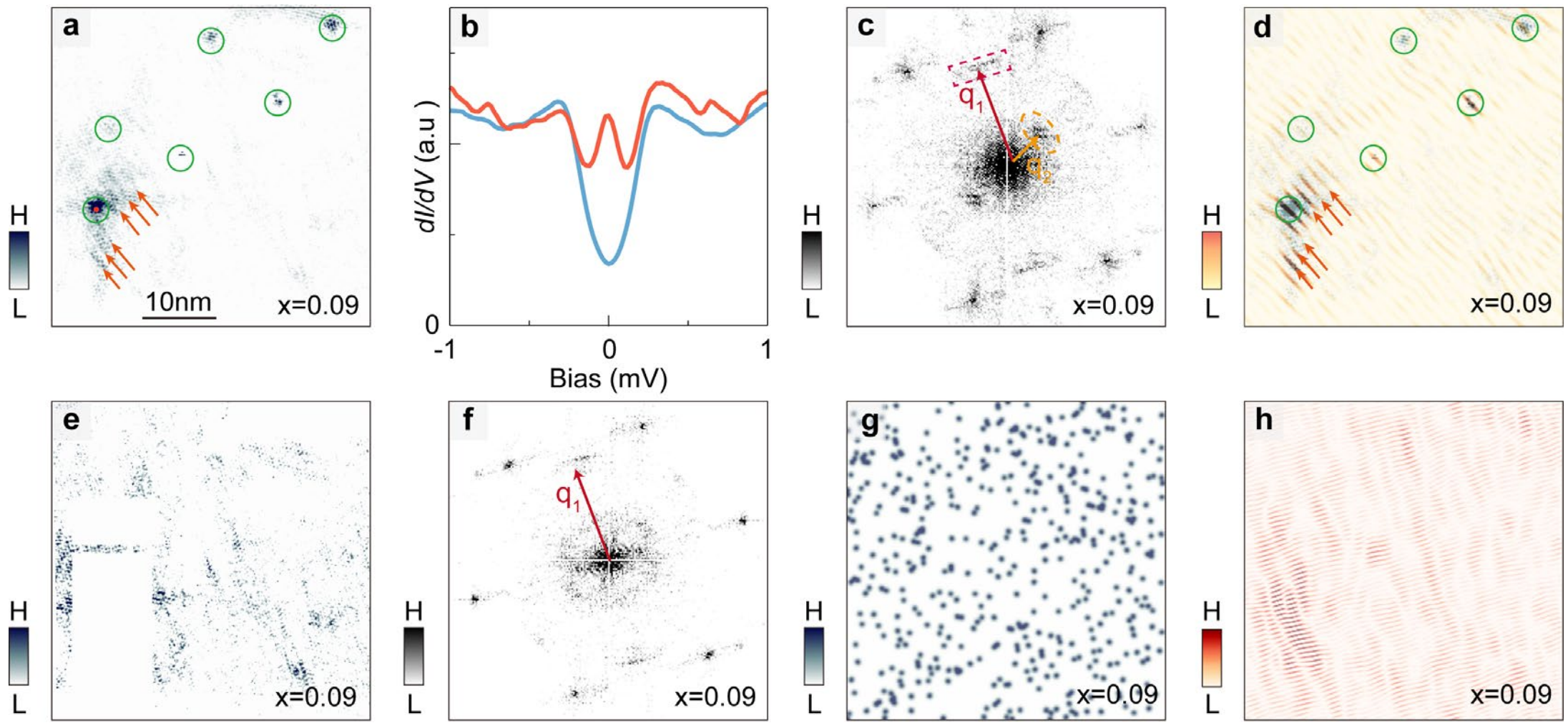

**Supplementary Fig. 20. Impurity state and real-space origins of $\mathbf{q_1}$ and $\mathbf{q_2}$ signals in the BQPI map. a,** Zero-bias *dI/dV* map of $CsV_3Sb_{4.91}Sn_{0.09}$ (Fig. 4 of the main text). Green circles mark the positions of the V-site defects which generate in-gap impurity states, and the orange arrows mark the standing wave patterns near one of these defects. **b**, Spectra showing the in-gap state (red curve) measured at the red spot in **a,** together with the spectrum far away from the defects (blue curve). **c**, The FT of **a**, showing two dominant sets of scattering vectors: $\mathbf{q_1}$ associated with the quasi-1D bands and the arc-like $\mathbf{q_2}$ feature. Note that their orientations are different. **d**, FT-filter map of the $\mathbf{q_2}$ signals in **c**. The green circles and orange arrows are shown in the same positions as in **a**. **e**, Zero-bias *dI/dV* map in which the regions showing in-gap states are masked. **f**, FT of **e**, where the $\mathbf{q_2}$ signal is significantly suppressed after masking. **g**, Spatial distribution of Sn dopants in the same FOV as **a**. **h**, FT-filter map of the $\mathbf{q_1}$ signals in **c**, showing negligible correlation with the Sn distribution in **g.**

Then, we masked the areas in the vicinity of the V-site impurities to eliminate their contribution to the total zero-energy residual DOS (Supplementary Fig. 20e). With this process, the signals at $\mathbf{q_2}$ significantly weaken, whereas those at $\mathbf{q_1}$ still persist (Supplementary Fig. 20f), indicating that the scattering associated with $\mathbf{q_1}$ is more extended in real space and gives rise to the global finite DOS at $E_F$. Thus, the pair-breaking effect near the V-site defect is not the main origin of the residual DOS at zero-energy. Furthermore, the Fourier filter result of the $\mathbf{q_1}$ feature (Supplementary Fig. 20h) displays a negligible correlation with the Sn dopant distribution (Supplementary Fig. 20g), indicating that pair breaking is more likely caused by scattering off the disorder potential associated with the short-range CDW (rather than the Sn dopant). We stress that while point-like disorder acts as a weak pair breaker for spin-singlet sign-changing superconductivity in the Cs-135 series according to a recent theoretical work [64] (owing to the sublattice degree of freedom), a spatially extended object that mixes electronic states of different sublattices, such as the short-range charge-stripe order in the current case, may generate remarkable pair-breaking scattering [27,66]. The effectiveness of such scattering potential in producing spatially inhomogeneous $\Delta$ lies in the fact that its correlation length ($\xi_{CDW}$ ~ 6-18 nm for $x$ = 0.15 [24], potentially larger at $x$ = 0.07 and 0.09) is comparable to $\xi_{ab}$ ($\simeq$ 30-40 nm, Supplementary Note 3) [27].

**Supplementary Note 16:**

**Nodeless gap structure in $x$ = 0.09 sample determined by BQPI measurement**

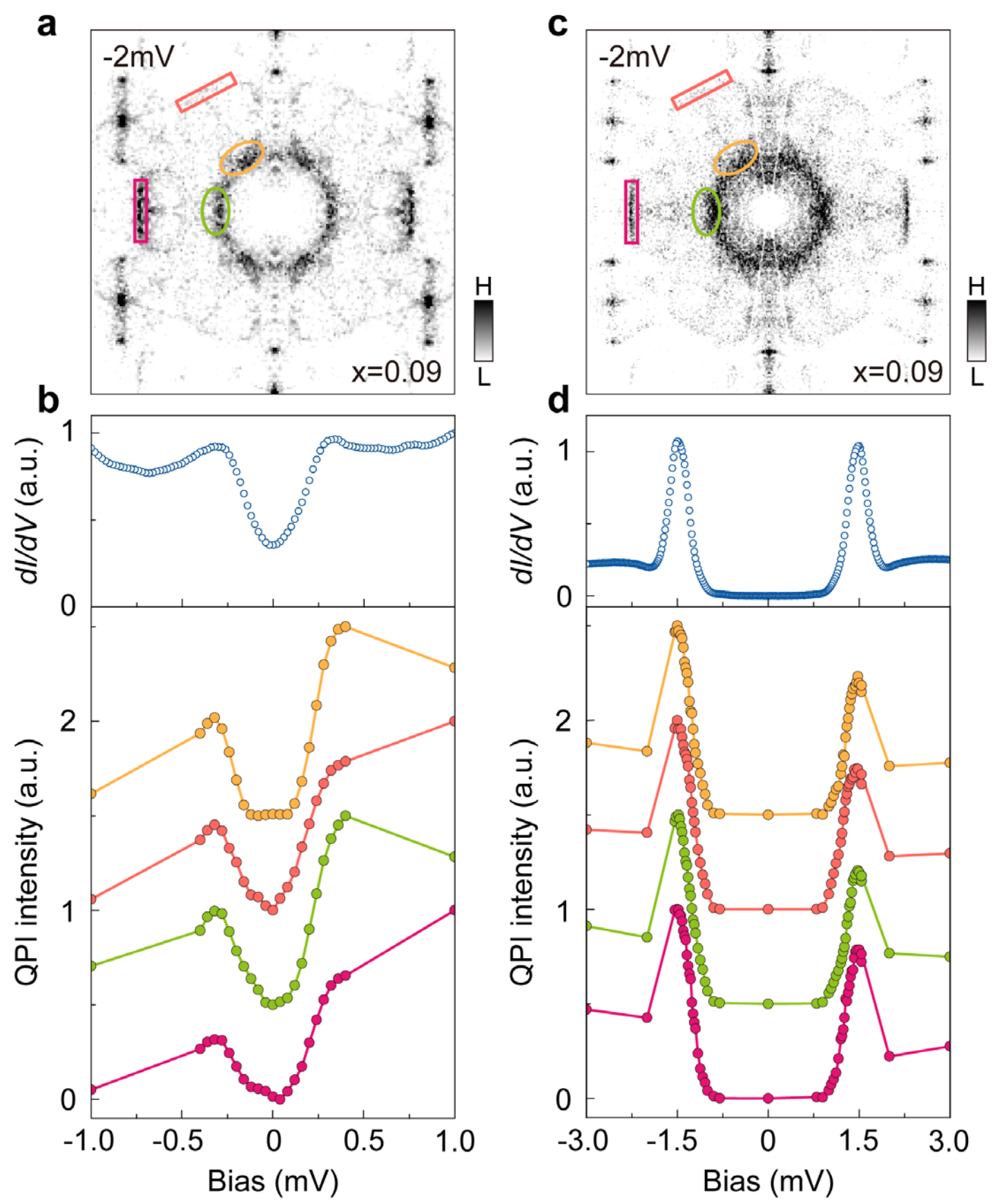

**Supplementary Fig. 21. Energy dependence of the BQPI intensity of $CsV_3Sb_{4.91}Sn_{0.09}$. a,** Quasiparticle interference pattern measured at −2 mV, outside the superconducting gap, using a normal tip at 20 mK. The colored boxes mark the momentum regions used to extract the BQPI intensities. **b,** Top panel: representative superconducting spectrum acquired under the same conditions. Bottom panel: energy dependence of the BQPI intensity at different scattering wavevectors marked in a. The BQPI intensities are strongly suppressed inside the superconducting gap, as expected from the depletion of quasiparticle spectral weight after gap opening. **c, d,** Same measurements and analysis as in **a** and **b**, but acquired using a superconducting Nb tip at 320 mK. In both datasets, the BQPI intensity is suppressed in all selected momentum regions upon entering the superconducting gap, indicating that the corresponding electronic states become gapped in the superconducting state. These results support a nodeless superconducting gap on the detected Fermi-surface sections of $CsV_3Sb_{4.91}Sn_{0.09}$.

**Supplementary Note 17:**

**Possible scenarios of the sign-reversal superconducting pairing**

The possibility that unconventional superconducting states emerge in a kagome lattice has been extensively investigated in existing theoretical efforts [51,52,67-73]. Surprisingly rich superconducting phase diagrams have been proposed for kagome superconductors, owing to their unique properties of (i) specific sublattice interference, (ii) appreciable intersite Coulomb interactions and (iii) the presence of multiple van Hove singularities in the vicinity of $E_F$ [51,52,67-69]. Considering the point group $C_{6v}$ of a 2D kagome lattice, the allowed pairing symmetries correspond to the irreps $A_1$, $A_2$ (spin-singlet), $B_1$, $B_2$, $E_1$ (spin-triplet) and $E_2$ (spin-singlet); here we focus on the most representative candidates of unconventional gap functions:

(1) The chiral triplet *p*-wave order ($p + ip$, $E_1$ irrep), see main text Fig. 5a,

(2) The chiral singlet *d*-wave order ($d + id$, $E_2$ irrep), see main text Fig. 5b.

Both (1) and (2) are time-reversal symmetry breaking (TRSB) and nodeless (i.e., exhibiting a fully-opened superconducting gap); both can be the leading pairing instabilities in kagome lattices under certain circumstances [52,68,69]. If the inter-sublattice interaction becomes dominant, a (3) nodal *f*-wave triplet pairing state (irreps $B_1$ or $B_2$, the latter shown in main text Fig. 5c) can be promoted [51,52,68,69]. Alternatively, if the electron-phonon coupling of the Sb *p* orbital (Fermi surface centered at the Γ point of the first Brillouin zone, distinguished from the hexagonal Fermi surfaces of the kagome V 3*d* electrons) is highlighted, the most likely scenarios of superconducting pairing include (4) a sign-changing $s^{+-}$ state (illustrated in main text Fig. 5d) [70].

Among pairing symmetries (1)-(4), we suggest that (3) can be largely precluded, because the *f*-wave superconductivity is nodal [51,52,68,69] with gap zeros occurring on the hexagonal Fermi surface along specific directions (Γ-K for the $B_2$ state in main text Fig. 5c), whereas our STM results demonstrate the absence of corresponding gap nodes (see Supplementary Note 16). Moreover, the most relevant scattering vectors (arrows in main text Fig. 5) inferred from the QPI analysis (main text Fig. 4) connect Fermi surface sections with the same sign of the superconducting order parameter for the *f*-wave pairing (main text Fig. 5c), thus become less effective in breaking Cooper pairs. The chiral *p*-wave state (1) is also less likely, since the triplet pairing ought to be sensitive to point-like disorder [64] and hence yield a prominent pair-breaking effect at the Sn dopants, which is not observed in our STM experiments (Supplementary Note 15).

The remaining two possibilities, i.e., (2) $d + id$ (main text Fig. 5b) and (4) $s^{+-}$ (main text Fig. 5d) pairing states, are more realistic for Cs-135 kagome superconductors [70] and both can explain our main observations at the qualitative level. As illustrated in main text Fig. 5b, the $d + id$ gap function shows a phase winding around the hexagonal Fermi surface, $\Delta(\varphi) = \Delta\exp(i2\varphi)$ ($\varphi$ is the polar angle with respect to the *x* direction) [71]. Consequently, the scattering process at vector $\mathbf{q_1}$ involves Fermi surfaces with remarkable phase shifts of the superconducting order parameter (main text Fig. 5b, except for the $\mathbf{k} \rightarrow \mathbf{-k}$ scattering denoted by the black arrow), effectively resulting in the breaking of Cooper pairs (note that the $\mathbf{k} \rightarrow \mathbf{-k}$ scattering induced by point-like impurities is only weakly pair-breaking [64]). For the $s^{+-}$-wave pairing (main text Fig. 5d), the Fermi surface segments connected by $\mathbf{q_1}$ host superconducting gap with the same sign, whilst the pair-breaking scattering may occur between the original and folded Fermi surfaces in the presence of charge-stripe order (red arrow in main text Fig. 5d); note that the folded bands produced by the $\lambda \sim 3a_0$ stripe order are close to the $p$-orbital Fermi surface located at Γ (main text Fig. 4e and Supplementary Fig. 15), therefore, $\Delta$ at these momenta might change its sign relative to the original kagome bands due to the $p$-$d$ hybridization.

To seek possible phase-sensitive evidence for such sign-changing pairing states, we further performed magnetic-field-dependent and phase-referenced QPI analyses (Supplementary Fig. 22). The dominant scattering vectors resolved in the QPI measurements are summarized schematically in Supplementary Fig. 22a, and the corresponding experimental QPI pattern measured at 2 mV is shown in Supplementary Fig. 22d. Their relation to two representative candidate pairing states, chiral $d + id$ and $s^{\pm}$, is illustrated in Supplementary Fig. 22b and c. Following Ref. [74], we examined the field-induced QPI change, defined as $\delta g(q,E,0.1\text{T}) = g(q, E, 0.1\text{T}) - g(q, E, 0\text{T})$, where sign-preserving and sign-reversing scattering channels are expected to respond differently to magnetic field. As displayed in Supplementary Fig. 22e-h, the field-induced QPI changes at representative in-gap energies are dominated by positive contrast, and no well-defined field-suppressed feature can be assigned to a sign-reversing scattering channel. To assess the possible influence of the set-point effect, we constructed conductance-ratio maps $Z(\mathbf{r}, E) = g(\mathbf{r}, +E)/g(\mathbf{r}, -E)$ for the $x$ = 0.09 sample. As shown in Supplementary Fig. 22i–l, the in-gap QPI features are too weak for a reliable field-dependent analysis. The origin of this weak QPI contrast remains unclear. Previous QPI/BQPI studies of the superconducting state in $CsV_3Sb_5$-based systems have also primarily analyzed $g(\mathbf{q}, E)$ maps [45,75]. Moreover, the normal-state conductance and integrated low-energy DOS in our $x$ = 0.09 sample show weak spatial variations (Fig. 3 and Supplementary Fig. 2), suggesting that spatial variations in the set-point normalization may be limited. We therefore retain the $g(\mathbf{q}, E)$ maps to examine the field response of the resolved QPI channels.

We also calculated the phase-referenced QPI signal based on the complex FT-QPI signal $g(q, E) = |g(q, E)|e^{i\varphi(q, E)}$, where $\varphi(q, E)$ is the Fourier phase. The phase-referenced signal is defined as $g_{PR}(q, E) = |g(q, +E)|\cos[\varphi(q, +E) - \varphi(q, -E)]$, for which sign-reversing scattering channels are expected to give negative contrast [76-78]. As shown in Supplementary Fig. 22m-p, the phase-referenced QPI maps of our data are also dominated by positive contrast, without a well-defined negative signal localized at the relevant QPI vectors. Thus, neither magnetic-field-dependent QPI nor phase-referenced QPI provides direct evidence for sign-reversing scattering channels in the present data.

The absence of such a phase-sensitive QPI signature does not rule out a momentum-dependent phase structure of the superconducting order parameter; instead, it can be attributed to the nature of the QPI channels resolved in $CsV_3Sb_{5-x}Sn_x$. The predominant QPI features in our measurements are mostly intra-pocket scatterings, rather than inter-pocket scatterings connecting two known opposite-sign momenta (which is the case for the previous successful applications of phase-sensitive QPI analysis for high-Tc superconductors). In Supplementary Fig. 22b and c, we illustrate this issue for the two most likely pairing symmetries, i.e., $d + id$ and $s_{\pm}$, respectively; the black double arrow $\mathbf{q}_1$ denote the predominant intra-pocket scattering vector that is captured by our QPI measurements, which *does not* connect Fermi surface regions with opposite signs of the superconducting order parameter. For complex pairing states such as chiral $d + id$ or chiral $p + ip$, or for the possible $s^{\pm}$ scenario discussed above, the dominant in-gap QPI signal $\mathbf{q}_1$ does not necessarily connect Fermi-surface regions with opposite signs of the superconducting order parameter (Supplementary Fig.

22b and c). Therefore, the present QPI data do not provide direct phase information or unambiguous evidence for sign-reversal pairing; determination of the exact superconducting phase structure will require future phase-sensitive experiments and theoretical modelling.

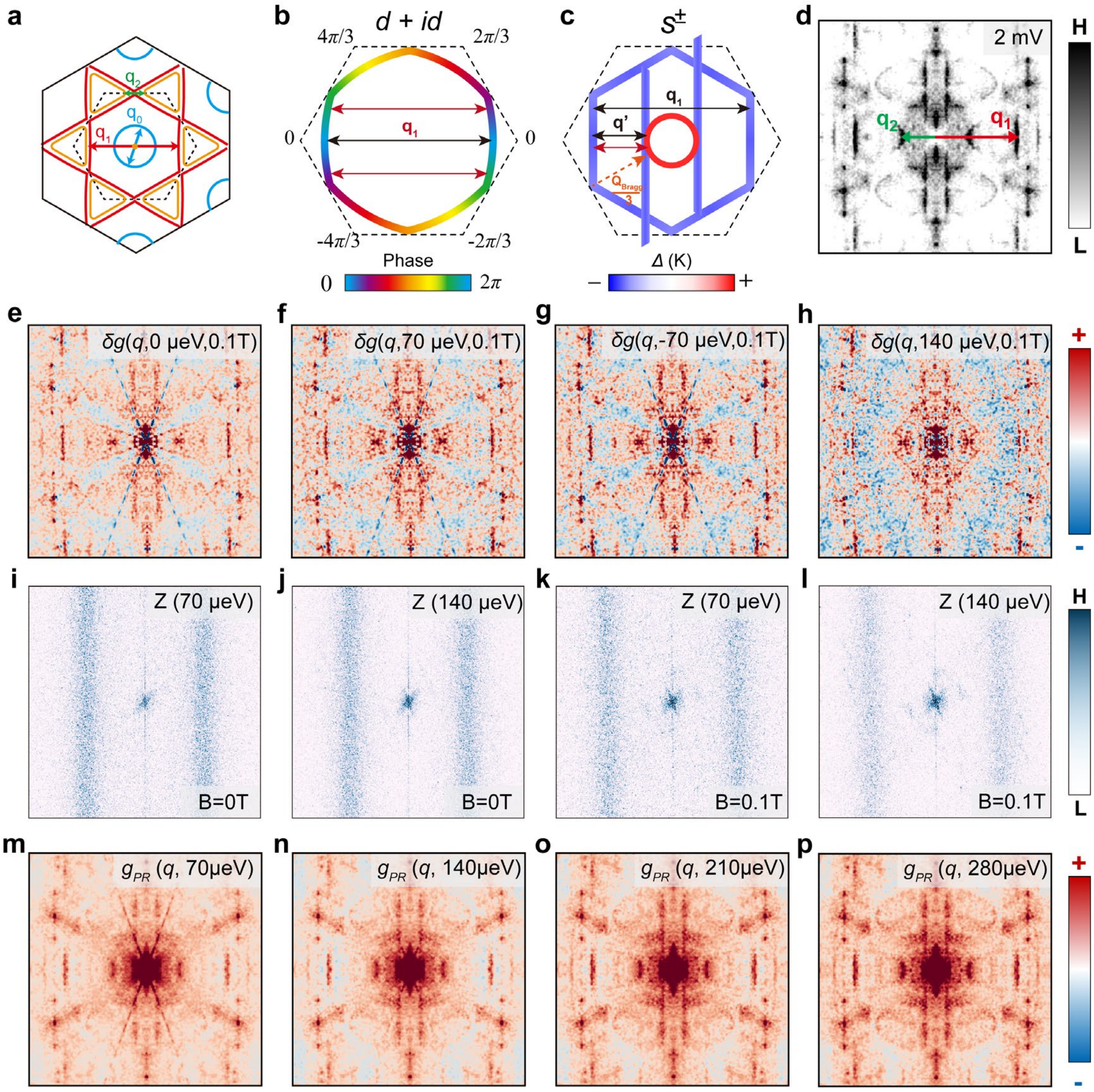

**Supplementary Fig. 22. Magnetic-field-dependent and phase-referenced QPI analyses. a**, Schematic Fermi surface and main scattering vectors resolved in the QPI measurements. **b**, **c,** Illustrations of chiral *d*-wave ($d + id$) pairing state (**b**) and the $s^{\pm}$-wave pairing state (**c**). The double-arrow vectors $\mathbf{q}_1$ correspond to those shown in **a**. **d**, Experimental QPI pattern measured at 2 mV, showing the dominant scattering features $\mathbf{q}_1$ and $\mathbf{q}_2$. **e-h**, Magnetic-field-induced QPI changes $\delta g(q, E, 0.1\ \mathrm{T})$ at representative in-gap energies. The field-induced signals are predominantly positive; no clear field-suppressed feature can be resolved at the relevant QPI vectors. **i, j** Fourier-transformed $Z(\mathbf{q}, E)$ maps of the $x = 0.09$ sample measured at 0 T, with $E = 70$ and 140 μeV respectively. **k, l,** Corresponding $Z(\mathbf{q}, E)$ maps measured at 0.1 T. The QPI features in these $Z$ maps are too weak for a reliable field-dependent comparison **m-p**, Phase-referenced QPI maps $g_{PR}(q, E)$ at representative in-gap energies. The phase-referenced signals are also dominated by positive contrast, without a well-defined negative signal localized at the relevant QPI vectors.

We emphasize that these arguments presented above are completely qualitative and phenomenological. The exact pairing symmetry of superconducting states in charge-stripe ordered $CsV_3Sb_{5-x}Sn_x$ will only be nailed down in future studies exploiting new phase-sensitive techniques. Some theoretical proposals [72,73] also predict the formation of pair density wave (PDW) order in kagome lattices, which possesses Cooper pairs characterized by specific finite momentum $q$ and may have intricate connections with the CDW order in the normal state. Nevertheless, we also mention that the unusual superconducting state in $CsV_3Sb_{5-x}Sn_x$ ($x$ = 0.09) can hardly be associated with the PDW state as discussed in the main text.